# Location Privacy Threats and Protections in 6G Vehicular Networks: A Comprehensive Review

Baihe Ma, Xu Wang, Xiaojie Lin, Yanna Jiang, Zhe Wang, Caijun Sun*, Guangsheng Yu, Suirui Zhu,

Ying He, *Senior Member, IEEE*, Wei Ni, *Fellow, IEEE*, and Ren Ping Liu, *Senior Member, IEEE*

*Abstract*—Location privacy is critical in vehicular networks, where drivers' trajectories and personal information can be exposed, allowing adversaries to launch data and physical attacks that threaten drivers' safety and personal security. This survey reviews comprehensively different localization techniques, including widely used ones like sensing infrastructure-based, optical vision-based, and cellular radio-based localization, and identifies inadequately addressed location privacy concerns. We classify Location Privacy Preserving Mechanisms (LPPMs) into user-side, server-side, and user-server-interface-based, and evaluate their effectiveness. Our analysis shows that the user-server-interface-based LPPMs have received insufficient attention in the literature, despite their paramount importance in vehicular networks. Further, we examine methods for balancing data utility and privacy protection for existing LPPMs in vehicular networks and highlight emerging challenges from future upper-layer location privacy attacks, wireless technologies, and network convergences. By providing insights into the relationship between localization techniques and location privacy, and evaluating the effectiveness of different LPPMs, this survey can help inform the development of future LPPMs in vehicular networks.

## I. INTRODUCTION

Vehicular networks, integral to the advancement of mobile systems like 5G and 6G, play a crucial role in delivering reliable, real-time information for enhancing driving safety and efficiency. Over the years, a broad range of Location-Based Services (LBSs) have been developed to improve the functionality of these networks. These services serve various purposes, including navigation, weather updates, venue locators, social media interactions, and crowd-sensing [1]. By utilizing these LBSs, drivers can access critical information such as traffic conditions, weather forecasts, and the most efficient routes to their destinations. This convenience, however, comes with the trade-off of sharing location data with LBS platforms, which may expose users to privacy risks [2]. As vehicular networks continue to evolve within the 5G and 6G landscapes, balancing the benefits of LBSs with robust privacy protections remains an essential challenge.

### A. Overview of Future Vehicular Networks

The architecture of 5G/6G vehicular networks is designed to facilitate seamless data exchange while supporting multiple communication environments, including ground, underwater, air, and space [3]. Vehicles in these networks frequently share data to access LBSs, which are essential for real-time operations. By incorporating edge computing, the networks are able to perform computational tasks closer to the end-side devices, reducing latency and network load. The architecture of these vehicular networks is illustrated in Fig. 1, demonstrating the integration of various communication modes and components. The core components of this architecture include entities, Vehicle-to-Everything (V2X) communications, and LBSs, as described below:

- **Entity:** In vehicular networks, entities include both infrastructure and vehicles. Infrastructure components such as sensors, Base Stations (BSs), and charging stations act as access points providing services to vehicles. The vehicles, including general vehicles, autonomous vehicles, and Unmanned Aerial Vehicles (UAVs), interact with other vehicles and infrastructure to access LBSs.
- **V2X Communication:** V2X communication enables vehicles to exchange location data with various entities to access LBSs. This communication occurs between Vehicle-to-Infrastructure (V2I), Vehicle-to-Network (V2N), Vehicle-to-Vehicle (V2V), Vehicle-to-Pedestrian (V2P), and Vehicle-to-Device (V2D).
- **LBS:** LBSs collect and store the location data transmitted by vehicles to offer various services, including real-time navigation and traffic management.

5G/6G-enabled vehicular networks are characterized by their ability to interconnect billions of devices, allowing seamless Device-to-Device (D2D) communications across heterogeneous environments. These networks promise significant advancements in communication capabilities, enabling improved data transmission rates, reduced latency, and higher scalability. As a result, it is essential to consider how these features impact the protection of location privacy in vehicular networks, as these networks introduce new challenges for maintaining security in real-time, high-speed environments.

- **High Data Rate:** The data rates in 5G/6G-enabled vehicular networks are projected to be hundreds to thousands

B. Ma (0000-0003-4167-2797), X. Wang (0000-0001-9439-6437), X. Lin (0000-0002-0133-4737), Y. Jiang (0000-0002-8176-6264), G. Yu (0000-0002-6111-1607), Y. He (0000-0003-1603-9375), and R. P. Liu (0000-0001-7001-6305) are with the Global Big Data Technologies Centre, University of Technology Sydney, Australia, 2007 (e-mail: Baihe.Ma@uts.edu.au; Xu.Wang@uts.edu.au; Xiaojie.Lin@uts.edu.au; Yanna.Jiang@uts.edu.au; Guangsheng.Yu@uts.edu.au Ying.He@uts.edu.au; RenPing.Liu@uts.edu.au)

C. Sun (0000-0003-1529-3179) is with Zhejiang Lab, Hangzhou, China, (e-mail: sun.cj@zhejianglab.com)

S. Zhu is with the University of New South Wales, NSW 2033, Australia. (e-mail: suirui.zhu@student.unsw.edu.au);

W. Ni (0000-0003-0780-4637) is with Data61, CSIRO, Sydney, Australia, 2122 (e-mail: Wei.Ni@data61.csiro.au)

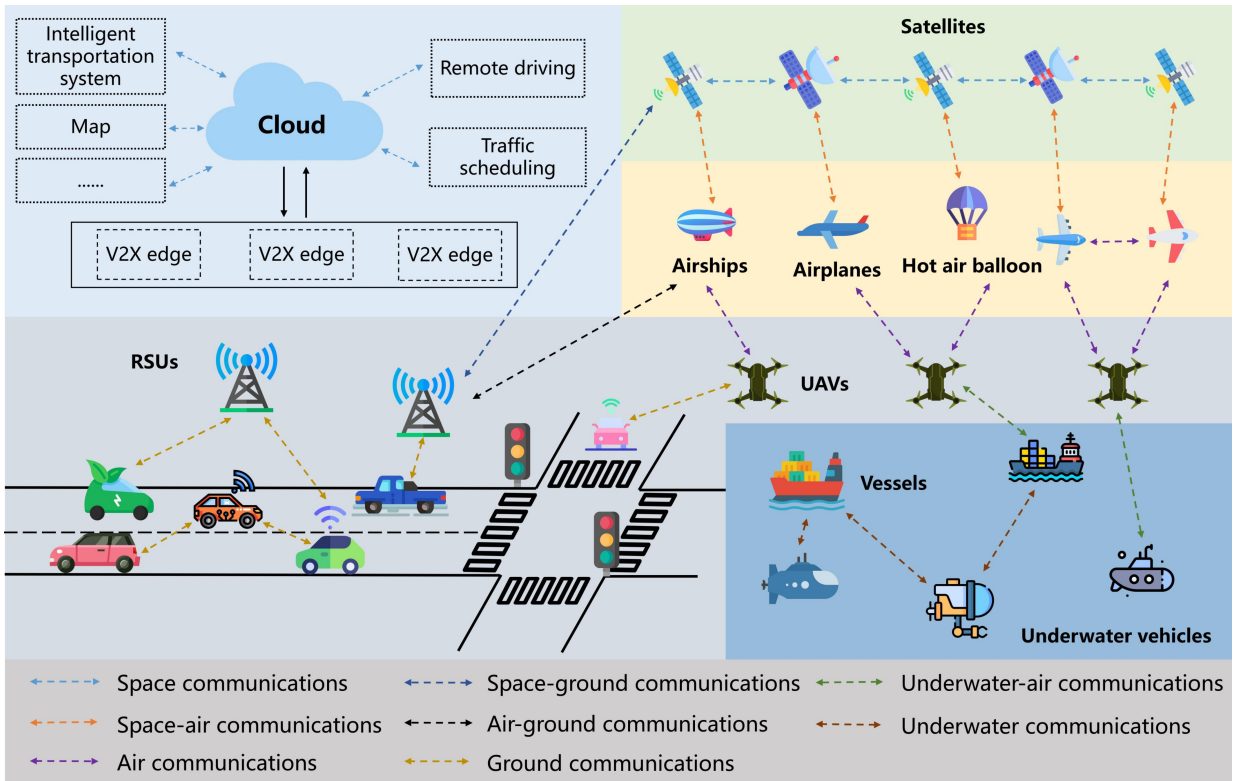

Fig. 1. 5G/6G-enabled Vehicular networks. With data sharing, vehicular networks gain benefits from advanced technologies. Vehicular networks with the Fifth Generation (5G) networks increase communication capacity, reduce communication delay and improve the connectivity of vehicular networks. Artificial Intelligence (AI)-enabled next-generation (6G) networks are proposed for future network intelligentization. Vehicular networks with AI-enabled 6G technologies can offer heterogeneous structures in three-dimensional environments (e.g., space, air, ground, and underwater).

of times faster than current networks [4]. Technologies such as D2D communications, massive Multiple-Input and Multiple-Output (MIMO), and millimeter-Wave (mmWave) are employed to achieve these high rates [5]. However, the increased data rates necessitate LPPMs that require low computational overhead to process location data efficiently.

- **Ultra-Low Latency:** 5G/6G-enabled networks aim to reduce latency to the millisecond or microsecond level, utilizing technologies like D2D communications, Software Defined Networks (SDN), and Cloud Radio Access Networks (C-RAN) [6]. Ensuring that location privacy is protected without adding significant latency is crucial to the functionality of these systems.

- **High Scalability:** With the ability to support a massive number of nodes, such as vehicles and other devices, 5G/6G-enabled networks rely on network function virtualization to scale effectively [7]. Although scalability enables cooperation for enhanced location privacy, detecting malicious nodes becomes increasingly difficult in such large, distributed networks.

To achieve flexibility, security, and low operational costs, 5G/6G-enabled vehicular networks can adopt various network management techniques [3]. These techniques help optimize network performance while ensuring effective handling of the complex communication requirements in modern vehicular systems.

- **SDN:** SDN allows for the decoupling of network control from data management, enabling more efficient and flexible management of vehicular networks. Through its layered architecture, where the control layer manages controllers and the data layer handles data forwarding and protocol processing, SDN facilitates cross-platform, cost-effective solutions for applications such as SDN-enabled vehicle path planning. When integrated with machine learning, SDN enables federated resource allocation across multiple tasks. For example, Gao et al. [8] proposed an architecture combining SDN and multipath techniques to enhance data transmission in high-mobility vehicular networks, while Sadreddini et al. [9] developed an SDN-based routing protocol that reduces link failure and enhances throughput.

- *Network Function Virtualization (NFV):* NFV allows network functions to be virtualized, replacing the need for traditional hardware-based network architectures [10]. Telecommunication Service Providers (TSPs) can benefit from this by eliminating physical middleboxes, thereby streamlining network operations and improving service quality. Although NFV networks face challenges, particularly in resource allocation, the development of 5G and 6G networks is expected to address these issues, enabling more efficient resource management in advanced vehicular networks [11].

- **Network Slicing:** Network slicing enables the creation of multiple virtual sub-networks over shared infrastructure to meet different Quality of Service (QoS) requirements in vehicular networks. Leveraging SDN and NFV, slicing allows for independent management and resource allocation for services like autonomous driving, infotainment, and emergency communications [10], [12]. As 6G networks evolve, network slicing becomes essential to manage the diverse traffic demands, such as Ultra-Reliable Low-Latency Communication (URLLC) and massive Machine-Type Communication (mMTC). This allows for optimized resource use in V2X communication, ensuring dynamic slice adjustments based on real-time data, traffic conditions, and vehicle density, enhancing the efficiency, safety, and cost-effectiveness of future vehicular applications.

### B. Location Data and Location Privacy Concerns

Effectively managing location data is crucial for ensuring both safety and efficiency in vehicular networks. With the integration of advanced communication and sensing technologies, such as V2V communication and cameras, vast amounts of real-time location data are generated continuously. This data is shared among vehicles, infrastructure, and pedestrians, characterized by its large volume, high correlation, dynamic nature, and varying levels of sensitivity [13]. Proper analysis and utilization of this location data are essential for optimizing traffic management, improving road safety, and supporting intelligent transportation systems:

- **Massive Data:** The sheer volume of location data generated by LBSs in vehicular networks creates significant privacy challenges. Each vehicle continuously produces streams of real-time location updates, and improper management of this data could lead to privacy risks. Unauthorized access to such data could expose sensitive information about a driver's habits or routines [14]. To mitigate these risks, advanced encryption techniques and data anonymization methods are needed to ensure that only necessary data is shared while individual identities are protected.

- **High Correlation:** Location data in vehicular networks is highly correlated, meaning that even small data can reveal extensive information. For instance, correlating location points could allow an observer to reconstruct a driver's route or identify frequently visited locations. Protecting privacy in this context requires robust aggregation and obfuscation techniques. Privacy-preserving analytics, such as differential privacy, can be applied to extract useful insights for traffic management without compromising individual drivers' privacy [15].

- **Dynamic Topology:** The ever-changing nature of vehicular networks, with constantly shifting locations as vehicles move, adds further complexity to privacy protection. Continuous updates in location data make it possible to track vehicles in real-time, posing a major privacy threat. Effective strategies to combat this include the use of frequently changing pseudonyms and secure communication channels to prevent eavesdropping [16]. Additionally, the implementation of location privacy zones, where data anonymization or restrictions are applied based on area sensitivity, can further protect real-time location data.

- **Uneven Significance:** Not all location data carry the same level of sensitivity [15]. For example, a driver's home or work address is far more sensitive than data relating to public places such as parking lots. Effective privacy protection must account for this uneven significance by deploying context-aware privacy policies. Such policies dynamically adjust the level of protection based on the sensitivity of the location, with selective sharing mechanisms—such as consent-based or multi-level privacy controls—ensuring that more sensitive information receives greater protection.

- **Driver Safety:** Sharing location data in vehicular networks is primarily aimed at enhancing driver safety, but privacy concerns must be carefully balanced. Real-time data sharing is critical for services like emergency response, naviga- tion, and collision avoidance, but it also presents privacy risks. Therefore, privacy-enhancing technologies, such as secure multi-party computation and blockchain, should be incorporated into data-sharing protocols to ensure that only authorized entities can access the data. This ensures data integrity without compromising privacy, fostering trust and promoting wider adoption of location-based services [17].

*1) Privacy Risks in Vehicular Networks:* Sharing location data in the LBSs of vehicular networks introduces significant privacy risks [15]. While LBSs rely on real-time location data to function effectively, untrusted service providers and potential adversaries pose substantial threats to the privacy and security of this information [13]. Adversaries can exploit vulnerabilities in LBS servers, intercept data through eavesdropping, or collaborate with untrusted servers to gain unauthorized access to location data [15]. These privacy breaches enable adversaries to analyze and misuse the shared data, highlighting the urgent need for robust privacy protection mechanisms in vehicular networks.

- **Exposure of Driver's Private Information:** Adversaries can use location data transmitted in vehicular networks to infer sensitive personal details, such as a driver's home address, religious affiliations, political beliefs, occupation, and workplace [18]. For instance, frequent visits to religious institutions or political offices can reveal private information about a driver's beliefs and affiliations. Regular trips to specific addresses may also expose personal details like home or work locations, which can then be used to profile individuals or groups. These profiles can be exploited for identity theft, discrimination, or targeted harassment. Moreover, adversaries can use these profiles to conduct personalized attacks, disrupt the activities of specific groups, or even sabotage entire vehicular network systems. Such unauthorized access to private information undermines trust in LBSs and vehicular networks, potentially reducing user engagement and impairing the functionality of these services.

- **Physical Attack:** Breaches in location privacy can lead to real-world physical threats, such as stalking, burglary, or even harm to the driver [19]. For example, if an adversary is aware that a driver is away from home, they may target the driver's property for burglary. Furthermore, attackers could manipulate navigation systems to mislead drivers into vulnerable areas, increasing their risk of harm. A notable example includes the spoofing of navigation data to create fake traffic jams or reroute drivers into dangerous locations[1]. Without adequate privacy protections, location data becomes a powerful tool for orchestrating physical attacks.

- **Exposure of Other Vehicles' Private Information:** Sharing location data in vehicular networks not only risks the privacy of individual drivers but can also compromise the privacy of other vehicles on the road. Adversaries can analyze shared data to track the trajectories and predict the movements of other vehicles, including sensitive or emergency vehicles [20]. For example, by studying encounter histories and movement patterns, adversaries could predict convoy routes, emergency vehicle paths, or even the movements of high-profile individuals, potentially endangering personal privacy and public safety. The ability to track or anticipate the movements of critical vehicles poses a significant risk to both individual and collective security.

*2) Challenges in Location Privacy in 5G/6G Vehicular Networks:* The rapid evolution of Vehicular Networks (VNs) introduces significant challenges for existing LPPMs. As VNs become more advanced, their architecture grows increasingly complex, leading to a greater volume of data exchanged across different network layers. This heightened complexity presents new opportunities for adversaries to exploit vulnerabilities, particularly through sophisticated cross-layer attacks. Traditional LPPMs, typically designed to address threats within a single network layer, are no longer sufficient for these emerging threats. Below are the primary challenges that must be addressed:

- **Complex Network Architecture:** Modern vehicular networks operate across multiple layers, including the physical, network, and application layers. This multi-layered structure amplifies the potential for adversaries to exploit vulnerabilities at various levels simultaneously.
- **Cross-Layer Attacks:** Adversaries can execute cross-layer attacks by combining information from different layers, leading to more accurate and effective location tracking. Traditional LPPMs, designed to protect a single layer, are ineffective against such coordinated, multi-layered threats.
- **Ineffectiveness of Traditional LPPMs:** Current LPPMs are not equipped to manage the complexities of cross-layer attacks. These mechanisms often fail to provide comprehensive protection, leaving vehicular networks exposed to privacy breaches from coordinated attacks spanning multiple layers.
- **Spyware and Location Tracking:** Spyware tools, like Pegasus, can access historical encounter data. When combined with real-time intercepted information, adversaries can achieve highly precise vehicle tracking, bypassing the existing LPPMs [21].
- **Increased Data Exchange in V2X Communications:** The

growing use of V2X communication in 5G/6G vehicular networks results in more frequent and extensive data sharing among vehicles, infrastructure, and other entities, increasing the risk of location privacy breaches.

- **Need for Multi-Layered Privacy Protection:** To address the emerging threats in 5G/6G vehicular networks, LPPMs must evolve to offer robust, multi-layered privacy protection. This includes the integration of advanced encryption, context-aware privacy controls, and real-time cross-layer anomaly detection systems.
- **Dynamic Adaptation to Evolving Threats:** Future LPPMs must be capable of dynamically adapting to real-time threats in highly dynamic vehicular environments. These mechanisms should effectively counter new types of cross-layer attacks, ensuring privacy protection in increasingly complex network scenarios.

*C. Contribution*

This survey investigates the impact of localization technologies and related applications on location privacy. Particularly, the threats of localization techniques at different network layers and related applications to the existing LPPMs have not been well considered in the existing studies, e.g., [22], [24]–[28]. As shown in Table I, existing works have only focused on LPPMs in the context 5G/6G vehicular networks. However, the balance between location privacy and data utility is important for the designs of LPPMs but has been to date overlooked in the literature, e.g., [22]–[29], with a detailed comparison provided in Table I.

In light of this, we review the latest localization and tracking methods and their potential threats to location privacy in vehicular networks. We also conduct an in-depth survey of the existing LPPMs in vehicular networks, and summarize their limitations in the face of the localization and tracking techniques from the aspects of the communication and localization requirements.

With a focus on the location privacy of vehicular networks, the key contributions of this survey are listed as follows.

1. We analyze existing LPPMs across user-side, server-side, and user-server interfaces within diverse 5G/6G localization and tracking paradigms, including considerations for multi-source data and cross-layer vulnerabilities.

2. We identify a new challenge: Adversaries can leverage multiple tracking data from different layers to evade the location privacy of vehicles. This threat is increasingly likely and devastating due to the proliferation of location-based applications and social network platforms. We provide a detailed examination of how adversaries might exploit multiple data sources and how upper- and lower-layer localization techniques can be exploited to compromise the location privacy of vehicles.

3. We also identify a substantial gap between theory and practical applications concerning location privacy. This is accomplished by analyzing recent literature and location-based apps. This gap arises primarily from the distinct reliability levels of data at various layers: Practical apps favor lower-layer data, while theoretical approaches often rely on upper-layer data. Addressing this gap requires a holistic approach that considers the integration and protection of data across different layers and sources.

---

[1] https://simonweckert.com/googlemapshacks.html

TABLE I
Comparison of related works.

| Reference | Localization | User-side[1] | Existing LPPMs | | | LPPMs in Future VNs[4] | Balance[5] |
|---|---|---|---|---|---|---|---|
| | | | Server-side[2] | User-server-interface[3] | Limitations | | |
| [22] | | | ✓ | | | | |
| [23] | | ✓ | ✓ | | ✓ | | |
| [24] | | | ✓ | | ✓ | | |
| [25] | | ✓ | ✓ | ✓ | ✓ | | |
| [26] | ✓ | | ✓ | | ✓ | ✓ | |
| [27] | | ✓ | ✓ | ✓ | | ✓ | |
| [28] | ✓ | | | | | | |
| [29] | | ✓ | ✓ | | | | |
| [30] | | ✓ | ✓ | | ✓ | | ✓ |
| [31] | | ✓ | ✓ | | ✓ | | |
| [32] | | | | | ✓ | ✓ | |
| [33] | | | | | ✓ | ✓ | |
| [34] | | | | ✓ | ✓ | | |
| [35] | ✓ | ✓ | ✓ | | ✓ | | |
| [36] | ✓ | ✓ | ✓ | ✓ | ✓ | | ✓ |
| [37] | ✓ | ✓ | ✓ | ✓ | ✓ | | |
| **This paper** | ✓ (90) | ✓ | ✓ | ✓ | ✓ | ✓ (28) | ✓ (80) |

[1] User-side: pass and run, certificate, secure computation, and data perturbation;
[2] Server-side: statistical disclosure control, homomorphic encryption, private information retrieval, and searchable encryption;
[3] User-server-interface: secure communication and trusted parties;
[4] VNs: Vehicular Networks;
[5] Balance: Balance between Location Privacy and Data Utility.

4. We highlight the challenges arising from the relationship between temporal and spatial information in trajectory data. Temporal information in trajectories allows for combining data from various layers and sources, which can lead to precise tracking and reconstruction of location information and breach location privacy. Our analysis underlines the need for robust privacy protection strategies that safeguard both location specifics and temporal information.

The rest of this paper is organized as follows. We compare this paper with the existing surveys and evaluate their contributions in Section II. We list the existing tracking techniques, threats and location privacy requirements in Section III. The existing LPPMs and the balance between location privacy and data utility are illustrated in Section IV, followed by the future location privacy challenges in Section V. We conclude our work in Section VI. The structure of this paper is shown in Fig. 2, with acronyms and abbreviations used listed in Table II.

## II. Existing Surveys and Gap Analysis

In this paper, we examine recent studies, particularly those published by IEEE and ACM over the past five years, focusing on the development of Location Privacy-Preserving Mechanisms (LPPMs) in vehicular networks. Our analysis is centered on localization technologies and explores the trade-off between protecting location privacy and maintaining data utility. This section compares our work with previous literature reviews [22]–[37], published between 2018 and 2024, to evaluate their contributions. As summarized in Table I, the comparison focuses on four key areas: localization techniques, existing LPPMs, future LPPMs in vehicular networks, and the balance between location privacy protection and data utility.

This paper reviews the localization techniques according to the development of vehicular networks and discusses localization techniques from the aspect of location privacy. Wang et al. [28] introduced future localization techniques with applications in 5G and 6G networks. /Hussain et al. [26] discussed the localization methods based on the architecture of 5G vehicular networks and attacks.

In this paper, the existing LPPMs are divided into user-side, server-side, and user-server-interface LPPMs according to the phase that the LPPMs are allocated, by considering the threats of localization techniques [38]. The user-side LPPMs allow the drivers to process data before sending it to the LBSs, while the server-side LPPMs process the aggregated location data in the dataset. The user-server-interface LPPMs use trusted third parties and secure communication to ensure that the location data is secure in transmission. We comprehensively compared related works of each category. The published surveys [22], [23], [29], [30] mainly focus on the existing user-side and server-side LPPMs in current vehicular networks.

Lu et al. [22] elaborated on the statistical disclosure control-based LPPMs, e.g., anonymity, in current vehicular networks, but other techniques are not discussed. Similarly, authors in [24] paid attention to the pseudonyms changing approaches of statistical disclosure control. The authors compared the cost of different pseudonyms-based LPPMs. Sheikh et al. [29] compared the user-side and server-side LPPMs, e,g, certificate, secure computation, and statistical disclosure control, under different adversaries and delineated the advantages and limitations of these LPPMs. The authors listed the location privacy issues and potential applications under the risk of adversaries in existing vehicular networks. Nevertheless, the study does not discuss the LPPMs in future vehicular networks.

By considering the same scenario, Ali et al. [23] classified the LPPMs into pseudonymous-based, group-based, hybrid, anonymous-based, and cryptography-based. The authors re-viewed the location privacy requirements, attacks, and the efficiency of each LPPM. The authors also compared the capability of each LPPM to resist different attacks, but the user-server-interface LPPMs were not well considered. In the same topic, Mundhe et al. [31] presented an overview of current vehicular networks and summarized the existing user-side and server-side LPPMs. The authors evaluated the advantages and limitations of the existing LPPMs in current vehicular networks, especially the cryptography-based LPPMs. Talat et al. [25] compared the location privacy-preserving capability of the LPPMs. The authors summarized that each LPPM could provide higher location privacy-preserving capability than its counterparts in specific cases. By focusing on applications of LBSs, the study in [30] discussed user-side and server-side LPPMs and qualitatively analyzed their performances.

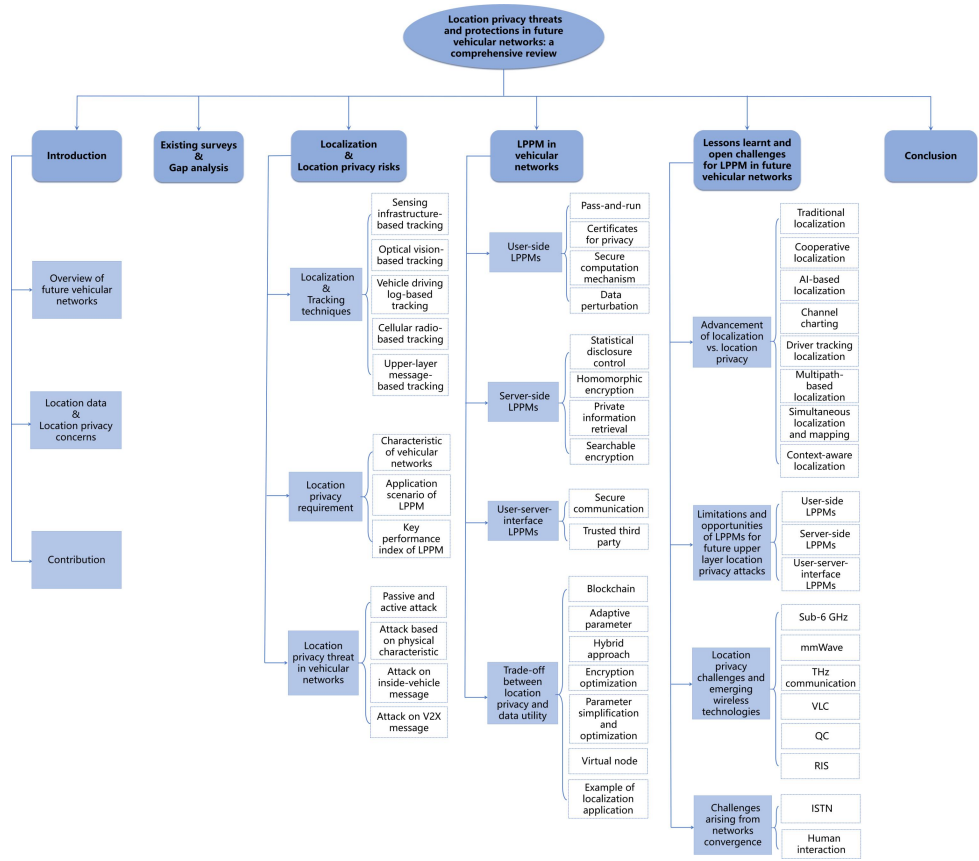

Fig. 2. The organization of this paper. Section I: the introduction of this paper; Section II: the existing surveys and gap analysis; Section III: the existing tracking techniques, threats and location privacy requirements; Section IV: the existing LPPMs and the balance between location privacy and data utility; Section V: the future location privacy challenges.

In this paper, we assess the advantages and limitations of the existing LPPMs from the aspects of localization requirements and future communication technology in vehicular networks.

In contrast, Ali et al. [23], [39] and Mundhe et al. [31] discussed the benefits and shortcomings of the existing user-side and server-side LPPMs according to their methodologies. Jiang et al. [30] highlighted the limitations of the existing LPPMs to the requirements of different LBSs. Boual et al. [24] evaluated the anonymity-based LPPMs, i.e., server-side LPPMs, with syntactic linking and semantic linking in existing vehicular networks. By considering more kinds of attacks and LPPMs, Talat et al. [25] compared the existing LPPMs in existing vehicular networks under various attacks, including denial of information/services, breach of information privacy, masquerading, and data modification. Hussain et al. [26], Dibaei et al. [32], Boualouache et al. [33], and Haddaji et al. [34] presented potential challenges to the existing LPPMs in 5G vehicular networks. Belal et al. [35] explored the integration of federated learning into vehicular networks to enhance LPPMs, while Rasheed et al. [36] systematically investigated privacy-preserving mechanisms focusing on the trade-off between location and query privacy. Mianji et al. [37] reviewed the role of multi-agent reinforcement learning in privacy and security for vehicular networks. Based on the listed threats, the authors discussed the performances of the existing LPPMs.

TABLE II
SUMMARY OF ABBREVIATIONS

| Abbreviation | Definition |
|---|---|
| 2D | Two-Dimensional |
| 3D | Three-Dimensional |
| 5G | The Fifth Generation (5G) networks |
| 6G | Artificial Intelligence (AI)-enabled next-generation |
| AI | Artificial Intelligence |
| AP | Access Point |
| AVI | Automatic Vehicle Identification |
| BS | Base Station |
| CAN | Controller Area Network |
| CSI | Channel State Information |
| D2D | Device-to-Device |
| ECU | Electronic Control Units |
| FHE | Fully Homomorphic Encryption |
| Geo-I | Geo-Indistinguishability |
| GNSS | Global Navigation Satellite System |
| GPA | Global Passive Adversaries |
| GPS | Global Positioning Systems |
| HE | Homomorphic Encryption |
| ID | Identification |
| LBSs | Location-Based Services |
| LED | Light-Emitting Diode |
| LPA | Local Passive Adversaries |
| LPPMs | Location Privacy Preserving Mechanisms |
| MAC | Medium Access Control |
| MIMO | Multiple-Input Multiple-Output |
| mmWave | millimeter-Wave |
| NFV | Network Function Virtualization |
| OBD | On-Board Diagnostic |
| OBU | On-Board Units |
| PHE | Partial Homomorphic Encryption |
| PIR | Private Information Retrieval |
| QC | Quantum Communication |
| QoS | Quality of Service |
| RFID | Radio Frequency Identification |
| RIS | Reconfigurable Intelligent Surface |
| RSU | Road-Side Units |
| SDC | Statistical Disclosure Control |
| SDN | Software Defined Networks |
| SE | Searchable Encryption |
| SHE | Somewhat Homomorphic Encryption |
| SLAM | Simultaneous Localization And Mapping |
| THz | Terahertz |
| TPMS | Tire Pressure Monitor System |
| TTP | Trust Third Party |
| UE | User Equipment |
| VN | Vehicular Networks |
| V2D | Vehicle-to-Device |
| V2I | Vehicle-to-Infrastructure |
| V2N | Vehicle-to-Network |
| V2P | Vehicle-to-Pedestrian |
| V2V | Vehicle-to-Vehicle |
| V2X | Vehicle-to-everything |
| VLC | Visible light communication |

In this paper, we also investigate the potential developments and cross-layer challenges of existing LPPMs in future vehicular networks from the aspect of future communication requirements in 5G and 6G vehicular networks. Hussain et al. [26] researched current privacy issues and solutions in 5G-enabled vehicular networks. An architecture for the 5G-enabled vehicular networks and potential machine learning-based LPPMs was developed by the authors to support 5G applications. Lu et al. [27] examined location privacy issues in the 5G V2X architecture. They analyzed the existing single-layer attacks and explored the future directions of LPPMs in 5G vehicular networks. By considering the technologies in future vehicular networks, Dibaei et al. [32] addressed the limitations of the existing LPPMs, based on which potential directions of future LPPMs were presented with blockchain and machine learning. Boualouache et al. [33] surveyed the existing server-side LPPMs, which were employed to protect location privacy by detecting misbehavior in future 5G and beyond vehicular networks. The authors presented potential developments of the AI-enabled LPPMs in future vehicular networks. Haddaji et al. [34] summarized the location privacy in AI-enabled vehicular networks. The authors covered the privacy and security issues in future vehicular networks and discussed AI-based solutions within future frameworks.

This paper further delineates the trade-off between location privacy and data utility with theory and practice examples, which is important to LPPM designing. In the compared surveys, only Jiang et al. [30] discussed methods to balance location privacy and data utility. They provided the trade-off based on the methodologies of different LPPMs in the existing LPPMs. This paper discusses the trade-off in theory and practice from the aspect of localization requirements and future communication techniques in vehicular networks.

### A. Key Takeaway

From the comprehensive analysis of existing studies, several critical insights have been identified regarding developing and applying LPPMs in vehicular networks as follows:

. First, while significant advancements have been made in user-side and server-side LPPMs, the lack of their integration in seamless user-server-interface mechanisms stands as a gap. The fragmented nature diminishes the ability to address holistic privacy and security concerns, especially in complex vehicular network environments.

. Second, most of the current studies focus on investigating the requirements and challenges involving existing network architectures, ignoring the cross-layer threats introduced by emerging 5G and 6G technologies, together with their advanced attack vectors. In other words, in the near future, much emphasis should be given to conducting studies that will lead toward adaptive and scalable LPPMs based on evolving technology.

. Third, although the balance between location privacy and data utility was identified by many as one of the important trade-offs, it cannot be developed thoroughly yet. Most discussions about such issues are theoretical studies with poor practical validation. They all use ad hoc metrics which might not be appropriate and relevant to different application situations.

### III. LOCALIZATION AND LOCATION PRIVACY RISKS

This section provides a summary of existing tracking techniques in various vehicular network scenarios from the perspective of location privacy, as shown in Table III. We then outline the key localization requirements, introduce common adversary models, and discuss potential privacy threats currently affecting vehicular networks. By examining the threats posed by localization technologies, we present the adversaries that have been well-documented in the literature.

### A. Localization and Tracking Techniques

In recent decades, various tracking techniques have been developed to improve the functionality of high-precision LBSs and anti-theft systems. These techniques offer real-time monitoring and navigation capabilities, which are vital for applications such as route optimization, emergency response, and stolen vehicle recovery [57]. However, despite their benefits, these same tracking methods can be exploited by adversaries to compromise drivers' location privacy.

Advanced tracking methods, such as Global Navigation Satellite System (GNSS) spoofing, cellular triangulation, and in-vehicle sensor data exploitation, enable adversaries to accurately deduce a driver's real-time location, daily routes, and habitual patterns. The misuse of these technologies presents significant privacy risks, allowing malicious actors to conduct surveillance, targeted attacks, or identity theft. To counter these threats, several LPPMs have been developed. These mechanisms utilize techniques like data anonymization [16], obfuscation [15], pseudonymization [58], and cryptographic protocols [59] to obscure or distort the precise location information exchanged within vehicular networks. However, as modern vehicular networks become more dynamic and interconnected, new vulnerabilities, such as cross-layer attacks, emerge, reducing the effectiveness of traditional LPPMs. As a result, there is a growing need for more robust and adaptive LPPMs that can provide comprehensive privacy protection against advanced tracking threats while balancing the need for data accuracy and utility in legitimate LBSs and vehicular applications.

Tracking techniques in vehicular networks can be classified into five major categories: sensing infrastructure-based, optical vision-based, vehicle driving log-based, cellular radio-based, and upper-layer message-based tracking. While most Location LPPMs focus on mitigating threats from upper-layer message-based tracking, such as protecting against attacks that exploit shared vehicular data packets, they often overlook the risks posed by vehicle driving log-based tracking. This form of tracking leverages data recorded directly by in-vehicle systems, such as Event Data Recorders (EDRs) or telematics systems, to reconstruct a vehicle's route, speed patterns, and stop locations, thereby exposing sensitive information without the need for external data interception.

Vehicular networks face significant privacy challenges due to the emergence of advanced tracking techniques that current LPPMs are not equipped to address. In particular, sensing infrastructure-based, optical vision-based, and cellular radio-based tracking methods present serious risks to location privacy. For example, roadside units with sensors and cameras can capture and analyze vehicle positions in real time, while optical vision-based systems, using surveillance cameras and computer vision algorithms, can track vehicles over large areas. Similarly, cellular radio-based tracking leverages mobile network signals to triangulate vehicle locations with high accuracy. Since these methods do not rely on upper-layer vehicular communications, they are challenging to detect or mitigate with traditional LPPMs. The widespread visibility of vehicles to these infrastructure-based tracking techniques underscores a critical gap in existing privacy protections, highlighting the need for innovative strategies to counter these increasingly pervasive threats.

*1) Proximity Detection:* Sensor-based tracking techniques utilize the physical characteristics of vehicles, detected through devices such as inductive loops and magnetic sensors, to monitor vehicle trajectories. These methods are primarily intended to improve traffic management and safety; however, they also pose significant risks to location privacy when exploited by adversaries. The data collected from these sensors can be misused to continuously track specific vehicles, leaving drivers vulnerable to privacy breaches, targeted surveillance, and potential physical threats.

- **Magnetic Sensor:** Magnetic sensors detect disruptions in the Earth's magnetic field caused by passing vehicles, enabling the determination of a vehicle's 3D position and 2D orientation. These sensors are not only used to identify vehicle presence but can also analyze travel times and even perform vehicle identification by mapping magnetic perturbations. Such tracking methods, including the localized magnetic perturbation models proposed by Feng et al. [60], offer cost-effective and environmentally friendly solutions. However, they also pose privacy risks, as they can uniquely identify vehicles based on magnetic signatures. Adversaries may exploit these systems to trace movement patterns or predict future locations, particularly when data from multiple sensors is combined. Su et al. [61] addressed issues like false positives and detection errors by using adjacent sensor data and graph optimization techniques, but these methods could also enhance the precision of tracking if exploited maliciously, increasing the risks to privacy by creating more accurate movement profiles.

- **Inductive Loop:** Inductive loop detectors, widely used in traffic management systems, collect data on vehicle speed, volume, and size. While essential for real-time traffic monitoring, these systems have proven to be effective across various regions, with advancements in vehicle queue modeling further enhancing their utility [62]. However, their proximity to vehicles and frequent installation at critical traffic points also makes them potential targets for privacy violations if compromised. Gheorghiu et al. [63] point out that while the susceptibility of these detectors to damage from traffic or roadwork poses maintenance challenges, it also creates vulnerabilities that adversaries could exploit to intercept unprotected data. Improper access to inductive loop data could allow malicious actors to track specific vehicles over time, exposing sensitive information regarding driver routines, frequented locations, and travel patterns. This information could be misused for malicious purposes, including stalking or planning criminal activities.

- **Tag and Signal Reader:** Techniques that leverage data from Automated Vehicle Identification (AVI) systems, Radio

TABLE III
Existing localization techniques.

| Technology | Description | Key Features | Advantages | Challenges |
|---|---|---|---|---|
| Magnetic Sensor | Catches vehicular magnetic field disturbances [40] | 3D position and 2D orientation detection [40] | Cost-effective, eco-friendly | Speed resolution limits; external field susceptibility |
| Inductive Loop Detector | Gathers vehicular traffic data [41] | Integrated in roads; measures speed, volume, size [41] | Proven utility | Maintenance; environmental effect |
| Tag and Signal Reader | Uses AVI, RFID, Wi-Fi/blacktooth MAC signals [42]–[45] | Vehicle localization via signal analysis [42]–[45] | Uses residential Wi-Fi for tracking | Privacy risks; RFID necessity |
| Optical Vision | Localizes vehicles using camera data [46] | High precision in diverse environments [47] | Advanced ML/AI tracking | Visual dependency; high computational load |
| ISAC | Radar-like cellular signal use for detection [48] | Passive detection; dual-purpose signals [48] | Innovative localization method | Privacy; unauthorized tracking risk |
| Vehicle-based Sensor | Accesses vehicular control unit data [49], [50] | Data from various sensors [51] | Sensor data amalgamation | Data breach risks; sensor dependence |
| Mobile-based Sensor | Uses GPS in mobiles for tracking | Side-channel attack facilitation | Enhances external localization | Privacy concerns; data transmission reliance |
| In-Vehicle Services | Enhances driving and experience [3] | Generates geolocation data | Essential for modern vehicles [3] | Privacy and security concerns |
| ToA | Locates vehicles via signal time delay | Used with vehicle anchors [52] | Error reduction in location estimation [53] | Transmission schedual; environmental factors |
| TDoA | Uses time differences in signal arrival at multiple receivers to locate RF sources [54] | Enhanced 3D localization accuracy [55] | Robustness in various scenarios | Limited by synchronization needs, signal interference, and computational demands |
| RSS | Based on signal strength attenuation | Near-field tracking [56] | Accurate near-field localization [56] | Environmental effects; high attenuation |
| Radio Fingerprint | Identifies unique radio wave signatures [56] | Hardware-specific signal analysis [56] | Database creation for RSS analysis | Privacy concerns; hardware limitation |
| DoA | Triangulates position from signal angles [52] | Precise 2D positioning | Accurate localization [52] | Large antenna arrays; signal processing complexity |

Frequency Identification (RFID) tags, and Wi-Fi/Bluetooth Media Access Control (MAC) signals are essential for vehicle localization but also introduce serious privacy risks when exploited by adversaries [42], [45]. The proliferation of home Wi-Fi networks has added a new layer to these risks. As vehicles pass through urban environments, they may be detected by numerous residential Wi-Fi devices, which broadcast their MAC addresses. This widespread connectivity allows adversaries to use the dense network of Wi-Fi devices to accurately track a vehicle's location. By monitoring the sequence and timing of MAC address broadcasts, they can reconstruct the vehicle's route, identify frequently visited locations, and predict future movements, thereby compromising the driver's privacy.

Near-field communication (NFC) devices, such as RFID tags, also pose tracking risks when densely deployed. Liu et al. [64] demonstrated how RFID readers can determine the location of tags by analyzing back-scattered signal phases.

While this method is highly effective for localization, it can be maliciously employed to track vehicles as they pass through multiple RFID readers along their routes. When combined with other data sources, such as video surveillance or GPS data, RFID-based tracking can provide even more detailed information about a vehicle's movements, increasing the privacy threat.

The privacy risks associated with MAC address tracking are particularly concerning due to the continuous broadcast of these unique identifiers. Spiess [65] highlighted how adversaries could exploit MAC address broadcasts to track vehicles over time, enabling unauthorized surveillance and potentially leading to targeted attacks or coercive actions. Liu et al. [66] further elaborated on vulnerabilities in using Wi-Fi MAC time slots for tracking vehicles. By observing the timing and sequence of broadcasts, adversaries can effectively locate and monitor vehicles without direct access to onboard systems. While some approaches suggest using pseudonym-changing schedules to disrupt continuous tracking, adversaries can still aggregate data and adapt to these changes using sophisticated pattern recognition techniques. Although these localization technologies offer significant benefits for applications such as traffic management, toll collection, and fleet monitoring, they also pose substantial risks to location privacy when misused. The potential for continuous, high-precision tracking through AVI, RFID, and Wi-Fi/Bluetooth MAC data calls for stronger privacy-preserving protocols, regulatory oversight, and technological safeguards to prevent unauthorized data collection and misuse. Furthermore, the development of advanced anonymization methods and encrypted communication protocols is critical in mitigating the growing threats to location privacy.

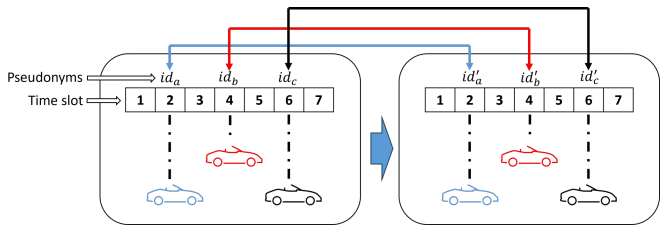

Fig. 3. An example of the MAC layer context data-based tracking attack [66].

- **Optical Vision:** Optical vision-based localization systems rely on camera-captured geometric frame data to accurately determine vehicle positions, even in environments with low texture or poor visibility [46]. Although these systems are highly effective for traffic management and monitoring, their ability to continuously track vehicles poses significant privacy risks when misused. By leveraging advanced machine learning (ML) and artificial intelligence (AI) techniques, these systems can extract vehicle appearance features, such as color and texture, and integrate this data with semantic information to enhance tracking precision [47].

This capability allows adversaries to track vehicles across multiple camera networks or infer vehicle identities based on distinctive characteristics. For example, Yang et al. [67] developed a collaborative sensing system that uses metric learning to re-identify vehicles by combining appearance data with traffic network connections, achieving highly accurate tracking. If maliciously exploited, such systems could enable continuous surveillance, allowing adversaries to monitor a target vehicle over extended distances or complex environments, thereby severely compromising location privacy.

*2) Cellular Sensing and Tracking:* For effective tracking and localization in vehicular networks, the integration of compact, omnidirectional sensor arrays with beamforming techniques offers significant advantages by capturing signals from multiple directions simultaneously, thus improving localization accuracy [68]. However, omnidirectional antennas face challenges, such as signal loss over long distances, limiting their utility in applications that require extended communication ranges. To overcome these challenges, researchers have adopted directional high-gain antennas, which concentrate signal energy in a specific direction, thereby improving the signal-to-noise ratio (SNR) and supporting more reliable long-range communication.

In addition to advancements in antenna design, signal-processing methods have been developed to enhance vehicular localization. For instance, Burghal et al. [69] proposed a localization approach that uses channel state information (CSI) from multiple-antenna transceivers, combined with feed-forward neural networks. This technique reduces the number of trainable parameters, resulting in efficient and accurate localization without requiring significant computational resources.

Millimeter Wave (mmWave) communications further advance vehicle localization capabilities, as phased arrays with electrically steerable directivity enable precise direction-finding [70]. Base stations equipped with mmWave antennas can employ multiple direction-finding algorithms, which, when used in a cross-bearing manner, allow for highly accurate vehicle triangulation. Cellular radio-based tracking techniques, such as Time Difference of Arrival (TDoA) and Received Signal Strength Indicator (RSSI), also contribute to localization by leveraging the widespread infrastructure of cellular networks. These methods are highly effective in both urban and rural environments, offering robust localization capabilities due to the extensive coverage provided by cellular networks.

- **Time of Arrival (ToA):** Time of Arrival (ToA) localization techniques determine a vehicle's position by measuring the time delay of signals transmitted from the vehicle to multiple anchor points. Although these methods are effective for determining vehicle locations, they pose significant privacy risks when exploited by adversaries. By analyzing ToA measurements and solving nonlinear equations based on time delays, adversaries can estimate a vehicle's precise location, particularly if they have prior knowledge of the vehicle's transmission schedule. With multiple reference points, this information allows for triangulation, enabling adversaries to track the vehicle's movements.

However, the accuracy of ToA-based localization can be impacted by various factors, such as errors in range measurements and suboptimal anchor placements. Additionally, inaccuracies may arise if incorrect assumptions about the transmission schedule are made. Rao et al. [53] addressed this issue by developing a method to minimize estimation errors, utilizing the correlation between error magnitude and distance to limit these errors within acceptable thresholds. Despite these enhancements, ToA-based tracking remains a privacy threat, as it requires relatively low-cost hardware and is resilient to signal attenuation over medium distances, making it accessible to adversaries.

Moreover, since ToA localization depends on knowledge of transmission schedules, any leakage of such information can lead to persistent and highly accurate tracking, severely compromising driver privacy. To counter these risks, it is essential to implement countermeasures, including secure transmission protocols and dynamic randomization of transmission schedules, to protect against the misuse of ToA-based localization by malicious actors.

- **TDoA:** TDoA is an effective method for locating Radio Frequency (RF) sources by using three or more synchro-

nized receivers to capture In-phase/Quadrature (I/Q) data and analyze the time differences in signal arrival across these receivers. The location of the signal source can be pinpointed at the intersection of the calculated distance curves derived from each receiver. While TDoA is highly accurate for localization, it also introduces significant privacy risks when exploited by adversaries, who can use these techniques to track vehicles or other targets without consent.

In a study by Sinha et al. [54], the authors explored the limitations of 3D localization and the influence of antenna characteristics on its accuracy. By deploying multiple RF sensors equipped with single or multiple dipole antennas, they extracted both Time of Arrival (ToA) and TDoA information, improving localization accuracy by accounting for varying signal radiation patterns in both line-of-sight (LoS) and non-line-of-sight (NLoS) scenarios. Although these advancements increase the robustness and precision of 3D localization, they also enhance the potential for misuse by adversaries. Malicious actors can use these techniques to track vehicles or even unmanned aerial vehicles (UAVs) by capturing RF emissions and calculating their location through TDoA, even in complex urban environments.

The ability to leverage TDoA techniques for tracking vehicles poses a severe threat to location privacy, as it enables adversaries to perform covert surveillance over large areas with minimal infrastructure. Moreover, by combining data from multiple receivers, attackers can reduce errors caused by signal obstructions or reflections, thereby improving their tracking capabilities. Such abuses could result in unauthorized monitoring, interception of communications, and potential targeting of vehicles or individuals. These risks underscore the urgent need for advanced countermeasures, such as secure communication protocols, dynamic frequency hopping, and signal encryption, to protect against these privacy threats.

- **Direction of Arrival (DoA):** DoA-based tracking determines a vehicle's 2D position by analyzing the angles at which multiple antennas receive signals. This technique is particularly useful in environments with multiple signal paths, where reflections could disrupt localization accuracy. By isolating and utilizing the angles of incoming signals, DoA methods can effectively filter out these reflections, resulting in more precise localization, especially in complex urban areas. Xu et al. [71] refined this method by employing large-scale uniform linear arrays (ULAs) instead of traditional small-scale antenna arrays. This improvement enhances spatial resolution and signal reception in high-dimensional environments. The study also addressed mutual coupling interference between antennas by applying a linear transformation to reduce signal interference. Additionally, Toeplitz rectification and phase transformation techniques corrected phase errors, improving accuracy. The proposed method achieves reliable, high-precision DoA tracking by incorporating SNR data and signal angles.

Despite these advancements, DoA-based tracking poses significant privacy risks when misused. Sophisticated DoA systems allow adversaries to deploy large-scale antenna arrays, capturing angular data from signals emitted by a vehicle to track its movements with high precision, even in densely populated urban areas. Advanced correction techniques used in DoA tracking make traditional privacy measures, like signal obfuscation and pseudonym changes, less effective in preventing unauthorized tracking. This capability could severely compromise security and personal safety in sensitive scenarios, such as the monitoring of law enforcement or military vehicles, underscoring the need for robust countermeasures to protect against such privacy threats.

- **Received Signal Strength (RSS):** RSS-based tracking relies on measuring the attenuation of signal strength due to factors like path loss, fading, and shadowing to estimate the location of a vehicle or target. It is widely used in wireless networks for localization because of its simplicity and the availability of signal-strength data. However, this method can be exploited by adversaries to track vehicles without the need for sophisticated infrastructure. Huang et al. [72] tackled the challenges of RSS-based tracking in near-field regions, where obstacles often disrupt localization accuracy. They introduced a method to create virtual Line-of-Sight (LOS) routes between nodes. They optimized the azimuth and phase parameters to maximize RSS, thereby improving accuracy and reducing computational demands.

While this advancement enhances localization in challenging conditions, it also poses significant privacy risks. Adversaries could adopt similar techniques to track vehicles even in signal obstructions or multipath interference environments. By simulating virtual LOS paths and fine-tuning RSS parameters, attackers could bypass common privacy protections such as signal jamming or pseudonym changes. This would enable continuous and covert tracking of vehicles, potentially leading to unauthorized surveillance, targeted attacks, or coercive monitoring, thus increasing privacy vulnerabilities.

- **Radio Fingerprint:** Radio waves emitted by base stations possess unique characteristics, known as "radio fingerprints," which can be systematically gathered to create a database of RSS information for tracking vehicles. Adversaries can exploit these unique RF fingerprints, associated with each vehicle's communication hardware, to accurately track them. Chen et al. [73] demonstrated this using Dedicated Short-Range Communications (DSRC) under the IEEE 802.11p protocol, commonly deployed in vehicular networks. By analyzing preamble field characteristics in the physical layer frames, they extracted RF fingerprints unique to each vehicle's hardware.

These RF fingerprints, generated by the inherent differences in hardware components, provide a means for adversaries to continuously monitor vehicles by correlating previously recorded fingerprints with a centralized database. Such an approach can circumvent privacy mechanisms like pseudonym changes or encryption at higher layers of communication. The ability to identify and track vehicles using their RF fingerprints presents a substantial privacy risk, enabling unauthorized surveillance and potential targeted attacks such as stalking, vehicle hijacking, or data breaches. The concern is heightened in scenarios where vehicles transmit signals without encryption or protection, leaving them vulnerable to passive eavesdropping and data collection. To mitigate these risks, advanced countermeasures like RF fingerprint randomization techniques, secure communication protocols, and hardware-level protections must be developed. These strategies would help obscure or modify the unique identifiers emitted by vehicular communication systems, thereby strengthening location privacy and safeguarding against sophisticated tracking techniques.

- **Integrated Sensing and Communication (ISAC):** ISAC is an advanced technique in vehicular networks that combines communication and passive vehicle detection using cellular radio signals. Functioning similarly to radar, ISAC utilizes the existing infrastructure of base stations to detect and track vehicles by analyzing reflected signals, eliminating the need for vehicles to emit active radio transmissions. This approach enables continuous monitoring of vehicles, and when data from multiple base stations is aggregated, it can accurately reconstruct vehicle trajectories.

  While ISAC provides significant advantages for traffic management and safety, it also introduces privacy concerns. Adversaries could exploit its passive tracking capabilities for covert surveillance, as vehicles are not actively transmitting signals, making it difficult for drivers to detect monitoring. By integrating data from various base stations, malicious entities could map real-time vehicle movements across extensive areas, revealing sensitive information such as driving patterns, frequently visited locations, and personal behaviors. This level of tracking poses serious risks to location privacy, undermining traditional privacy mechanisms like encryption and pseudonym changes that depend on active transmissions.

  To mitigate the misuse of ISAC, it is essential to develop privacy-preserving measures. These could include data anonymization techniques, secure data-sharing protocols, and regulatory frameworks to ensure that ISAC data is used only for legitimate purposes, preventing unauthorized tracking or surveillance.

*3) In-vehicle Sensing and Tracking:* The rapid evolution of vehicular technology has significantly increased the complexity of in-vehicle communication systems, introducing new privacy challenges. Systems such as the Controller Area Network (CAN), widely used to manage and record vehicle data, now track a broad range of parameters, including speed, engine status, and sensor activity, as highlighted by recent studies [74]. The growing use of mobile devices like smartphones and laptops as additional sensing platforms has further expanded the scope and detail of data collected within vehicles. This expansion heightens privacy risks, as sensitive information can be exposed through data recovery, pattern analysis, and trajectory matching techniques.

- **In-Vehicle Data Collection:** On-Board diagnostic (OBD) readers, which connect to the CAN, are crucial for accessing detailed data from a vehicle's Electronic Control Units (ECUs), providing insights into parameters such as engine performance, speed, and sensor outputs. To enhance security, manufacturers have divided CAN into sub-networks, aiming to reduce risks linked to tracking through single sensors. However, this defense is insufficient against advanced localization attacks that aggregate data from multiple OBD readers. By merging information from various sources—including inertial measurement units (IMUs), heading, pressure, and speed sensors—adversaries can accurately map a vehicle's trajectory and behavior [51].

  This ability significantly threatens location privacy, allowing malicious actors to track and reconstruct a vehicle's movements, even in areas where GPS signals are weak or unavailable. Furthermore, combining such data streams enables adversaries to predict future routes, analyze driving patterns, and identify frequently visited locations, making targeted surveillance or profiling possible. The precision achieved by integrating data from multiple sensors undermines the privacy benefits provided by CAN sub-networking, necessitating stronger privacy protections against these risks.

- **Tire Pressure Monitor System (TPMS):** The TPMS is a critical safety feature integrated into a vehicle's wireless network infrastructure, designed to monitor and report tire pressure levels. However, this system also introduces significant privacy risks due to its use of uniquely identifiable sensors and a communication protocol vulnerable to reverse engineering. Each TPMS sensor broadcasts a distinct identifier to inform the vehicle's central system about pressure changes. These broadcasts, which occur frequently and without the driver's knowledge, can be intercepted by Road-Side Units (RSUs) or other adversarial devices equipped with passive tracking capabilities. By capturing these TPMS packets, adversaries can exploit the unique identifiers to track a specific vehicle over time, reconstructing its routes and driving patterns [75].

  TPMS enables covert surveillance, as the continuous transmission of TPMS data allows attackers to gather location information unobtrusively. Additionally, the lack of encryption in the TPMS protocol and its vulnerability to reverse engineering make it easy for adversaries to decode the data, further heightening the threat to location privacy. In densely populated urban areas where vehicles pass through multiple RSUs, adversaries could compile a detailed log of a vehicle's movements, identifying regular routes, destinations, and behavioral patterns. This level of tracking poses severe risks, including unauthorized surveillance, targeted stalking, and potential criminal activities like vehicle theft or hijacking. Addressing these vulnerabilities requires the adoption of robust security measures, such as encrypted communication protocols for TPMS data, anonymization techniques to obscure sensor identifiers, and regulatory frameworks to prevent unauthorized data interception and misuse.

- **Mobile Terminal:** The widespread use of mobile phones equipped with GPS and other sensors has significantly increased the risk of side-channel attacks, which can deduce driving trajectories and compromise location privacy. Systems like Daimler Chrysler's Tegaron, which integrate advanced navigation features with data transmission to centralized control centers, exemplify how in-vehicle systems can be vulnerable to external localization attacks and breaches. By transmitting vehicle data to control centers, systems like

Tegaron expose sensitive information that adversaries could intercept and use to covertly monitor vehicular movements. In addition, adversaries have been known to exploit mobile sensor data, including accelerometers, gyroscopes, and magnetometers, to accurately track vehicle movements. Research by Guha et al. [76] demonstrated that accelerometer and gyroscope measurements, even without direct GPS input, could detect and analyze vehicular dynamics to reconstruct precise driving routes. The ability to passively collect sensor data from widely available mobile devices allows attackers to infer a vehicle's speed, direction, and movement patterns. When combined with external databases, such as those containing road network data, these attacks can disclose sensitive locations, frequent destinations, and driving habits, leading to unauthorized surveillance, profiling, and potential targeting of individuals or assets.

- **In-Vehicle Services and Applications:** The rapid expansion of in-vehicle services and applications to enhance autonomous driving features and improve passenger comfort has significantly complicated efforts to protect vehicular data privacy. These systems, which are now integral to modern vehicle operations, continuously collect and transmit vast quantities of data, including detailed information on vehicle states (such as speed, braking, and engine performance), geolocation, and user interactions with the vehicle interface [3].

  While crucial for optimizing vehicle performance, delivering real-time traffic updates, and supporting Advanced Driver Assistance Systems (ADAS), such data collection introduces numerous privacy risks. Adversaries can exploit vulnerabilities within these systems to gain unauthorized access to sensitive data streams, either through hacking connected applications or intercepting data transmissions to external servers. Navigation systems rely on frequent location updates and are particularly vulnerable, allowing malicious actors to track vehicles, reconstruct routes, and discern frequent destinations or behavioral patterns.

  Applications designed to provide personalized services—such as infotainment, predictive maintenance, or route optimization—typically require continuous data sharing with cloud-based platforms, heightening the risk of data breaches or unauthorized aggregation. Misuse of this data could lead to unauthorized profiling, targeted advertising, stalking, or more severe threats such as carjacking or vehicle theft. The extensive data generated by these services underscores the need for stringent data management practices, including robust encryption, secure authentication protocols, and strict access controls, to safeguard user privacy in intelligent transportation systems.

*4) Upper-Layer Message-based Tracking:* Upper-layer message-based tracking poses a substantial risk to location privacy in vehicular networks, particularly through the interception of V2X communications or direct attacks on entities involved in LBS. As shown in Fig. 4, LBS systems consist of three main components: users (drivers and their devices), servers (LBS and location privacy servers), and networks (Wi-Fi, cellular, and Internet). Each component is susceptible to exploitation, potentially allowing malicious actors to access sensitive location data and compromise privacy.

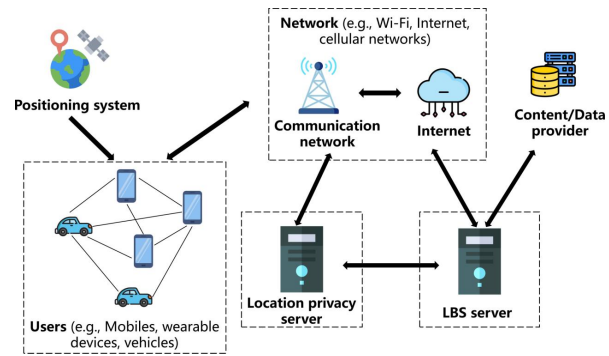

Fig. 4. System model of LBS. The LBS consists of drivers, servers, and networks. The driver accesses vehicular networks with access points, e.g., base stations and RSUs. Then, the LBS servers respond to the driver's enquires. All three components can be threatened by the adversaries.

Adversaries can eavesdrop on V2X communications to track vehicles' real-time positions or target servers that process or store location data. The increasing reliance on LBS for navigation, traffic management, and personalized services, which all require continuous exchange of location information, exacerbates these risks. Securing these components is, therefore, critical to preserving user privacy in connected vehicular networks.

- **User Component:** Drivers, vehicles, and associated devices such as mobile phones and wearables are vulnerable to privacy threats through data interception. Devices in vehicles frequently communicate with external networks via access points, including base stations and RSUs. If adversaries manage to eavesdrop on these communications, they could capture sensitive location data directly from the devices, exposing real-time positions, route histories, or frequent destinations.

- **Servers Component:** LBS and location privacy servers are responsible for processing location queries and managing data exchanges between users and service providers. These servers are prime targets for adversaries because breaching them can provide access to aggregated data, including user locations and movement patterns. For example, an attack on an LBS server could allow adversaries to intercept queries, track users over time, or modify location data to mislead or track individuals.

- **Network Component:** Communication between users and servers occurs over channels such as Wi-Fi, cellular networks, and the Internet, which are vulnerable to network eavesdropping. Adversaries may intercept data packets during transmission, potentially compromising the confidentiality and integrity of location data. For instance, man-in-the-middle (MitM) attacks can be used to capture or alter messages sent over the network, exposing sensitive location information to unauthorized parties.

Adversaries exploiting upper-layer message-based tracking techniques can be broadly categorized into active and passive attackersAdversaries exploiting upper-layer message-based tracking techniques can be broadly categorized into active and passive attackers [77]. Active adversaries actively disrupt network communications by injecting falsified data, launching spoofing attacks, and cloning or capturing legitimate drivers to insert fake messages into the network. In contrast, passive adversaries focus on monitoring and analyzing data traffic to infer drivers' positions, routes, and routines by intercepting communication streams without directly interfering with or altering them. Current LPPMs prioritize defense against passive adversaries, given their stealthy nature and the broader impact they can have on privacy. Active adversaries actively disrupt network communications by injecting falsified data, launching spoofing attacks, and cloning or capturing legitimate drivers to insert fake messages into the network. In contrast, passive adversaries focus on monitoring and analyzing data traffic to infer drivers' positions, routes, and routines by intercepting communication streams without directly interfering with or altering them. Current LPPMs prioritize defense against passive adversaries, given their stealthy nature and the broader impact they can have on privacy.

In the context of vehicular networks, passive adversaries present significant risks to location privacy. These adversaries are classified into two categories based on the extent of their monitoring capabilities: Global Passive Adversaries (GPAs) and Local Passive Adversaries (LPAs), as illustrated in Fig. 5. GPAs are the most potent class of passive attackers, with extensive surveillance reach. They possess the ability to intercept data transmissions across an entire network, leveraging comprehensive knowledge of the road infrastructure, legitimate access to authorized systems, or exploiting vulnerabilities in applications. GPAs can gather and analyze data over extended periods, ranging from hours to even years, enabling them to reconstruct detailed patterns of vehicular movement. Due to their far-reaching access and long-term monitoring capabilities, GPAs are considered the most dangerous adversarial model in numerous privacy-preserving frameworks [78]. However, the deployment of GPAs on a large scale is often limited by the high costs of maintaining such extensive infrastructure. In contrast, LPAs operate on a more restricted scale, typically focusing their monitoring efforts within a localized area and using fewer resources. Although their range is limited, LPAs still pose considerable privacy risks, particularly when targeting specific vehicles or geographical regions. Their ability to gather detailed location data within a limited scope can lead to the same privacy breaches as GPAs but with a more concentrated focus.

As shown in Fig. 6, the process of location attacks typically involves four main steps: data collection, analysis, infer-ence, and exposure. Initially, adversaries obtain location data through methods such as eavesdropping, data interception, or system compromise. Subsequently, they analyze this data using advanced techniques like context linking, probability theory, machine learning, or leveraging (fake) peer user scenarios to deduce a vehicle's location. The final step involves exposing the driver's identity, discretized trajectory points, and continuous movement paths, which adversaries can exploit for various malicious purposes, including unauthorized surveillance, profiling, or targeted attacks.

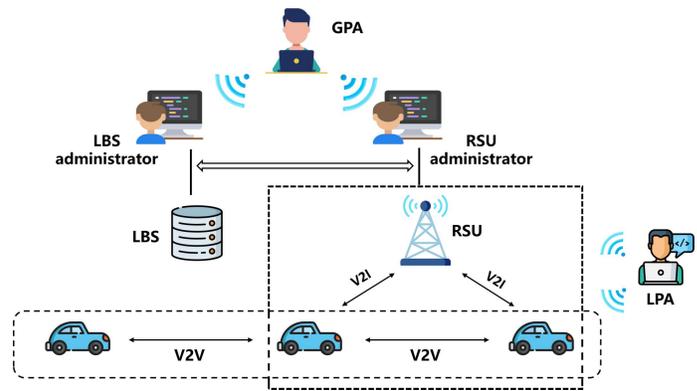

Fig. 5. GPA and LPA for vehicle location privacy. The GPA can eavesdrop on data transmitted in networks with full knowledge of the road, while the LPA eavesdrop on a relatively small range with limited equipment.

The key issue with upper-layer message-based tracking in vehicular networks is its susceptibility to various cyberattacks that can compromise privacy. Cyber adversaries exploit vulnerabilities in data transmission, using advanced big data mining techniques to aggregate information from lower layers. This includes sources such as proximity detections, transactional logs, parking locations, and Wi-Fi signal data, all of which can be combined to create a detailed profile of vehicular movements. When attackers intercept and mine this aggregated data, they can accurately reconstruct vehicle trajectories and infer sensitive location information, posing significant privacy threats.

To address the growing concerns around location privacy in 5G/6G VNs, various LPPMs have been developed, specifically targeting the protection of location and trajectory-related data against upper-layer message-based tracking. These mechanisms are designed to obscure the linkability of messages to individual users or locations, making it difficult for adversaries to accurately track vehicles. Methods such as context-linking obfuscation, probabilistic data alteration, machine learning-based anomaly detection, and the generation of indistinguishable fake user profiles have been employed to complicate adversarial efforts. These techniques are particularly effective at mitigating risks related to identity inference, exposure of segmented trajectories, and the reconstruction of continuous travel paths [79]. As discussed in Section IV, these strategies are crucial for preserving user privacy in the upper data layers, reinforcing their role in safeguarding against increasingly sophisticated location privacy threats.

### B. Location Privacy Requirement

Location privacy in vehicular networks has been defined in several ways by researchers [80], [81]. In this paper, location privacy refers to a subset of information privacy that focuses on safeguarding location data, allowing drivers to control when, what, and how their data is shared. This definition emphasizes the need for robust mechanisms to ensure that data shared within vehicular networks does not jeopardize the driver's security or expose them to potential threats. The core privacy requirements for vehicular networks can be grouped into four critical categories: confidentiality, anonymity, unlinkability, and contextual unobservability [82].

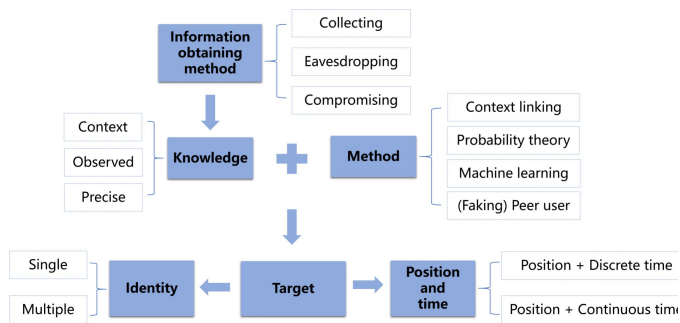

Fig. 6. Overview of the location attacks and adversaries. The adversaries can obtain the location information of drivers by collecting, eavesdropping, and compromising the vehicle's information. By analyzing the obtained data, the adversaries can gain knowledge of drivers with the obtained information.

- **Confidentiality:** Confidentiality is a cornerstone of location privacy, ensuring that the data exchanged between vehicles and external services, such as LBSs, remains secure and inaccessible to unauthorized entities [83]. This is achieved by employing robust encryption protocols to protect data during transmission, preventing eavesdropping or interception by adversaries. Without strong confidentiality measures, sensitive location data could be exposed, leading to severe privacy breaches, such as real-time tracking, profiling of driver behavior, or the malicious exploitation of data for stalking or theft. For example, an adversary could intercept location queries sent from a vehicle to an LBS server, revealing the vehicle's current position and route information.

- **Anonymity:** Anonymity is essential for protecting the identities of vehicles in vehicular networks, ensuring that even if location data is intercepted, the specific vehicle or driver cannot be identified [84]. This is typically achieved through pseudonymization, where vehicles use temporary identifiers that are frequently changed to prevent long-term tracking or profiling by adversaries. Anonymity is particularly critical in environments where vehicles frequently communicate with RSUs or other vehicles, as a lack of it would enable attackers to link location data to a particular vehicle or driver. This could lead to targeted attacks or significant privacy breaches.

- **Unlinkability:** Unlinkability ensures that adversaries or unauthorized entities cannot associate location data with a vehicle's identity, even if the data is intercepted [85]. This protection is vital for preventing longitudinal tracking, where attackers could accumulate data over time to deduce a vehicle's route, patterns, or frequent destinations. To maintain unlinkability, various techniques are employed, such as randomizing identifiers, introducing delays in transmission times, or utilizing mix-zones. In these zones, vehicles change their identifiers in a controlled environment, effectively severing any traceability links and enhancing privacy.

- **Contextual Unobservability:** Contextual unobservability is critical to ensuring that successive data transmissions from the same vehicle do not expose patterns that could compromise privacy [86]. This protection guards against inference attacks, where adversaries analyze repeated behavior in data releases to draw conclusions about a vehicle's movements and routines. For instance, if a vehicle consistently transmits location updates that reveal patterns like frequent trips to specific locations (e.g., home or work), adversaries could combine these data fragments to map out the vehicle's entire schedule. To counter this, techniques such as obfuscation, which alters the data to mask actual patterns, or differential privacy, which adds statistical noise, are often employed to prevent accurate inferences.

Confidentiality ensures that vehicle communication data is inaccessible to unauthorized parties, protecting the content from eavesdroppers, while anonymity hides the vehicle's identity within the data, preventing linkage to specific drivers. Unlinkability further prevents adversaries from associating different pieces of data with the same vehicle, making it difficult to track movement patterns over time. Contextual unobservability ensures that successive data releases remain uncorrelated, preventing adversaries from inferring trajectories or behaviors through repeated observations.

*1) Characteristic of Vehicular Networks:* Vehicular networks possess distinct characteristics that differentiate them from other communication environments, such as their unlimited transmission power, enhanced computational capabilities, and predictable mobility patterns. These factors introduce unique challenges to location privacy protection. The features of vehicular networks can be broadly categorized into three areas: topology features, node features, and transmission features, as depicted in Fig. 7. Understanding these characteristics is essential for designing effective LPPMs that address the specific threats and vulnerabilities inherent in vehicular networks.

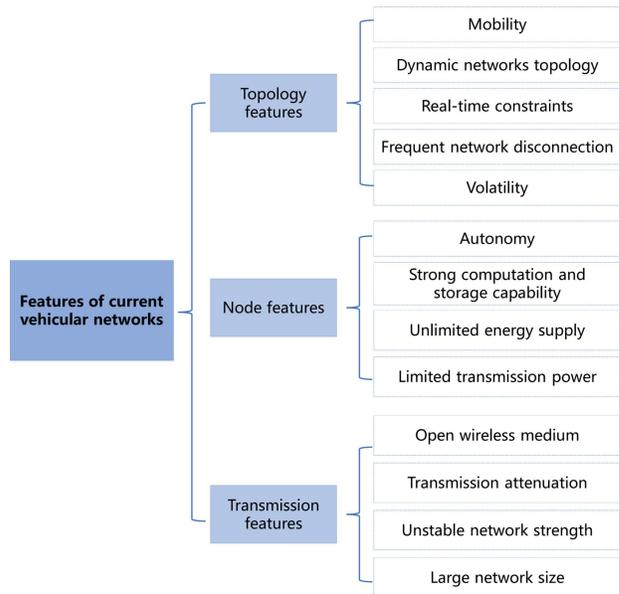

Fig. 7. Features of vehicular networks. Topology features, node features, and transmission features make vehicular networks to be significantly different from other scenarios.

**Topology Feature:** Vehicular networks exhibit distinct topology features that significantly impact location privacy,

including high mobility, dynamic network topology, real-time constraints, frequent disconnections, and connection volatility.

- **Mobility:** Vehicles in vehicular networks typically operate at high speeds, which poses significant challenges to maintaining location privacy through traditional LPPMs like handshake-based authentication technologies. These methods require multiple message exchanges between vehicles and network nodes to establish secure communication, which becomes impractical given the limited time available for such exchanges in high-speed scenarios [87].

As vehicles frequently change their network attachment points, such as moving from one base station or RSU to another, LPPMs must be designed with advanced mobility management techniques to ensure seamless handovers while preserving location privacy across different network segments. The high-speed mobility also creates brief communication windows that adversaries can exploit to intercept location data, particularly if the data is not promptly encrypted or adequately obfuscated. This necessitates the development of fast and efficient privacy protocols capable of providing robust protection even in the presence of rapid, frequent changes in network topology and connection points. These protocols must minimize latency and reduce the overhead associated with data exchanges to prevent adversaries from obtaining meaningful information while simultaneously ensuring that location data is anonymized and unlinkable across successive communications. Furthermore, adaptive techniques, such as dynamic pseudonym changes and context-aware privacy policies, should be integrated into LPPMs to counteract adversarial attempts to correlate intercepted data fragments with specific vehicles, thereby enhancing overall privacy in high-mobility vehicular environments.

- **Dynamic Network Topology:** The dynamic network topology of vehicular networks—characterized by frequent and unpredictable changes due to varying vehicle densities, speeds, and movement patterns—poses significant challenges to the effectiveness of LPPMs.

In environments with high vehicle density, such as urban city centers, cooperative LPPMs like pseudonym swaps can be highly effective, as numerous vehicles are available to exchange pseudonyms and disrupt the continuity of tracking by adversaries. However, in less dense areas, such as suburban or rural regions, the limited number of vehicles reduces the opportunities for such exchanges, thereby weakening the ability of these mechanisms to ensure privacy. This variability necessitates the development of adaptive LPPMs that can dynamically adjust to changes in network topology and vehicle density, ensuring that privacy protection remains robust even in sparse traffic conditions. For instance, in low-density scenarios, LPPMs may need to incorporate additional measures such as dummy message generation, opportunistic beaconing, or infrastructure-based pseudonymization (e.g., via RSU) to compensate for the lack of vehicle interaction. Moreover, the rapid changes in network topology, such as vehicles entering and exiting communication ranges, require that LPPMs employ rapid re-authentication and real-time decision-making processes to maintain privacy without compromising the timeliness of critical vehicular communications.

- **Real-time Constraint:** Vehicular networks are characterized by the need for rapid, low-latency communication to ensure the timely delivery of critical information, such as collision alerts, road hazard notifications, or emergency braking signals, which are essential for driver safety and traffic management [88].

However, maintaining location privacy while meeting these real-time constraints poses a significant challenge. Privacy-preserving mechanisms, such as encryption, pseudonym changes, and message verification, inherently introduce computational overhead and communication delays, slowing down the transmission of time-sensitive data. For instance, generating and verifying cryptographic signatures necessary to authenticate and protect messages can lead to latency that exceeds acceptable thresholds for safety-critical applications. Additionally, frequent pseudonym changes designed to enhance unlinkability can result in packet loss or increased transmission delays due to re-authentication processes. Consequently, LPPMs must be carefully designed to minimize these delays and prioritize the timely delivery of safety messages while protecting location data from eavesdropping and unauthorized access. This requires employing lightweight cryptographic algorithms, optimizing data packet structures, and employing adaptive privacy strategies that can dynamically adjust the level of privacy protection based on the context, such as the urgency of the transmitted data, the current network conditions, and the potential privacy risks involved. Balancing these competing requirements is critical to ensuring that vehicular networks maintain safety and privacy, preventing adversaries from exploiting real-time data exchanges to infer vehicle locations or track driver behaviors.

- **Frequent Network Disconnection:** Frequent disconnections in vehicular networks caused by high-speed movement, environmental factors like terrain and weather, or network congestion present a significant challenge to maintaining effective location privacy [89]. These interruptions lead to unreliable connectivity, complicating the implementation of continuous, real-time privacy protection. LPPMs must be resilient enough to safeguard sensitive location data even when network connections are lost or degraded. Solutions such as data caching strategies and delay-tolerant networking protocols are essential for preventing data exposure during disconnection. Moreover, decentralized LPPMs, which operate without relying on a central server, can locally manage location privacy within vehicles or clusters, enhancing robustness. Techniques like local data aggregation and intermittent encryption, which secure data even during sporadic exchanges, are vital for mitigating privacy risks associated with frequent disconnections. These approaches are crucial for defending against attacks that exploit network vulnerabilities during disconnection periods, such as capturing residual data or replaying intercepted data when the connection is restored. This ensures vehicular location data privacy under adverse network conditions.

- **Connection Volatility:** The inherent volatility of connections in vehicular networks, characterized by random and short-lived interactions between vehicles, poses significant challenges to maintaining location privacy using conventional long-duration mechanisms like password-based authentication [90]. In such a highly dynamic environment, vehicles frequently enter and exit the communication range of each other, resulting in frequent disconnections that disrupt sustained communication sessions necessary for traditional authentication methods.

This unpredictability makes establishing and maintaining secure channels difficult for extended periods, creating vulnerabilities that adversaries could exploit to intercept, analyze, or manipulate location data. As a result, LPPMs in vehicular networks must be designed to function effectively under these conditions, employing lightweight, rapid authentication processes and decentralized privacy strategies that do not rely on persistent connections. For instance, LPPMs may use ephemeral identifiers that frequently change or broadcast pseudonyms that cannot be easily linked to a particular vehicle, ensuring that location data remains untraceable even as connections fluctuate. Additionally, privacy mechanisms may incorporate opportunistic communication protocols that leverage short encounters between vehicles to exchange privacy-preserving information quickly, reducing the window of opportunity for adversaries to exploit network volatility for tracking purposes. By minimizing dependency on sustained communication links, these adaptive privacy solutions can provide robust protection against the unique privacy risks of the transient connectivity inherent in vehicular networks.

**Node Feature:** Transmission characteristics, such as limited transmission power, open wireless medium, transmission attenuation, unstable network capability, and large network size, are also critical to understanding location privacy challenges.

- **Autonomy:** Vehicles in vehicular networks exhibit significant autonomy, independently sending, routing, and receiving data with minimal reliance on centralized control [91]. This decentralized nature is both an asset and a challenge for maintaining location privacy. On the one hand, the absence of centralized oversight reduces the vulnerability to single points of failure or centralized attacks, which could otherwise expose the location data of all vehicles in the network. On the other hand, this autonomy requires the development of robust distributed LPPMs that can function effectively in a decentralized environment. These mechanisms must ensure that individual vehicles can manage their privacy while communicating and cooperating with others. For example, decentralized LPPMs must be capable of protecting location data without the constant support of a central server, often employing strategies such as pseudonym changes, local data encryption, and cooperative data obfuscation. However, the challenge arises in coordinating these privacy measures across a large, dynamic network, where each vehicle must autonomously decide when and how to implement privacy protections based on local conditions. Furthermore, the lack of centralized control can make it difficult to detect and mitigate sophisticated attacks, such as those involving coordinated eavesdropping or data aggregation by malicious entities. Thus, while autonomy enhances the flexibility and scalability of vehicular networks, it also necessitates advanced LPPMs that can dynamically adapt to the decentralized and rapidly changing nature of these environments to safeguard location privacy.

- **Strong Computation and Storage Capability:** Vehicles' strong computational and storage capabilities provide a unique opportunity for advanced LPPMs to leverage complex algorithms and large datasets for robust privacy protection [92]. These capabilities enable the implementation of sophisticated cryptographic protocols, machine learning models for anomaly detection, and distributed ledger technologies like blockchain to ensure the integrity and confidentiality of location data. For example, vehicles can utilize federated learning approaches to collaboratively train privacy-preserving models without sharing raw data, thereby reducing the risk of exposing sensitive information. However, these same computational resources also make vehicles attractive targets for adversaries.

Attackers may exploit the vehicles' powerful onboard systems to launch sophisticated attacks, such as deploying malware to harvest location data, executing data mining techniques to analyze stored information, or applying machine learning algorithms to infer patterns from intercepted communications. The ability to process large amounts of data locally increases the risk of adversaries compromising onboard systems to access and manipulate data for malicious purposes. This dual-edged nature of computational strength necessitates the development of LPPMs that not only utilize vehicles' computational resources for enhanced privacy but also incorporate robust security measures to protect these resources from being exploited by adversaries. Therefore, achieving location privacy in vehicular networks requires balancing advanced algorithms with ensuring the security and integrity of the vehicles' computational environments.

- **Unlimited Energy Supply:** The availability of substantial energy resources from vehicle batteries offers a significant advantage in vehicular networks, as it allows for the deployment of energy-intensive LPPMs without the usual constraints on power consumption [93]. This enables more sophisticated privacy measures, such as advanced encryption algorithms, real-time data obfuscation, and frequent pseudonym changes, which can significantly enhance location privacy. However, while the energy supply is virtually unlimited in modern vehicles, there is still a need to ensure that privacy mechanisms are optimized for efficiency.

Excessive use of computational resources could impact other critical vehicular functions, such as navigation, collision avoidance systems, or real-time communication in V2X networks. Furthermore, adversaries could exploit privacy mechanisms that consume large amounts of computational power by launching Denial-of-Service (DoS) attacks, forcing the system to overuse its resources and potentially causing privacy lapses or system failures. Therefore, while an unlimited energy supply allows for more robust pri-vacy protections, these mechanisms must balance security with the efficient allocation of computational resources to maintain overall system performance and ensure continuous protection against location privacy threats.

**Communication and Networks:** The transmission features of vehicular networks related to location privacy are open wireless medium, transmission attenuation, unstable network strength, and large network size.

. **Limited Transmission Power:** In vehicular networks, the inherent limitation on transmission power, dictated by communication protocols and the wireless environment, significantly affects the scope and effectiveness of privacy-preserving mechanisms [94]. Due to these constraints, vehicles are limited in their communication range, which poses challenges for implementing LPPMs that rely on inter-vehicle cooperation, such as pseudonym swapping or cooperative obfuscation.

The restricted communication range means that vehicles can only interact with nearby peers for short periods, making it difficult to execute complex privacy protocols that require sustained interaction between multiple vehicles. This limitation is particularly problematic in sparsely populated areas, such as rural roads or highways, where fewer vehicles are available to participate in privacy-preserving exchanges, thereby increasing the risk of location data exposure. Additionally, the limited range can hinder the effectiveness of decentralized privacy schemes that depend on wide network coverage for dispersing location data obfuscation across multiple nodes. As a result, LPPMs must account for these power and range constraints by incorporating adaptive techniques, such as adjusting transmission power dynamically based on the density of vehicles or using hybrid communication models that combine short-range and long-range communication technologies to maintain robust privacy protections across diverse vehicular environments.

. **Open Wireless Medium:** The use of an open wireless medium in vehicular networks presents inherent security challenges, particularly with location privacy. The open nature of the airwaves facilitates the interception of wireless communications—activities such as eavesdropping become significantly easier, posing a serious threat to the confidentiality of sensitive location data transmitted between vehicles and infrastructure [95].

To counteract these vulnerabilities, LPPMs must integrate robust security measures. Advanced encryption techniques, such as end-to-end encryption, ensure data remains indecipherable to unauthorized parties intercepting the transmissions. Moreover, data obfuscation methods can alter the precision or veracity of the location data being transmitted, adding an additional layer of security by making it difficult for potential adversaries to use intercepted data effectively. Implementing these strategies is crucial for maintaining location privacy within vehicular networks, as they mitigate the risks associated with the open transmission medium by safeguarding data against common wireless attacks, thereby preserving the integrity and confidentiality of sensitive in-formation shared in these highly dynamic environments.

. **Transmission Attenuation:** In vehicular networks, transmission attenuation caused by environmental factors such as diffraction, reflection, scattering, and refraction poses significant risks to location privacy. These physical layer vulnerabilities can degrade signal quality, leading to unintended leakage of location data, even when encryption techniques are employed to secure higher-layer communication. For instance, when signals reflect off buildings or vehicles, an adversary could exploit these reflections to infer a vehicle's position by analyzing the altered signal paths or signal strength fluctuations. In such cases, traditional encryption methods may be insufficient to safeguard against location-based attacks, as they focus on data content rather than transmission characteristics.

Effective LPPMs must address these physical-layer challenges by incorporating advanced signal processing techniques that mitigate the effects of attenuation. This could include techniques like multipath correction, adaptive modulation, and error correction codes to ensure that signal distortions do not provide adversaries with exploitable information about a vehicle's trajectory. Additionally, obfuscating transmission patterns and dynamically adjusting signal parameters could further protect against the reconstruction of vehicular trajectories based on signal behavior, thus preserving location privacy in environments where physical factors interfere with signal integrity.

. **Unstable Network Capability:** The communication and computation capabilities of vehicular networks are highly variable and subject to real-time traffic conditions, which directly impacts the effectiveness of LPPMs [96]. In dense traffic scenarios, such as traffic jams, the proximity of many vehicles can increase network strength, forming a robust vehicular network with high communication availability. This environment is conducive to cooperative LPPMs, such as pseudonym swaps, where vehicles frequently change their identifiers to prevent long-term tracking. However, in less congested scenarios, such as rural or suburban areas with fewer vehicles, the network becomes more sparse, reducing the opportunity for cooperation between vehicles and significantly weakening the privacy protection afforded by these mechanisms.

Adversaries can exploit these periods of network instability to track vehicles by focusing on the gaps between identifier changes, particularly in low-density areas where pseudonym swaps are less frequent. Therefore, LPPMs must be adaptive to network variability, ensuring robust privacy protection across high- and low-density traffic conditions by integrating mechanisms like dynamic pseudonym updates, local cooperation, or alternative privacy strategies that rely not solely on vehicle density or network strength. This adaptability is crucial to maintaining consistent location privacy, regardless of the fluctuating communication capabilities of the vehicular environment.

. **Large Network Size:** Vehicular networks typically span vast geographic areas, including densely populated urban environments and expansive highway systems, which introduces significant challenges for maintaining location pri-vacy. The large scale of these networks complicates the deployment of LPPMs because these mechanisms must effectively balance broad network coverage with the limitations of localized communication regions. In urban areas, the high density of vehicles enables frequent interactions between vehicles and infrastructure (e.g., RSUs), which can facilitate privacy-enhancing techniques such as pseudonym swapping or cooperative data obfuscation. However, fewer vehicles are available to participate in cooperative LPPMs in less populated regions, such as rural highways, making it more difficult to obscure location data through collaborative approaches.

The large network size increases the risk of prolonged tracking by adversaries who can exploit the sparse communication regions to monitor vehicle movements over longer distances without interference. This necessitates the development of adaptive LPPMs capable of dynamically adjusting to varying network densities, ensuring consistent location privacy protection regardless of the communication environment. Moreover, as vehicles move across different network regions, privacy mechanisms must seamlessly transition between densely and sparsely populated areas without compromising location privacy, requiring advanced mobility management and secure handoff protocols.

*2) Vehicular Communication Technologies:* As two widely studied vehicular communication technologies, Dedicated Short-Range Communication (DSRC) V2X and Cellular V2X (C-V2X) present distinct operational architectures, each with unique implications for location privacy and security.

. **DSRC V2X:** DSRC V2X operates on a decentralized, peer-to-peer communication model, enabling vehicles to exchange information directly with each other and roadside infrastructure without relying on a centralized control system. This decentralized architecture benefits low-latency communication, which is crucial for real-time applications like collision avoidance and traffic coordination, particularly in environments with limited cellular infrastructure, such as rural or remote areas [97].

This peer-to-peer model also introduces significant challenges to location privacy. DSRC communications are more vulnerable to eavesdropping and data injection attacks without a centralized authority overseeing security. Adversaries can intercept messages exchanged between vehicles or between vehicles and RSUs, gaining access to sensitive location data and tracking vehicles over time. The absence of robust, centralized security protocols means that DSRC networks often lack encryption or pseudonymization by default, making it easier for attackers to trace specific vehicles, profile drivers or infer their identities based on movement patterns. This exposes vehicles to long-term tracking, where adversaries can monitor a vehicle's routes, frequent destinations, and behavioral patterns, compromising both driver privacy and security. To address these vulnerabilities, advanced security mechanisms such as dynamic pseudonym change, end-to-end encryption, and periodic reauthentication should be incorporated into DSRC networks. These measures would prevent unauthorized tracking and reduce the risk of identity inference attacks, thereby strengthening location privacy in decentralized vehicular environments.

. **C-V2X:** C-V2X adopts a centralized communication architecture, utilizing existing cellular infrastructure, which service providers typically manage. This centralized structure provides enhanced control over communications, enabling more stringent security measures, such as advanced encryption protocols, robust data protection mechanisms, and secure authentication processes, essential for safeguarding location privacy [98].

Unlike decentralized models, C-V2X allows for more effective management of privacy-preserving techniques, including location obfuscation, which masks the exact coordinates of a vehicle, and pseudonymization, which frequently changes vehicle identifiers to prevent long-term tracking. Secure key management systems can also be centrally enforced, ensuring that only authorized entities can access sensitive location data. However, C-V2X's reliance on cellular infrastructure introduces certain limitations, particularly in remote or underdeveloped areas where cellular coverage may be weak or unavailable. In such scenarios, vehicles may struggle to maintain consistent communication with the network, potentially leaving them vulnerable to privacy breaches if they cannot implement real-time protective measures.

Despite these limitations, the centralized control inherent in C-V2X enables service providers to detect and mitigate unauthorized tracking attempts or data interception more effectively than decentralized models like DSRC. The network operators can monitor for suspicious activity, apply timely countermeasures, and adjust security protocols as needed, making C-V2X generally more secure for protecting location privacy. However, its overall effectiveness is contingent on the quality and extent of cellular network coverage. This capability highlights C-V2X's potential to offer stronger location privacy protections, particularly in urban areas with comprehensive cellular infrastructure while remaining vulnerable in less connected regions.

*3) Application Scenario of LPPM:* LPPMs are employed at various stages in the lifecycle of location data in vehicular networks to ensure privacy is maintained from the point of data collection to the point of data publication. As shown in Fig. 8, the protection mechanisms can be divided into two main phases: the collection phase (online) and the publication phase (offline). Different LPPMs are used in each phase depending on the sensitivity of the data and the privacy requirements of the application [99].

. **Collection Phase (Online Protection):** During the collection phase, real-time or batch-wise LPPMs are applied to ensure that location data remains secure as transmitted from the vehicle to the LBS servers. In real-time protection, location privacy is preserved. At the same time, data is being sent to time-sensitive LBSs, such as traffic management systems, emergency services, or navigation systems, which require immediate and continuous updates. Real-time LPPMs must ensure low latency while encrypting or anonymizing the data to prevent adversaries from tracking the vehicle's location in real time. Techniques such as dynamic pseudonymization, real-time encryption, and spatial cloaking are commonly employed to safeguard the driver's location data during transmission. These methods reduce the risk of interception by adversaries and ensure that the driver's location is not revealed, even when communicating with time-critical LBSs.

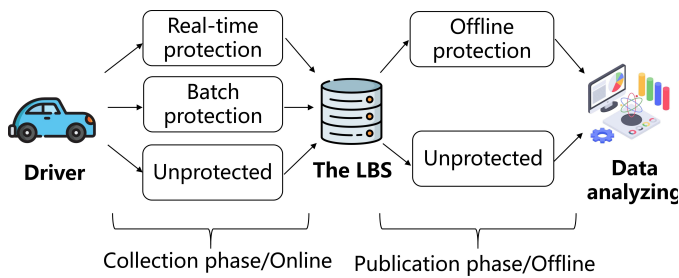

Fig. 8. The cases of employing LPPM to protect location privacy. The LPPM can be utilized in two phases, i.e., the collection and publication phases, to satisfy different privacy requirements.

On the other hand, batch protection is used for less time-sensitive LBSs, where the location data of multiple drivers is aggregated before being sent to the server. This method offers stronger privacy guarantees since data from multiple sources is grouped, making it harder for adversaries to trace specific individuals. Batch LPPMs are especially useful for applications that collect large volumes of location data for statistical analysis, transportation planning, or route optimization, where immediate real-time data is not required. In these cases, techniques such as data aggregation, k-anonymity, or homomorphic encryption can be employed to ensure that even if the data is intercepted, it cannot be linked to individual drivers. The batch-wise approach ensures that the privacy of each driver is protected while maintaining the utility of the aggregated data.

- **Publication Phase (Offline Protection):** Once location data reaches the LBS server, it may be processed and eventually published for various analytical purposes, such as urban planning, traffic pattern analysis, or academic research. In the publication phase, the location data is no longer tied to real-time services and offline LPPMs are employed to protect the entire dataset before it is shared publicly or with third-party entities [100]. Offline LPPMs focus on preventing the re-identification of individuals within published datasets, where adversaries might attempt to link anonymized location trajectories back to specific vehicles or drivers by cross-referencing external data sources.
Offline LPPMs utilize techniques such as differential privacy, which adds statistical noise to the dataset to prevent exact re-identification, or trajectory obfuscation, which alters the dataset to reduce the risk of privacy breaches while maintaining the overall utility for analytical purposes. These methods are particularly important in scenarios where large-scale datasets, including vehicle trajectories over extended periods, are published for research or commercial purposes. Without strong offline protections, adversaries could use sophisticated data mining techniques to infer sensitive in-formation about drivers, such as their frequent destinations, travel habits, or even home addresses, posing significant risks to location privacy.

The effectiveness of LPPMs lies in their ability to balance privacy with data utility. While the goal is to ensure that location data remains private and inaccessible to unauthorized parties, LPPMs must also ensure that the data can still be used for its intended purpose, whether for navigation, traffic management, or urban planning [15]. Real-time LPPMs prioritize low latency and security to ensure that drivers receive accurate and timely information without compromising privacy, while offline LPPMs focus on protecting data during analysis, where re-identification risks are higher. This dual approach ensures that privacy is maintained across the entire lifecycle of location data in vehicular networks. LPPMs play a vital role in protecting the location privacy of drivers in vehicular networks by addressing the vulnerabilities present at various stages of data handling. During the collection phase, real-time and batch LPPMs ensure that sensitive location data is protected from interception or unauthorized access, while offline LPPMs safeguard the integrity of published datasets, preventing adversaries from performing post-hoc attacks aimed at re-identifying individuals from anonymized data. This multi-layered approach to privacy protection is crucial in vehicular networks, where the constant movement of vehicles and the transmission of sensitive location data across open wireless channels make it easy for adversaries to track, monitor, or compromise drivers' privacy.

*4) Key Performance Index of LPPM:* LPPMs are evaluated based on several key performance metrics that reflect their effectiveness in balancing privacy protection with usability and system efficiency. These metrics—privacy, data utility, and efficiency—provide a comprehensive framework for comparing different LPPMs, particularly in location privacy in vehicular networks. Understanding and optimizing these metrics is essential for developing LPPMs that can protect sensitive location data without significantly compromising the performance or functionality of LBSs [1], [101].

- **Privacy:** The privacy metric measures the extent to which an LPPM protects the location data of drivers from adversarial attacks. In vehicular networks, adversaries may infer a vehicle's location, identity, or travel patterns through eavesdropping, data mining, or re-identification attacks on anonymized data [19]. A robust LPPM should ensure that sensitive location data remains inaccessible to unauthorized parties, even in the presence of sophisticated attacks. This can be achieved through techniques like encryption, dynamic pseudonymization, spatial cloaking, or differential privacy, each designed to obscure the true location of the vehicle while still enabling communication with the LBS. However, the challenge in designing effective LPPMs lies in the dynamic nature of vehicular networks. Vehicles frequently move in and out of communication zones, and adversaries can exploit these transitions to track movements over time, making it crucial for LPPMs to continuously protect data across different network environments and interactions. The privacy metric, therefore, is a critical measure of how well the LPPM can adapt to these scenarios while thwarting various privacy threats.

• **Data Utility:** While protecting privacy is paramount, LPPMs must also maintain a high level of data utility to ensure the functionality of LBSs. Data utility refers to the accuracy, completeness, and timeliness of the location data used by LBSs, which in turn affects the quality of services such as navigation, traffic management, and emergency response [101]. Privacy and data utility are inherently conflicting metrics—stronger privacy protection often leads to a reduction in data utility, as more aggressive anonymization or encryption methods obscure valuable information needed by the LBS. For example, applying high levels of spatial cloaking to protect location data may result in less accurate navigation routes or delays in traffic alerts, reducing the effectiveness of the LBS for drivers. In extreme cases, when location data is fully anonymized or obfuscated, the LBS may become unusable, negating its utility entirely. The challenge for LPPMs is to strike a balance between privacy and utility, providing sufficient protection without degrading the service quality. Techniques such as selective obfuscation, context-aware privacy controls, and adjustable privacy levels based on the sensitivity of the service can help LPPMs manage this trade-off, ensuring that drivers receive accurate, timely information while still protecting their location privacy.

• **Efficiency:** The efficiency metric evaluates the computational and storage overheads associated with implementing LPPMs, which is particularly important in resource-constrained environments like vehicular networks [102]. Efficiency is measured in terms of computational time, storage requirements, scalability, and error tolerance. Since vehicular networks require real-time data processing to ensure the safety and coordination of vehicles, LPPMs must operate with minimal delay to avoid disrupting communication between vehicles and LBSs. High computational overheads can lead to latency issues, reducing the responsiveness of time-critical applications such as collision avoidance systems or emergency notifications. Additionally, LPPMs that require large amounts of storage for cryptographic keys, pseudonym management, or trajectory data can strain the onboard systems of vehicles, which may have limited memory and processing power. Therefore, efficient LPPMs must be designed to minimize these overheads, ensuring they can scale to accommodate large numbers of vehicles and communication events without degrading system performance. This is especially important in large urban areas or highway networks, where the density of vehicles can vary dramatically, requiring the LPPM to adjust dynamically to the available network resources. Techniques such as lightweight encryption algorithms, decentralized processing, and efficient key management can help improve the efficiency of LPPMs, making them more suitable for deployment in real-world vehicular networks.

Achieving an optimal balance between privacy, data utility, and efficiency is one of the central challenges in designing LPPMs for vehicular networks. In many cases, improving one metric can negatively affect the others [15]. For instance, enhancing privacy by applying stronger encryption may reduce data utility by making the location information less precise, or it may increase computational overhead, reducing the efficiency of the system. Conversely, prioritizing data utility by minimizing obfuscation or encryption may leave location data vulnerable to attacks. The key to developing effective LPPMs lies in finding an equilibrium where privacy is protected to a satisfactory level without significantly compromising the usability or performance of the system. This balance can vary depending on the specific application or context—LPPMs for time-sensitive LBSs, such as emergency services, may prioritize data utility and efficiency, while those for non-critical applications, such as route planning or parking assistance, may emphasize stronger privacy protections. Additionally, user-controlled privacy settings can allow drivers to choose their preferred balance between privacy and utility based on their individual needs or concerns.

### C. Location Privacy Threat in Vehicular Networks

Vehicular networks, due to their dynamic and open communication environment, are increasingly vulnerable to a range of location privacy threats. These attacks target sensitive location data, aiming to compromise the privacy of drivers by tracking their movements, inferring their identities, or analyzing their behavior over time. One of the simplest, yet impractical, methods to track a vehicle is by illegally installing a Global Navigation Satellite System (GNSS) on the vehicle [103]. While this could provide adversaries with real-time location data, it is not feasible for large-scale surveillance because it requires physically placing GNSS equipment on each target vehicle. Therefore, adversaries typically resort to more sophisticated digital attacks that exploit the weaknesses inherent in the wireless communication systems of vehicular networks.

*1) Passive and Active Attack:* Location privacy attacks in vehicular networks can be conceptualized as a Multiple Target Tracking (MTT) problem, where adversaries gather a set of noisy measurements or observations detected periodically by a sensor, with the goal of inferring the most accurate estimate of the driver's location or trajectory. Through these techniques, adversaries can systematically reduce the uncertainty surrounding the vehicle's location, enabling precise tracking over time. These location privacy attacks are typically categorized into passive and active attacks, each posing unique risks and challenges to vehicular network security.

Passive attacks involve adversaries silently monitoring and analyzing the data traffic in vehicular networks to infer location information. These attacks are stealthy in nature, as they do not directly interfere with communication between vehicles or disrupt network services, making them difficult to detect. The goal of passive attackers is to collect as much location-related data as possible by exploiting the inherent vulnerabilities of the network. The following are common passive attack types in vehicular networks [104]:

• **Wireless Eavesdropping Attack:** The open and decentralized nature of vehicular networks, coupled with the use of wireless communication protocols, makes them highly vulnerable to wireless eavesdropping attacks. In such attacks, adversaries can intercept communication signals exchanged between vehicles or between vehicles and RSUs over the shared wireless medium. Given that vehicular networks are designed to broadcast information to facilitate road safety and navigation, much of the data transmitted is unencrypted or insufficiently protected, exposing sensitive information like real-time location, navigation requests, and travel routes to unauthorized interception. Eavesdropping allows adversaries to compile detailed movement profiles of vehicles, which can be further exploited to track drivers' routines, identify frequent destinations (e.g., home or workplace), and potentially launch more targeted attacks.

The privacy risks associated with wireless eavesdropping are exacerbated in scenarios where vehicles frequently communicate in public spaces, making it easier for adversaries to set up listening devices or deploy passive monitoring equipment. Moreover, adversaries can use advanced analytical techniques to correlate the captured data with other information sources, thereby reconstructing entire vehicular trajectories or inferring identities, even if the original data is anonymized.

- **Tracing Back Attack:** In tracing back attacks, adversaries exploit triangulation techniques to calculate the precise location of a vehicle by measuring the signal strength, direction, or TDoA from multiple antennas deployed at different locations [105]. By using at least two strategically positioned antennas, the adversary can capture and analyze communication signals, such as those from V2X communication systems, enabling them to estimate a vehicle's position with high accuracy. This form of tracking becomes particularly effective in dense urban environments, where the proliferation of wireless infrastructure and the prevalence of signal reflection, refraction, and multipath propagation can actually enhance the accuracy of triangulation.

Multipath propagation, typically considered a challenge in wireless communication, may provide adversaries with additional data points to refine their estimates, especially when the adversary employs advanced signal processing techniques. The triangulation method poses a significant threat to location privacy because it does not require direct access to the vehicle's communication system or any cooperation from the vehicle itself, making it difficult to detect or mitigate. Moreover, the passive nature of this attack allows adversaries to track vehicles continuously over extended distances without raising suspicion, leading to long-term profiling, route prediction, and possible exposure of sensitive destinations such as home addresses, workplaces, or frequently visited locations.

- **Traffic Analysis Attack:** In a traffic analysis attack, adversaries exploit the data flow patterns in vehicular networks to infer the location and movement of vehicles by monitoring the volume and frequency of communication. Core nodes, such as RSUs and network hubs, play a pivotal role in vehicular networks by handling substantial amounts of data traffic between vehicles and LBSs [106]. These nodes are responsible for facilitating communication, making them prime targets for adversaries seeking to gather location-related information.

By analyzing traffic flow to and from these core nodes, adversaries can identify high-priority vehicles, such as emergency responders or commercial delivery trucks, that generate more frequent or voluminous data exchanges. The high volume of traffic to these vehicles, compared to ordinary vehicles, allows adversaries to track their movements or infer their locations. Additionally, the temporal and spatial correlation of traffic can reveal patterns, such as frequent visits to specific areas, allowing adversaries to predict future movements or identify routine destinations. This poses a significant risk to location privacy, as adversaries can use this information for long-term surveillance, profiling, or targeted attacks.

- **Packet Analysis Attack:** Even when vehicular networks employ encryption to secure communication, they remain susceptible to sophisticated packet analysis attacks. Such attacks exploit the metadata associated with transmitted packets rather than the encrypted payload itself. Adversaries capture and analyze various packet features, such as timestamps, packet sizes, transmission intervals, and packet frequencies, to infer sensitive location information [107]. For instance, by monitoring the appearance of packets with the same ID at different geographical locations with consistent timestamp correlations, attackers can map out the trajectory of a vehicle, thereby compromising its location privacy.

This is particularly concerning in scenarios where vehicles are repeatedly transmitting similar packets while navigating through various checkpoints or RSUs. Even slight variations in packet timing or sizes, often resulting from vehicle speed changes or congestion, can provide adversaries with enough information to deduce real-time movements and predict future routes. The attack becomes more effective in dense urban environments, where multiple observation points can be used to triangulate positions and refine trajectory estimates.

- **Back-rolling Attack:** Adversaries leverage substantial storage and computational capabilities to archive and analyze extensive historical data on vehicular movements, which subsequently correlate with newly captured location data to track a vehicle's current position and anticipate its future trajectories. By maintaining a comprehensive database of movement patterns, adversaries can construct detailed behavioral profiles of drivers, enabling them to identify routine patterns such as daily commutes, frequent visits to specific locations, and preferred routes. This granular insight into a driver's habits significantly heightens the risk of privacy invasion, as adversaries can pinpoint sensitive locations like home and work addresses with high precision.

The predictive aspect of back-rolling attacks enables adversaries to foresee potential future locations, thereby facilitating even more intrusive surveillance measures, such as targeted tracking or physical stalking. The severity of this threat is further exacerbated by the adversaries' ability to exploit temporal data for pattern recognition, making it challenging for drivers to conceal their movements over time through conventional privacy-preserving techniques like pseudonym changes or encryption.

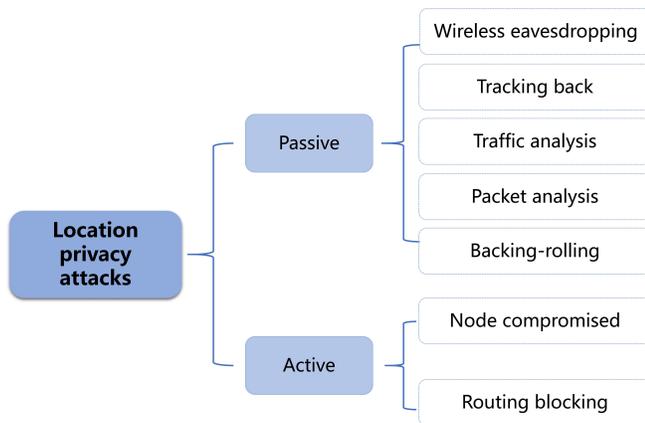

Fig. 9. Major privacy attacks. The attacks on location privacy can be classified into passive and active attacks. Passive attacks aim to monitor and analyze the data traffic, while active attacks interfere with vehicular networks.

Unlike passive attacks, active attacks involve direct interference with the vehicular network, such as disrupting communication channels, injecting false data, or impersonating legitimate nodes to deceive vehicles or RSUs. These attacks are more aggressive but also easier to detect compared to passive attacks. Active attackers can significantly compromise location privacy by manipulating the network to gather location data or prevent vehicles from maintaining their privacy. The following are common active attack types in vehicular networks [104]:

- **Node Compromised Attack:** Node-compromised attacks present a critical threat to location privacy in vehicular networks by compromising the trust and integrity of legitimate network components, such as RSUs and On-Board Units (OBUs) [108]. When adversaries gain control over these nodes, they can intercept, manipulate, and analyze the data transmitted between vehicles and the network. This enables attackers to extract sensitive location information from captured data packets and map out the network topology, including commonly used routes and crucial network nodes. The impact of such attacks goes beyond mere data breaches, as they can lead to long-term privacy violations and further exploitation of compromised network components.

  Compromised nodes can be key entry points for launching more advanced attacks, such as node cloning, where adversaries replicate legitimate node identities to create deceptive communication paths. This can mislead vehicles into sharing sensitive information with these cloned nodes, enabling real-time tracking or even the manipulation of navigation instructions, potentially directing vehicles into vulnerable or hazardous locations. Additionally, compromised nodes can undermine the effectiveness of LPPMs by disrupting privacy strategies such as pseudonym changes, which depend on the trustworthiness of network infrastructure. As a result, the exploitation of compromised nodes poses a significant challenge to maintaining location privacy in vehicular networks.

- **Routing Blocking Attack:** Routing-blocking attacks are a serious threat to location privacy in vehicular networks as they disrupt the normal flow of data transmissions, jeopardizing both the integrity and confidentiality of location information [109]. In such attacks, adversaries selectively intercept or block data packets along specific communication routes, particularly those used by the target vehicle to communicate with RSUs or other vehicles. By blocking these routes, adversaries can force the target vehicle to reroute its communication through less secure or compromised nodes, enabling them to monitor these alternative paths. This allows attackers to trace the origin and destination of communications, potentially revealing the vehicle's location and movement patterns. Additionally, these disruptions can cause delays in data transmission, creating gaps in communication patterns that adversaries can exploit to infer the vehicle's position. The dynamic nature of vehicular networks further complicates this issue, as vehicles constantly enter and leave the network, making it difficult to establish secure and reliable communication routes. The adversary's ability to force vehicles onto alternative paths not only increases the risk of data interception and unauthorized access but also opens up opportunities for more sophisticated attacks, such as location spoofing, where false location data is injected into the network to mislead vehicles and RSUs.

*2) Attack based on Physical Characteristic:* Attacks based on physical characteristics represent a significant threat to location privacy in vehicular networks, as they exploit the inherent identifiability of vehicles through attributes such as shape, color, license plates, and other distinguishing features (as shown in Fig. 10[2], ). Advanced tracking technologies like Automatic Number Plate Recognition (ANPR) systems have made it feasible for adversaries to systematically monitor and record vehicles' movements across multiple locations [110].

Once a vehicle's physical characteristics are identified at a single location, the adversary can utilize this information to establish a temporal-spatial profile of the vehicle's trajectory. As the vehicle is observed at more locations, the accuracy of the trajectory estimation increases, reducing the adversary's estimation error and enhancing their ability to predict future movements. This continuous monitoring not only compromises the driver's real-time location privacy but also enables long-term tracking, profiling, and behavioral analysis. Such data can be further leveraged for subsequent, more invasive attacks, including identity inference or even targeted surveillance. The risk is amplified in scenarios where multiple sensors, such as roadside cameras and toll booths, are networked together to form a comprehensive surveillance grid, providing adversaries with a holistic view of the vehicle's movements.

*3) Attack on Inside-vehicle Message:* Inside-vehicle communication systems, such as Electronic Control Units (ECUs), tire pressure monitoring systems (TPMS), and remote keyless entry (RKE) technologies, play a critical role in modern vehicles by controlling various functions and enhancing driver safety. However, these systems also pose significant risks to location privacy due to their susceptibility to unauthorized access and data breaches. Adversaries can exploit these vulnerabilities to gain access to sensitive location-related data. For instance, by connecting an illegal OBD reader to the OBD port, attackers can retrieve real-time information about the vehicle's speed, acceleration, steering angle, and even GPS coordinates stored within the vehicle's network [108]. Such data, when combined, can reveal precise location trajectories and driving patterns.

---

[2]The vehicle in the figure is owned by the authors

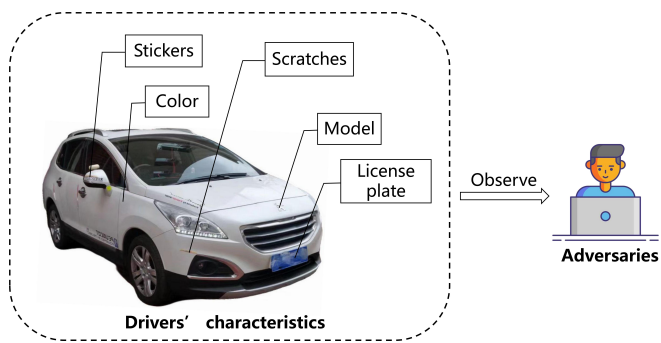

Fig. 10. The identifiable characteristics of the vehicles for tracking. The adversaries can track the vehicle through the appearances like stickers, color, model, scratches, and license plate. The optical vision-based tracking can almost not be prevented.

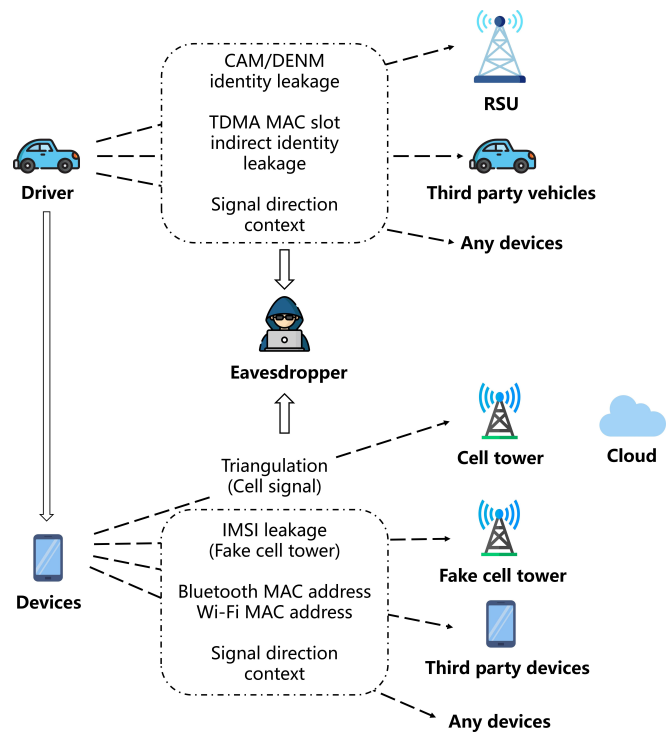

Fig. 11. V2X and in-vehicle communication threats. The adversaries can obtain the data by eavesdropping on the V2X communication channel since the vehicles broadcast data periodically. The adversaries can also use the in-vehicle communications to track the vehicles easily.

The wireless data transmission within the vehicle's network further exacerbates these privacy risks. For example, the TPMS often broadcasts unencrypted data, including the vehicle's unique sensor identifiers, which can be intercepted from a distance of up to 40 meters, allowing adversaries to track the vehicle's movement without direct access to the vehicle itself. Similarly, RKE systems emit unique identifiers over short-range broadcasts to unlock doors, and if these identifiers are captured, they can not only compromise vehicle security but also serve as a digital fingerprint to trace the vehicle's location [111].

*4) Attack on V2X Message:* V2X communication is a cornerstone of modern intelligent transport systems, facilitating real-time exchange of information between vehicles and infrastructure to enhance road safety and traffic efficiency. However, this continuous broadcasting of location-sensitive information exposes significant privacy vulnerabilities. Key V2X message types such as Cooperative Awareness Messages (CAM) and Decentralized Environmental Notification Messages (DENM) contain detailed vehicle status data, including precise timestamps, geographical coordinates, and speed information [112]. These messages are essential for safety-critical applications like collision avoidance and hazard warnings but are transmitted in plaintext to minimize latency and ensure timely delivery of alerts. Unfortunately, this lack of encryption makes it easy for malicious entities to intercept and analyze these broadcasts, enabling them to reconstruct the vehicle's real-time trajectory and identify potential patterns in its movement.

The digital signatures appended to CAM and DENM messages, intended for authentication and integrity verification, can be exploited by attackers to pinpoint the vehicle's precise location and track it over time. This is exacerbated by the high temporal resolution of these messages, broadcast at frequencies ranging from 1 to 10 Hz, providing a continuous stream of highly granular location data. Moreover, vulnerabilities in the

MAC layer protocols, such as using Time-Division Multiple Access (TDMA), further compromise privacy. TDMA assigns unique time slots to each vehicle, effectively functioning as a digital identifier. Adversaries can leverage this information to monitor slot allocations and track the movement of vehicles across the network [113].

In addition to these direct V2X threats, location privacy can be compromised through vehicle peripheral devices. For example, telecommunication signals from drivers' mobile phones can be triangulated using nearby cellular towers, enabling adversaries to determine the vehicle's approximate location. Advanced localization techniques, employing data from multiple towers, can achieve high-precision tracking. Moreover, the unique International Mobile Subscriber Identity (IMSI) associated with each mobile device can be intercepted by IMSI catchers, allowing persistent tracking of a vehicle as long as the driver's phone remains within [27]. Similarly, Bluetooth devices within vehicles, including those used for in-car entertainment and driver assistance, frequently broadcast advertisement packets containing their MAC addresses. Adversaries can monitor these broadcasts to deduce the vehicle's location and movement history [114].

The adversaries can track vehicles without the content of the transmitted data. As shown in Fig. 9 and Fig. 10, the signal direction context is sufficient for the adversaries to realize traceback attacks within the wireless sensor networks [115]. The adversaries can also infer the position and velocity changes with received CAM in vehicular networks.

*D. Key Takeaway*

The analysis of location privacy threats and protection mechanisms in 5G/6G vehicular networks reveals several pivotal insights about the state of the field and the challenges faced.

Vehicular networks are inherently vulnerable due to their open communication environment, high mobility, and reliance on continuous data exchange. This exposes them to adversarial techniques, including passive eavesdropping, active attacks, and physical-layer exploits. Even advanced privacy-preserving mechanisms, such as pseudonym changes or encryption, struggle to provide comprehensive protection due to the adversaries' ability to exploit metadata, signal characteristics, or cross-layer vulnerabilities. The challenges are particularly pronounced in dense urban environments where adversaries can easily correlate data from multiple sources and in rural areas where limited network density undermines cooperative privacy strategies.

The interplay between privacy, utility, and efficiency is a fundamental challenge in designing LPPMs. High privacy levels often reduce the utility of location-based services, while excessive resource demands compromise real-time functionality. Striking this balance is critical for ensuring that LPPMs protect location data and support vehicular applications' operational requirements, such as collision avoidance or emergency response. This underscores the importance of designing LPPMs that integrate seamlessly into the dynamic and resource-constrained environment of vehicular networks.

The key observations of this section are as follows:

- **Evolving Threat Landscape:** Adversaries are becoming increasingly sophisticated, employing advanced techniques such as metadata correlation and physical-layer attacks that bypass traditional privacy defenses. Thus, LPPMs need to address threats across multiple attack vectors.
- **Intrinsic Network Challenges:** High mobility, dynamic topology, and varying vehicle density create unique vulnerabilities in vehicular networks. These characteristics complicate the implementation of static privacy mechanisms, emphasizing the need for adaptive and context-sensitive solutions.
- **Privacy and Utility Balance:** The inherent tension between maintaining location privacy and preserving the utility of location-based services remains a central challenge. Solutions must ensure that privacy protections do not compromise critical vehicular applications' accuracy, timeliness, or functionality.

## IV. LPPM IN VEHICULAR NETWORKS

In this section, we assess the existing LPPMs in three categories and discuss the limitations of each category, as shown in Table IV. The LPPMs are classified into user-side, server-side, and user-server-interface LPPMs. User-side LPPMs process location data during the collection phase, while server-side LPPMs protect location privacy during publication. The user-server-interface LPPMs use trusted third parties and secure communications to realize location privacy protection. We also summarize the limitations of LPPMs from the aspect of localization and communication requirements and review the methods for balancing location privacy and data utility from theory and practice. The theoretical methods include the blockchain, adaptive parameters, hybrid, encryption, element simplification, and virtual nodes. The real-world in-vehicle tracking techniques, i.e., COVID-19 tracking applications, are used as examples to discuss the balance between location privacy and quality of service in practice.

*A. User-Side LPPMs*

The user-side LPPMs aim to protect location privacy on the driver side in the collection phase. The popular user-side LPPMs include pass-and-run, certificate, secure computation, and data perturbation. The comparison of the existing user-side LPPMs is shown in Table V.

*1) Pass-and-Run:* The pass-and-run strategy is a privacy-preserving technique designed to enhance location privacy in vehicular networks by leveraging the inherent mobility of vehicles. Instead of directly transmitting sensitive location data to LBSs, vehicles relay the information through intermediate vehicles, reducing the direct exposure of the data. Initially conceptualized by Dunbar et al. [116] for use in delay-tolerant networks, this method allows a vehicle to dynamically decide whether to pass its message to nearby vehicles or to upload it directly to LBSs based on network conditions and privacy requirements.

Building upon this concept, Lu et al. [117] introduced a lightweight pass-and-run method that incorporates data perturbation techniques aligned with vehicle mobility and transmission delay constraints. This method employs two adaptive strategies: a greedy strategy, where vehicles prioritize minimizing delay by choosing the closest relay, and a random strategy, which maximizes privacy by randomly selecting a relay vehicle. The use of asymmetric encryption further protects the integrity and confidentiality of the transmitted messages during relaying. Despite these advancements, the method incurs significant transmission delays, especially in sparse vehicular networks or under high mobility conditions, where the likelihood of finding suitable relay vehicles is reduced. This latency can potentially hinder real-time applications and limit the effectiveness of the pass-and-run strategy in scenarios requiring prompt data delivery, thus necessitating a balance between privacy preservation and communication efficiency.

*2) Certificates for Privacy:* Certificates, or privacy-preserving attribute-based credentials, are pivotal in maintaining location privacy in vehicular networks. These cryptographic mechanisms enable drivers to authenticate themselves and prove possession of certain attributes without revealing unnecessary personal information to LBSs. The flexibility of these certificates allows for selective disclosure of only the required information that meets the LBSs' predetermined predicates, thereby mitigating risks associated with over-sharing sensitive data such as real-time location [139]–[142]. The robustness of certificate-based systems often relies on malleable signatures, allowing for controlled manipulation of certain message parts to prevent the disclosure of specific attributes, including location.

TABLE IV
AN OVERVIEW OF THE EXISTING LPPMs.

| Category | Techniques | Description | Benefits | Limitations |
|---|---|---|---|---|
| User-side LPPMs | Pass-and-Run | Transmit location data through other vehicles | Transmission delay breaks the linkage of location information | High communication complexity |
| | Certificates for Privacy | Authentication | Provide both location privacy and authentication | High computational and storage consumption |
| | Secure Computation | Operate encrypted data directly | Operation is flexible and does not reveal private data | High computational and transmission delay |
| | Data perturbation | Apply LBSs with fake location data rather than the actual version | Considers the prior knowledge of the adversaries and does not need a trusted third party | Low data utility |
| Server-side LPPMs | Statistical Disclosure Control | Anonymize or obfuscates data in the dataset | Keep the general statistical features of the dataset | Need a trusted third party |
| | Homomorphic Encryption | Operate encrypted data directly on the server side | The operation could be flexible and does not need to decrypt data | High computational consumption and transmission delay |
| | Private Information Retrieval | Hide the requested items | The drivers can apply for LBSs without disclosing their requirements | High computational consumption |
| | Searchable Encryption | Hide plaintext keywords in searching. | Can be combined with other methods | Low accuracy |
| User-server-interface LPPMs | Secure Communication | Use traditional protocols or end-to-end encrypted services | Techniques have been developed in past decades | There are some limitations of the traditional protocols |
| | Trusted Third Party | Introduce trusted third parties to assist communication | High efficiency | It is an ideal environment |

• **Attribute-based Signature (ABS):** Shahandashti et al. [118] introduced the concept of ABS, providing drivers with the ability to sign messages using specific attributes without disclosing personal identifiers. These signatures rely on the attribute value, which can be expressed as binary-bit strings or based on complex data structures, allowing for selective disclosure in vehicular networks. While ABS offers fine-grained control over the information shared, enabling users to conceal sensitive data such as their identity or location, the system's effectiveness hinges on maintaining strong anonymity.

El Kaafarani et al. [120] formalized this requirement by emphasizing that the signature should neither reveal the driver's identity nor allow adversaries to infer location through the identification of attribute patterns. This is critical for preserving location privacy, as even seemingly benign attributes, if linked over time, could be exploited to trace a driver's movement. However, one limitation of ABS lies in its potential vulnerability to advanced correlation attacks, where adversaries cross-reference multiple data points or attributes to infer location. To address this, Kaaniche et al. [121] proposed an improved ABS model using concrete mathematical constructions and the random oracle model, significantly enhancing both location and identity privacy. While this improvement strengthens privacy guarantees, the computational complexity associated with generating and verifying signatures remains a concern, particularly in high-mobility environments like vehicular networks where real-time communication is essential. Despite these limitations, ABS represents a powerful tool for balancing the need for location privacy with the utility of shared data in vehicular communications, offering flexible and scalable solutions for privacy preservation.

• **Group Signature:** Group signatures offer a strong mechanism for preserving location privacy by allowing a member of a group (e.g., a driver) to sign messages anonymously on behalf of the group without revealing their identity. This ensures that while the message is verified as coming from a legitimate member, the actual identity of the signer remains concealed, which is essential for protecting drivers' location privacy in vehicular networks.

Enhancements like the linkable group signature proposed by Zheng et al. [122] add functionalities such as auditing and tracing, improving the framework's security by incorporating advanced cryptographic techniques like blind signatures and trapdoor commitments. This allows authorities to audit or trace suspicious activities without compromising general anonymity. The use of bilinear pair cryptography, as explored by Hakeem et al. [59], reduces the size of signatures and increases the efficiency of message verification across multiple zones. However, bilinear pair cryptography can be computationally intensive for OBUs and RSUs, as pointed out by Wu et al. [123], which can limit the scalability and practicality of these systems in real-time vehicular environments. To address this, elliptic curve cryptography is suggested to reduce computational complexity, making group signatures more feasible for devices with limited resources. Ring signatures, as developed by Mundhe et al. [124], offer unconditional privacy by passing messages through verified legal vehicles, further enhancing location privacy by ensuring anonymity within a group. Additionally, the use of pseudonyms and allowing RSUs to assist in signature generation, as in ring-signature-based LPPMs, significantly reduces authentication delays, improving real-time efficiency. Nonetheless, the trade-offs include increased overhead in signature management and potential challenges in revocation, where compromised keys or pseudonyms may need to be efficiently invalidated.

TABLE V
COMPREHENSIVE OVERVIEW OF USER-SIDE LPPMS IN VEHICULAR NETWORKS

| LPPM Type | Mechanism Description | Key Reference | Primary Advantage | Notable Disadvantage |
|---|---|---|---|---|
| Pass and Run | Treating vehicular networks as delay-tolerant networks, routing messages via other vehicles | [116] | Disrupting spatial-temporal correlation | High complexity in communication and transmission |
| | | [117] | Disrupting spatial-temporal correlation; Personalized protection | |
| Attribute-based Signature | Sign messages with fine-grained control over identifying information. | [118] | Fine-grained control; Personalized Protection | The signature can reveal the driver High storage consumption High computational consumption |
| | | [119], [120] | Anonymity attribute-based signature | |
| | | [121] | Anonymity attribute-based signature; Low computational consumption | High storage consumption |
| Group Signature | Group members collectively authenticate anonymously through group signatures | [122] | Traceable group signature; Increase communication efficiency and security | High signature generation time and verification time |
| | | [59] | Short-size signature; Low signature generation time and verification time | Bilinear pair cryptography is complex for OBUs and RSUs |
| | | [123] | Use an elliptic curve cryptosystem; Decrease computational complexity | Ignore the cooperation of vehicles in the authentication |
| | | [124] | Achieve unconditional privacy with anonymous ring signature | High authentication delay |
| | | [125] | Introduce RSUs in signature; Decrease authentication delay | High bandwidth resources overhead in transmission |
| | | [126] | Certificeless signature; Low bandwidth resource overhead in transmission; Low storage consumption | The participate of vehicles influences privacy protection less |
| Sanitizable Signature | Semi-trusted sensors, when authorized, can alter parts of a signed message | [127] | First develop sanitizable signature | Log information can expose the sensitive data |
| | | [128] | Hide the authenticated identification; Anonymized log information | Overlook the prior knowledge of the vehicle |
| Blind Signature | Message content is obscured prior to being signed | [129] | Use zero-knowledge to improve the security of the blind signature | Bilinear pair cryptography is complex for OBUs and RSUs |
| Secure Computation | A scheme for secure outsourced computation utilizing multiple keys | [130] | Low authentication and communication consumption; Avoid duplicating and useless encrypted LBS messages | No key agreement and mutual authentication |
| Data Perturbation | Perturb location before sending the it to LBS servers | [131] | Use dummy locations to achieve $k$-anonymity | Overlook spatio-temporal correlation |
| | | [132] | Combine the local differential privacy and anonymity | |
| | | [133] | First improve Geo-I | |
| | | [134] | Consider spatio-temporal correlation | |
| | | [135] | Balance utility and privacy in edge computing | High computational consumpation |
| | | [136] | Personalized Protection | Ignore the short-distance location shift |
| | | [137] | High data utility for location-sensitive LBS | High privacy cost |
| | | [138] | Use long-term identifier to reduce the computation and storage consumption | Less security due to the long track window |
| | | [58] | First use DP to assist pseudonym swap | Overlook the non-repeatability of the pseudonyms |

- **Sanitizable Signature:** Sanitizable signatures, introduced by Ateniese et al. [127], allow authorized semi-trusted parties, or "censors," to modify specific parts of a signed message without needing to interact with the original signer. This capability is particularly useful in situations where sensitive data, such as a driver's identity or location, needs to be concealed or altered before the data is shared with third parties. For instance, Pamies et al. [128] demonstrate how combining log anonymity with sanitizable signatures can protect identity and location data when transferring information from local nodes to remote storage servers. By allowing selective modification, sanitizable signatures help maintain privacy while ensuring the message's integrity. However, this method has limitations, especially concerning trust: the semi-trusted censors could potentially abuse their power, altering more information than necessary or even introducing malicious modifications. Moreover, the computational complexity of verifying modified signatures and maintaining control over which parts of the message can be altered adds overhead, which may impact the scalability and efficiency of vehicular networks. Sanitizable signatures offer significant privacy benefits by enabling flexible data sharing, particularly in scenarios where dynamic data sanitization is necessary to protect location privacy without compromising message authenticity.

- **Blind Signature:** Blind signatures offer a robust privacy-preserving mechanism, particularly relevant for protecting location privacy in vehicular networks. In a blind signature scheme, the content of a message is hidden or "blinded" before it is signed, meaning that the signer—typically a trusted entity—cannot see the actual message content, which could include sensitive location data. This separation between the message author and the signer ensures that even though the message is authenticated, the signer remains unaware of its specific content, providing strong anonymity for the driver. Sun et al. [129] enhanced this approach in a fog-computing-based crowdsensing architecture, where drivers' identity and location privacy are safeguarded through partially blind signature authentication. By incorporating zero-knowledge verification, the system ensures that the signer can verify the authenticity of the request without learning anything about the driver's location, strengthening both security and privacy. Building on this concept, recent advancements in blind signature schemes have introduced variants that incorporate public metadata [143]. These allow issuers to tie anonymous tokens to specific metadata without compromising user unlinkability, as demonstrated by a real-world deployment in Google's VPN service. This approach maintains the privacy and scalability of traditional blind signatures while enhancing their utility in specific applications.

Blind signatures face several limitations. One key issue is the increased computational overhead, especially in resource-constrained vehicular environments where real-time responses are critical. Additionally, while zero-knowledge proofs bolster privacy, they also add complexity to the verification process, potentially slowing down authentication in high-traffic scenarios. Moreover, if not managed properly, the need for trusted third parties in blind signature schemes could introduce potential vulnerabilities or single points of failure. The reliance on strong RSA moduli and the assumption of cryptographic hash functions as random oracles further limits the flexibility of these schemes, which could pose challenges in environments requiring modular adaptability.

In summary, while signature-based methods such as attribute-based signatures, group signatures, and blind signatures offer significant advantages for preserving location privacy in vehicular networks, they are not without their limitations. One of the key challenges lies in the certificate management process, which incurs substantial computational and storage overhead. Managing certificates, particularly in large-scale, dynamic vehicular networks, requires constant updating, revocation, and verification, all of which can strain the computational resources of both OBUs and RSUs. This becomes especially problematic when handling high volumes of data in real-time, where delays in certificate verification could lead to latency in critical applications such as collision avoidance systems. Additionally, the storage demands are amplified by the need to maintain multiple pseudonyms or cryptographic keys to ensure ongoing privacy, increasing the risk of inefficiency in resource-constrained environments. The complexity of operations such as bilinear pairing, elliptic curve cryptography, and zero-knowledge proofs further exacerbate these issues, making real-time processing challenging. Despite these concerns, signature-based methods remain essential for enhancing location privacy by ensuring that messages can be authenticated without revealing sensitive information. Future developments in lightweight cryptographic techniques and decentralized certificate management protocols could help mitigate these issues by reducing the computational burden while maintaining strong privacy protections across vehicular networks. The adoption of more scalable approaches, such as certificateless cryptography or aggregated signatures, could also offer potential solutions to balance security, privacy, and efficiency.

*3) Secure Computation Mechanism:* Secure computation mechanisms play a pivotal role in protecting location privacy by enabling the processing of sensitive location data without directly revealing it to third parties, including LBSs. These techniques, originally formalized in 1982 by Yao's protocol based on the millionaire problem [144], allow multiple parties to jointly compute a function over their inputs while keeping those inputs private. In vehicular networks, secure computation ensures drivers can use LBSs without exposing their precise location data. By processing encrypted location information, secure computation mechanisms allow LBS providers to deliver services like route optimization or traffic alerts without gaining access to the raw data, thus preserving the driver's privacy.

Zhuo et al. [130] extend this concept by proposing a multi-key secure outsourced computation scheme that enhances location privacy without requiring direct interaction between LBS servers and drivers. Their approach minimizes unnecessary communication overhead by eliminating redundant or unusable encrypted messages before identity verification. This is critical in maintaining the quality of service in LBS applications, where excessive communication delays can degrade user experience. The multi-key scheme leverages homomorphic encryption, allowing computations to be performed on encrypted data, ensuring that only the final output—such as navigation instructions or traffic alerts—needs to be decrypted. This model significantly enhances privacy and efficiency, reducing the computational burden on the driver's device and ensuring that the LBS cannot infer sensitive location details.

While secure computation mechanisms provide robust privacy guarantees, they are often computationally expensive, particularly when dealing with complex vehicular applications that require real-time responsiveness. The cryptographic operations involved, such as homomorphic encryption and secure multi-party computation, can introduce latency, which may not be acceptable in time-sensitive scenarios like emergency vehicle routing or collision avoidance systems. Moreover, secure computation schemes assume that all participating entities adhere to the protocol, meaning they are vulnerable to adversaries who may attempt to disrupt the computation process or manipulate the encrypted inputs. Future work could focus on optimizing these cryptographic protocols to reduce latency and computational complexity, making secure computation more practical for real-time vehicular applications.

*4) Data Perturbation:* As a branch of obfuscation-based LPPMs, data perturbation techniques play a critical role in protecting drivers' location privacy by distorting the actual location information before it is transmitted to LBSs [15]. These techniques can be applied on the user side, where drivers manipulate their data prior to transmission, or on the server side. In user-side data perturbation, drivers do not need to trust external entities, as they control the obfuscation of their location information. This independence from centralized trust models enhances privacy by reducing the risks associated with potential data breaches on LBS servers. The challenge for user-side perturbation lies in balancing the balance between privacy and data utility—ensuring that location data remains useful for LBSs, while simultaneously providing robust privacy protection.

- **Dummy-Based Method:** Dummy-based methods offer a simple yet effective approach to preserving location privacy by generating multiple fake locations (dummies) alongside the real ones, making it difficult for adversaries to distinguish the true location. These methods do not require the involvement of a trusted third party or key-sharing mechanism, allowing drivers to maintain full control over their location data. Niu et al. [131] improved this technique by introducing a dummy-location selection algorithm that leverages $k$-anonymity, where the real location is hidden among at least $k$ other dummy locations, thereby increasing the uncertainty for potential attackers. This method selects dummy locations based on the entropy of anonymity, maximizing privacy by ensuring the dummies are not easily distinguishable. Building on these methods, Tadakaluru and Qin [145] proposed a Voronoi-based semantically balanced dummy generation framework, which generates dummy locations that account for both spatial and semantic similarity. Their approach effectively resists temporal constraint attacks, a common limitation of traditional dummy methods, by ensuring that dummy locations are semantically balanced and geographically dispersed. However, the framework introduces additional computational overhead, which may limit its suitability for real-time applications.

  One significant limitation of dummy-based methods lies in their vulnerability to spatio-temporal correlation attacks, particularly in sequential LBS requests. Over time, adversaries can analyze patterns such as time reachability and direction similarity to filter out the dummies, gradually exposing the real location [134]. To address this, Liu et al. proposed filtering strategies that improve privacy by ensuring that the dummies behave in a realistic manner over time, such as maintaining similar trajectories or travel times as the real location. While these strategies enhance privacy, they also increase computation delay, which may hinder real-time applications in vehicular networks. Despite this, the storage cost remains acceptable, making dummy-based methods a practical choice for applications where real-time response is less critical, and privacy concerns are paramount. In summary, while dummy-based methods provide a relatively low-cost and decentralized approach to location privacy, they face limitations in high-frequency data environments where temporal and spatial correlations can reduce their effectiveness.

- **Local Differential Privacy (LDP):** Differential privacy can bind the knowledge obtained by the adversaries but could decrease the quality of LBSs. From the LBS providers' perspective, applications can provide a high quality of services if high accuracy is received, but most LBSs can accept location data that is not entirely accurate. The differential privacy can be defined as follows [146],

*Definition 1 (Differential Privacy (DP)):* A mechanism $\mathcal{M}$ satisfies $\varepsilon$-DP if and only if, for any pair of data $x_i$ and $x_j$,

we have

$$\frac{\Pr[\mathsf{M}(x_i) \;\rightarrow\; y]}{\Pr[\mathsf{M}(x_j) \;\rightarrow\; y]} \leq e^{\varepsilon}, \tag{1}$$

where $y$ is the output of the mechanism. $\varepsilon$ depends on the sensitivity of the dataset. Typically, $\varepsilon$ is a small positive number.

LDP is a distributed variant of traditional differential privacy that allows the drivers to perturb their location information before sending it to servers. Erlingsson et al. [132] amplify the privacy to achieve high privacy-preserving capability from local differential privacy by combining differential privacy and anonymity. The authors point out that location privacy security can be achieved without adding any significant noise if the method employs LDP on the client side and a shuffling strategy on the server side. Nevertheless, the authors ignore the spatio-temporal correlation in the road network.

- **Geo-Indistinguishability (Geo-I):** Geo-I is first proposed by André [133]. Geo-I allows the drivers to enjoy *Er*-differential privacy in the given obfuscate radius $r$ with a privacy budget $E$, which is as given by

*Definition 2 (Geo-I):* A mechanism $\mathsf{M}$ satisfies $\varepsilon$-Geo-I if and only if, for any two locations $x_i$ and $x_j$, the following holds

$$\frac{\Pr[\mathsf{M}(x_i) \;\rightarrow\; y]}{\Pr[\mathsf{M}(x_j) \;\rightarrow\; y]} \leq e^{\varepsilon \, d\{x_i, x_j\}}, \tag{2}$$

where $y$ is the output of the mechanism and $d(x_i, x_j)$ is the distance of $x_i$ and $x_j$.

Based on the Geo-I, Zhou et al. [135] design a framework to balance the utility and privacy in edge computing. The framework includes two parts: privacy-preserving location-based service usage method and privacy-based service adjustment. The authors add two-dimensional Gaussian noise to shift actual locations. Although the authors consider the balance between privacy and service quality, the framework can lead to a high calculation consumption. Based on the background knowledge of trusty serves, Li et al. [147] employ correlation probabilities and correlation transition probabilities to realize Geo-I. The proposed method can provide different privacy-preserving levels for various requirements. The location shift decides the level of privacy protection. The shift is invalid when the driver's location shift is shorter than the threshold. Although the method can provide a different protection level, the method ignores the road condition. If the traffic jams and the driver applies LBSs frequently, the obfuscated location, by which the adversaries can obtain the driving state, will never change. Li et al. [137] improve an enhanced Geo-I definition named Perturbation-Hidden to ensure perturbed locations are valid. The Perturbation-Hidden method transforms the map of road networks into a grid where the acceptable locations are used as the candidate set. Furthermore, dynamic programming is employed to determine the retrieval area to provide accurate LBSs. The authors employ the dynamic programming method to provide the drivers with the shortest radius of retrieval radius. The limitation of the Perturbation-Hidden method is that it may lead to high privacy costs in privacy-limited regions.

- **Pseudonym:** Pseudonyms are employed as temporary anonymous certificates generated and distributed by the certificate authority. The pseudonym-based LPPMs aim to ensure the unlinkability between the driver's identity and pseudonyms in communication [148]. Wang et al. [138] treat pseudonyms as a long-term identifier to decrease the computation and storage consumption. The authors design a trigger of pseudonym exchange requests to assist the certificate authority in the pseudonym changing. Vehicles change their pseudonyms when meeting the trigger. The method has a limitation, that is, a long period of existing pseudonyms may provide a longer track window for the adversaries. Pseudonym-Indistinguishability is first proposed in [58] to ensure strict unlinkability in the pseudonym swap process. The pseudonym swap process satisfies differential privacy. The adversaries cannot link the pseudonyms after swapping, even if the driving states of the two vehicles are similar. The Pseudonym-Indistinguishability method can provide a high swap complete probability with fewer pseudonyms. The limitation of the method is that the authors ignore the conflict in the pseudonym swap. There are two weaknesses of the pseudonym-based LPPMs. One weakness is that the pseudonym-based LPPMs must manage the vehicles' pseudonyms, leading to high computation and storage consumption. Another weakness is that the pseudonym-based LPPMs cannot ensure unlinkability in the tracking attack.

*5) Summary of User-Side LPPMs:* This part gives an overview of user-side LPPMs, showing how signature-based methods and data perturbation methods can be combined to bring additional protection for location privacy.

The key feature of ABS is the capability for vehicles to selectively sign messages with specified attributes without leakage of personal identifiers, enabled to prove, for instance, that a vehicle is part of an authorized group, such as emergency services, without revealing its identity. The resultant fine granularity of ABS makes it ideal for situations where verification for some attributes is needed while sensitive information needs protection. ABS offers personalized privacy protection and flexibility, whereby an attribute can be selectively disclosed about the need felt by the service and enhancement of privacy is done through reduced exposure of data on necessity during message authentication. However, the computational and verification overhead of ABS remains a significant challenge, particularly for resource-constrained vehicular devices, and the repeated use of certain attributes could expose patterns, making the system susceptible to correlation attacks.

Group signatures enable a vehicle to sign a message on behalf of a group in an anonymous way. Verifiers can verify the legitimacy of messages without tracing the sender's identity and, at the same time, retain traceability for auditing misbehaving vehicles. Providing strong anonymity, group signature schemes are quite suitable for privacy-preserving authentication in dynamic vehicular networks. However, the generation and verification of signatures with high computational complexity are not conducive to real-time performance. Besides, there are big scalability challenges in large-scale environments: managing group membership and revoking compromised keys.

Blind signatures allow the vehicle to verify messages without revealing the content to the signer; thus, they provide strong location and sensitive content privacy in communications. Indeed, they offer strong anonymity and content privacy, especially when combined with other techniques for preserving privacy, but their dependence on trusted third-party entities opens up possible security vulnerabilities. Besides this, high computational overheads in blind signature schemes may result in verification delays; hence, they are not suitable for time-critical vehicular networks.

Data perturbation techniques maintain location privacy by adding some obscurity to the information before it is transmitted, be it by adding noise to or generating dummy locations within data. This is further supported through the use of differential privacy that guarantees formal privacy for users. These are lightweight and inexpensive in computation, hence appropriate on user-side devices. Thus, these methods work actively against adversarial tracking, efficiently creating uncertainty in location information. However, excessive noise can degrade data utility, resulting in inaccurate LBS outcomes, while perturbation methods remain vulnerable to spatio-temporal correlation attacks that exploit location patterns over time.

Secure computation allows a vehicle to process encrypted location data without decryption, maintaining confidentiality for operations like trajectory analysis or route optimization by securely outsourcing computation. In such a way, this will ensure data confidentiality in an effective manner; therefore, it is the correct fit for privacy-preserving outsourcing tasks such as encrypted traffic analysis or route queries. However, secure computation adds significant computational overhead that affects real-time performance, while complex key management procedures reduce scalability, especially in highly dynamic environments involving vehicular participation.

By leveraging their complementary strengths, a multi-layered framework can be constructed to address the challenges of anonymity, integrity, and computational efficiency in user-side LPPMs for vehicular networks.

. **ABS and Group Signatures** can work together to balance fine-grained privacy control with anonymous group-level authentication. ABS allows vehicles to sign messages selectively using attributes without revealing personal identifiers, while group signatures ensure that the message source remains anonymous within a group. For example, drivers can use ABS to prove their credentials (e.g., "authorized user" or "emergency vehicle") without disclosing their identities, and group signatures ensure the vehicle's presence cannot be traced. However, both methods introduce computational overhead and verification delays. To address this, the adoption of elliptic curve cryptography can reduce signature size and verification time, making the combination practical for resource-constrained vehicular devices.

. **Blind Signatures Combined with Data Perturbation** provide an effective solution for ensuring both content and location anonymity. Blind signatures enable drivers to sign messages without revealing their content to the signer, ensuring strong privacy protection during authentication.

To further obscure location data, differential privacy-based perturbation techniques can add controlled noise or dummy locations before the data is sent to LBSs. This two-tier approach ensures that both the message and the vehicle's location remain protected. One challenge is the computational complexity of blind signatures, especially in real-time environments. This can be mitigated by integrating multi-signature aggregation, which allows multiple messages to be verified simultaneously, reducing overhead without compromising privacy.

. **Secure Computation Integrated with Group Signatures** offers a robust solution for maintaining data confidentiality while enabling secure message authentication. Group signatures provide anonymity for vehicles when signing messages, while secure computation ensures that encrypted data can be processed without revealing the plaintext location. For example, vehicles can outsource trajectory analysis to a secure computation scheme, allowing LBSs to process location-based queries without accessing raw data. The limitation of key management in secure computation can be addressed through decentralized certificateless cryptographic protocols, which simplify the distribution and verification of keys, improving system scalability and reducing reliance on centralized authorities.

Integrating these methods into a hybrid framework can enable a multi-faceted solution that enhances privacy, reduces communication overhead, and ensures real-time performance. The key challenge in implementing this hybrid approach lies in managing computational complexity and ensuring scalability. Techniques such as edge computing can be employed to offload computational tasks, such as signature verification and data perturbation, to nearby edge servers, reducing latency and enabling real-time performance. Additionally, using certificateless cryptographic schemes can streamline certificate management, minimizing storage overhead and simplifying revocation processes.

### B. Server-Side LPPMs

By using server-side LPPMs, service providers must perform additional processing on their clients' hosted data [149]. Service providers can achieve this by anonymizing databases, removing identifying traces or encrypting data contents. The comparison of the recent server-side LPPMs is shown in Table VI.

*1) Statistical Disclosure Control (SDC):* SDC mechanisms mainly protect data within statistical databases, balancing data utility and drivers' location privacy. Generally, the output of the SDC mechanisms ensures that the databases do not reveal information related to a specific driver. Database anonymity and differential privacy are the two most popular techniques in the SDC mechanism.

. *k*-**Anonymity:** $k$-anonymity-based LPPMs protect a driver's location by ensuring that the driver's data is indistinguishable from that of at least $k-1$ other vehicles, thereby increasing adversarial uncertainty. This method enhances location privacy by forcing adversaries to guess between multiple potential locations, effectively increasing their estimation error. The $k$ vehicles are typically selected using the maximum entropy principle, which chooses vehicles with similar historical request patterns to maximize anonymity. $k$-anonymity LPPMs can be implemented in both centralized and distributed models. Centralized $k$-anonymity requires trusted cloaking servers to aggregate and anonymize location data, which introduces concerns about server trustworthiness and potential single points of failure. In contrast, distributed $k$-anonymity methods rely on cooperation among participants, eliminating the need for a central authority but adding complexity in terms of coordination.

TABLE VI
Comprehensive Overview of Server-side LPPMs in Vehicular Networks

| LPPM Type | Mechanism Description | Key Reference | Primary Advantage | Notable Disadvantage |
|---|---|---|---|---|
| SDC | The output guarantees non-disclosure of driver-specific information by the databases | [150] | Low storage consumption; Low computational consumption; No need a trusted third party | No authentication and key-agreement protocols |
| | | [151] | Consider historical trust information | |
| | | [152] | Use fixed transceivers to decrease the influence of traffic conditions | Require special hardware |
| | | [153] | Allow vehicles to transmit encrypted information in the silent region | Need a trusted third party; High computational consumption |
| | | [154] | Extends anonymous authentication that supports a request-limiting property | Privacy protection can be breached by the identity disclosure |
| | | [155] | Cost-effective anonymity | Need a trusted third party |
| | | [156] | Low communication consumption | Overlook the prior knowledge |
| | | [157] | Consider the data correlation; reduce the required noise | The compressed data can disclose the original data |
| | | [158] | Improve the efficiency of location data publishing | Overlook spatio-temporal correlation |
| | | [159] | Employ private dual decomposition technology; Improve the scheduling performance; High scalability | Only provide privacy protection for the scheduling process |
| HE | Analyze ciphertexts directly by mirroring the corresponding operations performed on the plaintexts | [160] | Avoid privacy breaches in frequency scheduling process | The considered semi-trusted fog nodes are not reliable |
| | | [161] | Reduce the encryption consumption of the sensor data | High transmission delay |
| | | [162] | Allow a group of passengers to share a vehicle with a minimum aggregate distance | High computational consumption; High transmission delay |
| | | [163] | Avoid the linkage between different pseudonyms of the same vehicle | |
| | | [164] | Low computational consumption | Overlook the knowledge of adversary |
| | | [2] | Outsource encrypted location data to the cloud server to obtain accurate LBS query results securely | The data of the LBS provider is highly coup with the mechanism |
| PIR | Allows drivers to request data items without disclosing the specific item being retrieved | [165] | Decrease the computational cost | Privacy protection can be breached by a long tracking window |
| SE | Enables drivers to securely maintain plaintext keywords while searching in LBSs | [166] | High accuracy of searching results | High computational consumption |

Recent advancements, such as the blockchain-based distributed $k$-anonymity model proposed by Li et al. [150], leverage blockchain technology to record and verify hashed safety beacon messages, reducing both storage consumption and processing time. This decentralized approach also enhances data integrity by ensuring that location data cannot be tampered with, all without requiring a trusted third party. Similarly, Luo et al. [151] improve this method by incorporating a trust management system, where the trust level of vehicles is calculated based on historical interactions and recorded in a blockchain, further bolstering the method's reliability. In addition, Liu et al. [167] introduced a location usability measure method for distributed $k$-anonymity privacy protection, addressing the issue of malicious cooperative users providing fake locations. By evaluating the usability of received locations based on historical data, including visited time, distance, and frequency, their method successfully filters out fake locations, ensuring the effectiveness of the anonymous cloaking region. Despite its effectiveness, the method relies on accurate historical data, and the additional computation overhead may limit its real-time applicability in certain scenarios.

$k$-anonymity methods have limitations. For example, they are vulnerable to spatiotemporal correlation attacks where adversaries can link multiple queries overtime to filter out dummy or anonymized locations. Additionally, maintaining a sufficient number of nearby vehicles to ensure $k$-anonymity can be challenging in low-density traffic areas, reducing its effectiveness. Nonetheless, $k$-anonymity offers significant privacy benefits, particularly in high-traffic scenarios where a large pool of vehicles enhances anonymity, and blockchain integration further strengthens trust and integrity without relying on centralized entities [168].

- **Mix-Zone:** Mix-zone methods enhance location privacy by allowing drivers to change their pseudonyms in designated areas, known as mix-zones, where their entry and exit points are decoupled, creating uncertainty for adversaries attempting to track them [169]. Drivers enter the zone in one order and leave in a different order, making it harder for adversaries to link old and new pseudonyms. This approach can be effective in disrupting location tracking, particularly in dense urban environments where many vehicles are present. However, mix-zones face significant limitations, particularly against timing and transition attacks. Adversaries can correlate the timing of vehicles entering and exiting the mix-zone to re-identify pseudonyms, reducing the effectiveness of the mix-zone in protecting location privacy. Furthermore, continuous query correlation attacks pose another challenge, where adversaries link old and new pseudonyms based on patterns of LBS requests, traffic conditions, or time constraints. Some mappings between pseudonyms can be ruled out based on the known waiting times and observed vehicle behaviors.

To address these issues, Amro [152] introduced a mix-zone method that uses fixed transceivers acting as virtual vehicles to participate in the pseudonym swap process, compensating for a lack of physical vehicles in low-traffic areas. While this innovation mitigates the impact of traffic conditions, it introduces potential security vulnerabilities, as adversaries who obtain the transceivers' pseudonyms could compromise the swap process. Moreover, the reliance on additional infrastructure increases the complexity and potential attack surface of the system. Another improvement is the group-based dynamic mix-zone method, which allows vehicles to exchange encrypted information in a silent region, considering factors such as pseudonym expiration time and vehicle personalization needs. This dynamic approach enhances flexibility and privacy, particularly in resource-limited areas.

Despite these advances, mix-zones still struggle with timing-related vulnerabilities and require careful design to ensure robust pseudonym unlinkability while maintaining low computational and infrastructural overhead.

- **Other Anonymity-Based LPPM:** Meng et al. [154] propose a method that extends anonymous authentication for navigation services by incorporating a request-limiting feature, which effectively safeguards location, route, and identity privacy. This approach ensures that continuous LBS queries do not expose the driver's real-time movements, thereby preserving unlinkability and confidentiality across multiple sessions. However, while this method excels in protecting against identity correlation attacks, it may introduce latency in real-time applications due to the overhead of request-limiting mechanisms.

Zhu et al. [155] introduce the Anonymous Smart-parking And Payment (ASAP) method, which uses a combination of cloaking techniques, hashing, and encryption to anonymize drivers' locations for smart parking services. ASAP effectively hides the actual location, but its reliance on cloaking regions can lead to decreased accuracy in services, especially when precise navigation or parking is required.

Singh et al. [156] propose the Masqueraded Probabilistic Flooding for Source-Location Privacy (MPFSLP) method, which ensures non-traceability by using probabilistic flooding. Instead of generating dummy packets, the vehicles re-send previously transmitted messages, thus complicating adversarial tracking efforts. The key benefit of MPFSLP is its ability to reduce communication costs while maintaining message integrity and authentication. However, as with many anonymity-based approaches, the method faces challenges related to spatio-temporal correlation, as adversaries may be able to infer patterns over time. Moreover, probabilistic flooding may introduce inefficiencies in large networks due to potential redundancies in message transmission.

- **General Differential Privacy:** In vehicular networks, differential privacy can be applied on the server side to anonymize location information before it is used by Location-Based Services (LBSs) or other entities. Soheila et al. [157] propose a Differentially Private Data Streaming (DPDS) system that aggregates data from vehicular networks, where groups of vehicles compress and send location data through a group leader, reducing the amount of noise required for privacy preservation. While this aggregation reduces the noise, its key limitation is that the compressed data size remains similar to the original, which may not provide significant storage or bandwidth savings. Moreover, in real-time applications, such as traffic monitoring or accident prevention, the delay introduced by aggregation and compression could impact system performance. Miao et al. [158] further enhance DP by introducing a noise quadtree and Hilbert curve-based method, which efficiently organizes spatial data to improve location data publishing. However, this method's limitation lies in its focus on two-dimensional space, which does not fully capture the complexity of real-world vehicular environments that involve more dynamic and multidimensional data, such as elevation or time-sensitive routing information.

- **Joint Differential Privacy:** Joint differential privacy extends traditional differential privacy by ensuring that the privacy guarantees hold for groups of individuals rather than just single entities. This is particularly useful in vehicular networks, where joint actions—such as shared rides or collective data reporting—are common. By limiting the manipulation power of a single driver, joint differential privacy reduces the risk of adversaries inferring sensitive information from false reports or outliers. In ride-sharing services, joint differential privacy offers enhanced protection by making it difficult for adversaries to single out individual drivers or passengers from aggregated data.

The study by Han et al. [159] leverages this concept to propose a scheduling protocol that balances location privacy with operational efficiency in ride-sharing systems. The method employs private dual decomposition and driver clustering to minimize travel miles while maintaining privacy. By grouping drivers and vehicles in ways that obscure individual routes and preferences, the system protects location data while optimizing ride assignments. Additionally, the private ride assignment protocol ensures that sensitive information about ride requests and driver locations is protected throughout the scheduling process. However, despite its strong privacy guarantees, joint differential privacy can introduce efficiency challenges. The need to balance privacy and utility—especially in time-sensitive contexts like ride-sharing—can lead to computational overhead and delayed decision-making, particularly when dealing with large-scale vehicular networks. Additionally, while joint differential privacy reduces individual manipulation power, it may still be vulnerable to collusion among multiple parties, where adversaries work together to infer location data from shared reports.

*2) Homomorphic Encryption:* Homomorphic Encryption (HE) is a powerful cryptographic technique that enables computations to be performed directly on encrypted data without requiring decryption. This property is especially valuable for protecting location privacy in vehicular networks, as it allows sensitive location data to remain encrypted throughout the entire process of transmission, storage, and computation. HE schemes mirror operations on ciphertexts to corresponding operations on plaintexts, ensuring that when the encrypted data is processed, the results are consistent with those that would have been obtained from the original, unencrypted data. This capability is particularly useful in contexts like LBSs, where drivers may need to submit their location information for real-time services, such as traffic updates or navigation assistance, but wish to keep their actual location hidden from the service provider.

The most common definition of HE is as follows.

*Definition 3:* Let $P$ and $C$ be a set of plaintexts and ciphertexts, respectively. An encryption mechanism $M$ satisfies homomorphic if and only if, for any given encryption key $k$, and any pair of data $x_i$ and $x_j$, the following holds: $\forall x_i, x_j$

$$M(x_i \ O_{\sigma} \ x_j) \leftarrow M(x_i) \ O_C \ M(x_j), \quad (3)$$

for some operators $O_{\sigma}$ in $P$ and $O_C$ in $C$, where $\rightarrow$ means that the left-hand side can be directly computed from the right-hand side.

The HE schemes can be classified into fully homomorphic encryption and partially homomorphic encryption, as follows.

- **Partial Homomorphic Encryption (PHE):** Partial Homomorphic Encryption (PHE) is one of the earliest and simplest forms of homomorphic encryption, supporting either homomorphic addition or multiplication, but not both simultaneously [170]. This makes PHE particularly useful for specific, limited computations on encrypted data without requiring decryption. In the context of location privacy, Yucel et al. [160] applied PHE to protect drivers' location data in charging station scheduling scenarios. Frequent scheduling requests, which reveal real-time location data, pose a privacy risk for drivers. By hiding the drivers' location information using PHE, the system ensures that sensitive location data is never exposed during scheduling operations, allowing drivers to dynamically join and leave the charging station network while maintaining privacy. The encryption is maintained throughout the process, meaning only the necessary computations (such as determining scheduling slots) are performed on encrypted data, and the actual location information remains secure.

The limitations of PHE stem from its restricted functionality—it can only perform either homomorphic addition or multiplication, not both. This limits its applicability in more complex scenarios where both types of operations are needed for processing, such as route optimization or multi-step location queries in vehicular networks. Additionally, while PHE is computationally efficient and more lightweight compared to fully homomorphic encryption (FHE), its simplicity makes it less versatile for dynamic, real-time applications. The benefit of PHE lies in its efficiency, making it a viable option for applications where privacy protection is needed for specific, well-defined tasks with minimal computational overhead. Despite its limitations, PHE is a practical solution for scenarios like charging station scheduling, where limited but effective encryption can still provide robust privacy protection without introducing significant delays or resource consumption.

- **Somewhat Homomorphic Encryption (SHE):** SHE schemes offer a balance between privacy protection and computational efficiency by allowing a limited number of homomorphic additions and multiplications on encrypted data. This makes SHE well-suited for applications in vehicular networks where privacy is critical, but the data processing requirements are not too extensive. For instance, Subramaniyaswamy et al. [161] introduced an integer-based SHE scheme that reduces encryption overhead by optimizing the data transmission intervals between sensors, thereby enhancing the efficiency of SHE algorithms. In the context of ride-hailing services, Yu et al. [162] applied SHE to calculate the aggregate distance traveled by multiple passengers in a shared vehicle. By using ciphertext packing, they ensured that the encrypted sum of distances could be computed without revealing the actual locations of the passengers, thereby protecting location privacy.

While SHE provides a good balance between security and performance, it has limitations. The finite number of operations that SHE supports makes it less suitable for complex or long-running computations, where the encryption scheme could break down or become inefficient. Additionally, like other homomorphic encryption schemes, SHE still incurs higher computational costs compared to traditional encryption methods, making real-time applications in vehicular networks challenging. Despite these drawbacks, SHE is valuable in scenarios like ride-sharing or sensor data aggregation, where the scope of computation is moderate, and location privacy must be maintained without requiring excessive computational resources.

- **Fully Homomorphic Encryption (FHE):** FHE offers a powerful solution for preserving location privacy in vehicular networks by enabling unlimited operations on encrypted data without requiring decryption This makes FHE ideal for applications like LBSs, where sensitive location data needs to be processed continuously while ensuring privacy. One of the major benefits of FHE is that it allows vehicles to perform complex computations—such as route optimization or traffic management—on encrypted location data, protecting drivers from location tracking.

Perma et al. [163] addressed the challenge of linking pseudonyms by combining FHE with pseudonym encryption, ensuring that the location data remains unlinkable even when multiple pseudonyms are used. In real-time applications like vehicular fog cloud computing, Mohammed et al. [164] leveraged FHE to provide secure, cost-efficient computations that respect mobility constraints and deadlines. Similarly, Farouk et al. [2] demonstrated the feasibility of using FHE for urban mobility simulations and LBS queries, ensuring that encrypted location data can be processed securely in the cloud. However, despite its strong privacy guarantees, FHE suffers from high computational consumption and low efficiency, making it difficult to apply in real-time, large-scale vehicular networks. The encryption and decryption processes are computationally intensive, leading to delays and increased overhead, which are particularly problematic in dynamic vehicular environments where rapid responses are essential. Thus, while FHE provides robust location privacy, its practical deployment is limited by its significant performance drawbacks. Future advancements in optimizing FHE's computational efficiency and reducing overhead could make it more suitable for real-time vehicular applications.

*3) Private Information Retrieval:* Private Information Retrieval (PIR) is a cryptographic technique that allows drivers to query specific data items from a database—such as location-based services—without revealing to the service provider which data item is being requested. This mechanism is highly beneficial in protecting location privacy, as it prevents LBS servers from linking a query to the driver's real-time location or personal data. By enabling queries without disclosure, PIR ensures that even if the LBS server is compromised, the adversary cannot determine the driver's position based on the requested data. However, traditional PIR protocols impose significant computational overhead, making them challenging to implement in real-time applications like vehicular networks. The high cost of computation and communication, often stemming from the need to process large amounts of encrypted data, can lead to inefficiencies, especially in dynamic environments where speed and responsiveness are crucial.

To address these concerns, Tan et al. [165] proposed an optimized PIR approach that leverages prior knowledge of road networks to reduce computational costs. By integrating the transportation information (e.g., traffic patterns and road topology) into the PIR process, the system can narrow down the scope of data queries and reduce the amount of computation required for each retrieval. This significantly enhances the practicality of PIR in vehicular networks by improving its efficiency without sacrificing privacy protection. However, while this optimization reduces the computational burden, it introduces a dependency on accurate and up-to-date transportation data, which could limit its applicability in rapidly changing traffic conditions. Moreover, although PIR ensures that the data request remains private, it does not prevent spatio-temporal correlation attacks over time, where adversaries might infer a driver's location based on repeated queries.

*4) Searchable Encryption (SE):* SE is a server-side privacy-enhancing technique that allows drivers to encrypt their search queries (e.g., location-based keywords) while enabling efficient searches within LBSs without revealing the plaintext of the queries. This mechanism is crucial in vehicular networks for preserving drivers' privacy when searching for nearby services such as gas stations or restaurants, as the queries may reveal sensitive location data. SE schemes typically encrypt keywords and enable servers to process search queries without decryption, ensuring that sensitive information remains hidden from the service provider. Moreover, SE schemes can be enhanced by integrating additional tools such as group signatures, Cuckoo filters, Pederson commitments, smart contracts, and proxy re-encryption, which can improve search precision, efficiency, and security. For instance, public-key encryption combined with SE allows users to perform searches on encrypted data while maintaining strong security guarantees.

However, traditional SE schemes face challenges in handling geometric range searches, essential for location-based services that depend on spatial data. Most SE schemes yield incorrect or imprecise results when searching for locations within a defined geometric area, as they struggle to process the spatial relationships between points. To address this, Chen et al. [166] developed a novel SE method based on a public-key system that supports arbitrary geometric area searches with 100

Despite these advancements, SE schemes, especially those that support complex geometric searches, may introduce computational overhead and latency in real-time applications, as the search process involves cryptographic operations on both the encrypted queries and data. Additionally, maintaining the balance between search efficiency and privacy is challenging, particularly in large-scale vehicular networks where rapid, real-time responses are critical.

*5) Summary of Server-side LPPMs:* This part covers a detailed review of the server-side LPPMs that strengthen location privacy through anonymization, encryption, or selective data management at the server level. While server-side LPPMs normally offer strong privacy guarantees, their effectiveness can be further enhanced by the judicious integration of complementary methods to overcome some individual limitations and enhance scalability, efficiency, and practicality in vehicular networks.

SDC mechanisms, including $k$-anonymity and mix-zone techniques, focus on protecting location data within statistical databases. $k$-anonymity ensures that a driver's location is indistinguishable from a group of nearby vehicles, effectively increasing adversarial uncertainty. Recent innovations, such as blockchain-based distributed $k$-anonymity, reduce reliance on trusted third parties and enhance data integrity. However, $k$-anonymity is vulnerable to spatio-temporal correlation attacks and becomes less effective in low-density traffic areas. Mix-zone methods, on the other hand, allow pseudonym changes within designated regions, disrupting location tracking. Improvements like virtual transceivers compensate for sparse traffic but introduce security risks and infrastructural complexity. To strengthen these techniques, combining blockchain for decentralized trust and group-based pseudonym exchanges can address vulnerabilities while maintaining scalability in dynamic traffic conditions.

HE offers a powerful solution for server-side location privacy by enabling computations on encrypted data without decryption. PHE provides lightweight encryption for specific tasks, such as charging station scheduling, while SHE balances computational efficiency and privacy for moderate data processing, as seen in ride-sharing applications. FHE allows unlimited operations on encrypted data, ensuring strong privacy for complex computations like route optimization or LBS queries. However, FHE suffers from significant computational overhead and latency, limiting its real-time applicability. To address this, hybrid encryption schemes that combine PHE/SHE for simpler operations and offload intensive FHE tasks to edge servers can optimize performance without sacrificing privacy guarantees.

PIR allows drivers to query location-based information without revealing the target object to the server. Integrating road network knowledge into PIR processes has contributed to the significant reduction in computational costs by recent techniques, hence making PIR applicable in practical vehicular use. However, PIR is still vulnerable to certain attacks based on spatio-temporal correlation, given repeated queries over time, which reveals the driver's location. Thus, enhancing PIR using dummy query mechanisms or query obfuscation techniques can further disrupt adversarial tracking while retaining query efficiency.

SE enables drivers to conduct encrypted keyword searches on LBSs without leaking query content. More recently, new enhancements in the form of geometric range search-based SE were proposed, which guarantee 100% search accuracy for any arbitrary spatial region—an inherent limitation of traditional schemes of SE. However, computational delays arise in SE due to the complex cryptographic operations involved. Indexing techniques and proxy re-encryption may be employed to optimize such trade-offs between search efficiency and privacy for fast, real-time responses in large-scale vehicular networks.

To address their individual limitations and leverage complementary strengths, server-side LPPMs can be integrated into a hybrid privacy-preserving framework as follows:

. **SDC with HE:** SDC methods, such as $k$-anonymity, anonymize location data before storage, while HE ensures that computations occur securely on encrypted data. This combination protects location data during storage and processing, preventing adversarial inference attacks. By integrating HE, vulnerabilities in SDC, such as susceptibility to spatio-temporal correlation attacks, are mitigated, as encrypted data remains inaccessible even if statistical anonymity fails.

. **PIR and HE:** PIR allows drivers to query data privately without revealing the requested content, and HE ensures that the computations on these queries occur in encrypted form. This integration provides end-to-end privacy for location queries without compromising LBS functionality. By offloading computationally intensive tasks to edge servers, PIR's overhead is significantly reduced, while HE guarantees data confidentiality throughout the entire query process.

. **SE with SDC:** SE enables secure and efficient encrypted searches within anonymized databases created using SDC methods, such as mix-zones or $k$-anonymity. This combination ensures accurate spatial data retrieval while preserving privacy. Pre-processing the anonymized data and indexing encrypted search queries reduces SE's computational delays, enabling rapid and efficient lookups even in large-scale systems.

By strategically combining these methods into a hybrid framework, it is possible to protect sensitive data during the storage, query, and computation phases. Future research should focus on optimizing the computational efficiency of HE, enhancing the scalability of PIR and SE, and leveraging edge computing to ensure real-time performance. Such advancements will enable a robust, privacy-preserving infrastructure that balances security, utility, and operational efficiency.

### C. User-Server-Interface LPPMs

User-server-interface LPPMs focus on securing the communication channel between drivers and Location-Based Services (LBSs) by protecting the data exchange process to maintain the confidentiality and integrity of location information. These mechanisms rely on channel-side techniques that ensure secure communication and trust in third parties. While they provide essential protections against eavesdropping and tampering, they also face limitations in terms of scalability and trust dependency.

*1) Secure Communication:* Secure communication can be classified as follows.

. **Encrypted Communication Protocol:** Encrypted communication protocols, such as Transport Layer Security (TLS) and Secure Shell (SSH), play a crucial role in safeguarding location privacy in vehicular networks by encrypting data exchanged between drivers and LBSs. These protocols utilize public key cryptography to establish secure channels, preventing adversaries from intercepting sensitive location data during transmission. TLS and SSH ensure that even if the communication is monitored, the transmitted data remains unintelligible without the corresponding decryption key, thus providing strong protection against pervasive communication surveillance.

Traditional encrypted protocols have their limitations. For instance, SSH requires periodic user verification, which increases computational overhead and can drain the resources of drivers' devices, especially in real-time applications where frequent location updates are necessary [171]. Additionally, the encryption process itself introduces latency, which may hinder the performance of time-sensitive vehicular applications. Despite these drawbacks, encrypted communication protocols remain a cornerstone for securing data in vehicular networks, as they effectively reduce the risk of location data leakage during transit. The challenge lies in optimizing these protocols for high-speed, real-time vehicular environments without compromising the robust privacy guarantees they provide.

- **End-to-end Encrypted Services:** End-to-end encrypted (E2EE) services offer robust privacy protection by ensuring that only the intended recipient can decrypt the data transmitted between end-users, making it impossible for adversaries or intermediaries to access sensitive location information during transmission [172]. In the context of vehicular networks, E2EE can be a powerful tool for safeguarding location privacy, particularly in scenarios where vehicles need to exchange data directly, such as in cooperative driving, traffic coordination, or ride-sharing. The major advantage of E2EE is that it eliminates the need for trusting intermediaries or third-party service providers, thereby minimizing the risk of data breaches and unauthorized access.

However, while E2EE is effective at protecting data in transit, it is seldom used in current vehicular networks due to challenges related to key management and scalability. Managing encryption keys dynamically across a vast, distributed network of vehicles is complex, and ensuring low-latency communication in real-time vehicular applications while maintaining strong encryption can be resource-intensive. Additionally, E2EE does not inherently protect against endpoint vulnerabilities, meaning that if an adversary gains access to the sender's or receiver's device, the encrypted data may still be exposed. Despite these limitations, as vehicular networks evolve with the integration of 5G and 6G technologies, E2EE is expected to play a larger role in ensuring secure, private communication between vehicles and infrastructure, especially in tasks that require high levels of data integrity and privacy, such as autonomous driving or secure traffic management.

*2) Trusted Third Party (TTP):* TTP techniques serve as intermediaries in vehicular networks to protect drivers' location privacy by anonymizing and pseudonymizing their data. The primary goal of a TTP is to prevent adversaries from linking drivers' real identities to their location data, thus protecting sensitive information. TTPs help drivers interact with LBSs without exposing their locations or identities by issuing temporary pseudonyms or managing encrypted communications. However, relying on a TTP introduces several challenges. Malicious entities, especially in collaborative attacks, may possess the resources and capabilities to breach a TTP or trace drivers through accumulated pseudonym data, thereby undermining privacy [173].

Moreover, TTPs pose a single point of failure: if compromised, vast amounts of sensitive information could be exposed. For this reason, recent studies advocate for privacy-preserving methods that do not require a centralized trusted third party. Decentralized systems using blockchain, for instance, offer an alternative by distributing trust across multiple parties, thereby reducing reliance on a single entity [174], [175]. These decentralized approaches enhance privacy and resilience but introduce new challenges, such as higher computational costs, latency, and storage requirements, which can limit scalability in resource-constrained environments.

### D. Balance between Location Privacy and Data Utility

The balance between location privacy and data utility is critical in vehicular networks because both objectives must be balanced to ensure the system's effectiveness. Location privacy is essential for protecting drivers from potential tracking, profiling, or malicious attacks, which can arise from the misuse of sensitive location data. However, the data utility of LBSs depends on the accuracy and availability of real-time location information to provide services such as navigation, traffic updates, and personalized recommendations. Over-protection of privacy, such as through excessive data perturbation or obfuscation, can degrade the quality of these services, making them less effective or even unusable. Conversely, insufficient privacy protections expose users to significant risks, reducing trust in LBSs and leading to reluctance in sharing data.

This balance allows users to benefit from LBSs while having confidence that their location data is adequately protected. Moreover, failing to achieve this balance could result in diminished user engagement and a loss of value for both service providers and users. Privacy-preserving techniques must thus be designed to optimize data utility while minimizing privacy risks, tailoring protection levels to the specific needs of different applications and user preferences.

Many LPPMs focus on the balance between location privacy and data utility, as shown in Table VII. There are six major methods to achieve the balance between location privacy and data utility, as follows.

*1) Blockchain:* Blockchain technology offers a powerful solution for distributed management in vehicular networks by enabling decentralized data storage and trust management without the need for a central authority. This is particularly beneficial for LPPMs, as blockchain allows vehicles to access necessary information securely while preserving anonymity and location privacy. For example, Luo et al. [151] leverage blockchain to store historical trust data using Dirichlet distribution, enabling vehicles to cooperate with others anonymously. By doing so, they reduce communication delays while maintaining a secure and trustworthy environment. Similarly, Li et al. [150] introduce two key metrics—connectivity and average distance—to measure the balance between data utility and privacy in a blockchain-based $k$-anonymity framework.

TABLE VII
EXISTING LPPMs IN THE BALANCE BETWEEN LOCATION PRIVACY AND DATA UTILITY.

| Methods | Techniques | LPPM category | Papers |
|---|---|---|---|
| Blockchain | Distribution Management | $k$-anonymity | [150], [151], [176] |
| | | others | [17], [177]–[183] |
| Adaptive parameters | Tuning parameters | Pass-and-run | [117] |
| | | Attribute-based signature | [118], [121] |
| | | Geo-I | [15], [135], [137], [147] |
| | | Pseudonym | [58] |
| | | Other anonymity | [155] |
| | | General differential privacy | [157], [158] |
| | | SHE | [161] |
| | | FHE | [164] |
| Hybrid | Combining multiple methodologies | Group signature | [122], [124] |
| | | Sanitizable signature | [128] |
| | | Dummy-based | [131] |
| | | LDP | [132] |
| | | Joint differential privacy | [159] |
| | | SE | [166] |
| Encryption optimization | Reducing computational and communication consumption | Group signature | [123] |
| | | Blind signature | [129] |
| | | Secure computation | [130] |
| | | Mix-zone | [153] |
| | | PHE | [160] |
| | | FHE | [163] |
| Parameter simplification and optimization | Simplifying process | Group signature | [59], [126] |
| | | Blind signature | [129] |
| | | Dummy-based | [134] |
| | | Pseudonym | [138] |
| | | Other anonymity | [156] |
| | | FHE | [2] |
| | | Private information retrieval | [165] |
| Virtual nodes | Introducing cooperators | Group signature | [125] |
| | | Pseudonym | [138] |
| | | $k$-anonymity | [152] |
| | | TTP | [174], [175], [184], [185] |

Blockchain technology has emerged as a powerful tool for achieving a balance between location privacy and data utility in vehicular networks by providing decentralized solutions that eliminate the need for a trusted third party [186]. Malik et al. [17] developed a blockchain-based authentication and revocation method that reduces computational and communication overhead, thereby enhancing the efficiency of privacy-preserving mechanisms. Similarly, Wang et al. [180] combined blind signatures with blockchain to create a high-effectiveness authentication scheme, using blockchain to store the vehicle's public key and comparing it with a calculated Merkle root value for secure verification. Tang et al. [181] extended this concept by integrating group signatures and other cryptographic tools into a blockchain-based privacy-preserving searchable encryption scheme for parking lot sharing, balancing privacy protection with service accuracy. Further, Chaudhary et al. [176] and Liang et al. [182] used blockchain to address the limitations of $k$-anonymity, with the latter introducing cloaking regions to protect location data better. Wang et al. [183] enhanced trust management through blockchain-based Roadside Units (RSUs), and Lu et al. [177] protected location privacy by using blockchain proofs of presence and absence to avoid identity leakage. These solutions demonstrate blockchain's ability to secure location data while maintaining data utility by ensuring trustworthiness and reducing vulnerabilities to semantic, linking, and data mining attacks. However, achieving an optimal balance remains crucial, as over-reliance on privacy mechanisms can reduce the effectiveness of real-time services in vehicular networks.

Their approach demonstrates that blockchain can optimize the balance between location privacy and data utility by improving the efficiency of safety beacon message exchanges. The blockchain-based system not only ensures that vehicles' real-time data remains private but also reduces data processing time, ensuring timely communication without compromising privacy [187]. In summary, blockchain enhances privacy and efficiency, but it must be carefully calibrated to avoid performance degradation in data-dependent applications.

*2) Adaptive Parameter:* Adaptive Parameter-Based LPPMs offer a dynamic approach to achieving a balance between location privacy and data utility by adjusting privacy parameters based on the specific requirements of drivers and the contextual needs of the system. This adaptability reduces unnecessary resource consumption and minimizes the loss of data utility, which is often a challenge in static privacy-preserving methods. For instance, Lu et al. [117] quantify location privacy and "perfect privacy" to help vehicles route data through other vehicles while maintaining efficiency and accuracy in Location-Based Services (LBSs). Similarly, Shahandashti et al. [118] and Kaaniche et al. [121] permit vehicles to reveal only necessary data based on predefined attributes, ensuring privacy without compromising utility. Moreover, Ma et al. [15], Zhou et al. [135], and Li et al. [147] introduce LPPMs that allow vehicles to adjust privacy settings based on protection and utility needs, improving the effectiveness of both privacy and service quality. Li et al. [58] further optimize this by enabling vehicles to swap pseudonyms according to driving conditions, enhancing privacy while maintaining the utility of real-time data. Techniques like differential privacy are also adapted by controlling the noise level to balance privacy protection with retrieval accuracy, as demonstrated by Ghane et al. [157] and Miao et al. [158]. Additionally, edge computing and encryption techniques reduce time delay and resource consumption, ensuring real-time applications remain efficient [161], [164]. The key advantage of adaptive parameters is their ability to flexibly meet both privacy and utility needs, offering a practical solution to the inherent balance between protecting sensitive location data and providing accurate, efficient services in vehicular networks.

*3) Hybrid Approach:* Hybrid Approaches in LPPMs leverage a combination of methodologies to optimize both privacy protection and system efficiency. These approaches blend the strengths of different techniques to overcome limitations and achieve a more balanced solution. For instance, Zheng et al. [122] and Mundhe et al. [125] combine group signatures with pseudonyms to ensure secure identity verification while reducing computational overhead. By using pseudonyms, the length of signatures can be shortened, which decreases both computation time and resource consumption. Similarly, Pamies et al. [128] employ a hybrid of log anonymization and sanitizable signatures to further mitigate privacy risks, ensuring that sensitive information such as location data remains protected during data transfers. Additionally, Niu et al. [131] enhance *k*-anonymity with the use of dummy locations, considering side information about vehicles to increase the robustness of anonymity, while Erlingsson et al. [132] use differential privacy to obscure location signals at the user side, preventing adversaries from linking pseudonyms. Tong et al. [159] integrate joint differential privacy with spatial indexing and distributed optimization to secure location data while optimizing computational efficiency. Similarly, Chen et al. [166] combine SE with computational private information retrieval, reducing computational burden while maintaining location privacy in road networks. These hybrid approaches offer a nuanced solution to the fundamental balance between location privacy and data utility, allowing for context-specific adjustments that provide robust privacy protections without compromising the quality or timeliness of LBSs.

*4) Encryption Optimization:* Encryption Optimization in vehicular networks focuses on reducing the communication and computational overhead associated with traditional encryption methods while preserving location privacy and maintaining data utility. Advanced cryptographic techniques, such as bilinear pair cryptography, often result in high computational costs and time delays, which can degrade the performance of LBSs. To address this, researchers like Wu et al. [123] have turned to more efficient cryptographic schemes, such as the elliptic curve cryptosystem, which reduces the computational burden of message signing and verification compared to bilinear pairs. Additionally, Sun et al. [129] combine encryption-based methods, including zero-knowledge proofs and homomorphic encryption, to improve network response times while ensuring strong privacy protection. Using non-colluding servers by Zhou et al. [130] eliminates unnecessary encrypted data exchanges between vehicles and LBS servers, further reducing latency and improving practicality without compromising security. Li et al. [153] proposes transmitting encrypted data in mix zones to reduce the storage costs of pseudonym management, enhancing privacy while ensuring data utility remains intact. Yucel et al. [160] protect location privacy by incorporating homomorphic encryption into bichromatic nearest neighbor assignments, achieving low resource consumption and fast convergence. FHE, while offering robust privacy, can lead to significant communication overhead, as noted by Prema et al. [163], who mitigate this by combining FHE with pseudonyms to control message frequency. The challenge remains in striking the right balance between strong location privacy and maintaining high data utility with minimal delays, as overly complex encryption schemes can reduce service responsiveness, whereas overly simplified methods risk compromising privacy. By optimizing encryption techniques, it is possible to protect location privacy without excessively compromising the efficiency and utility of vehicular services.

*5) Parameter Simplification and Optimization:* The time delay can be significantly decreased when the LPPMs can simplify the process. For example, the non-certificate LPPMs can eliminate the certificate authentication time delay and certificate storage consumption. Hakeem et al. [59] achieve authentication over multiple zones of large-scale BSs by using a single message and short signature with bilinear pairing cryptography and short-size signature. The authors significantly decrease the generation and verification time of the signature. Mei et al. [126] use a certificateless aggregate signature scheme with full aggregation technology to reduce resource consumption. The balance between privacy protection and data utility is achieved by considering random oracles under the computational Diffie-Hellman assumption. Sun et al. [129] optimize a fog-bus-based vehicle crowdsensing framework that severs the relationship between the identity and location data. The authors simplify the data process in data reporting, reputation management, and reward issuing to improve effectiveness.

Liu et al. [134] focus on the time reachability, direction similarity, and in-degree/out-degree of the location data. The developed spatiotemporal correlation-aware LPPM simplifies neighboring location sets for personalized location privacy protection. Wang et al. [138] improve the data utility of pseudonyms by introducing a trigger-based structure, avoiding the frequent pseudonym changing. Singh et al. [156] simplify the required characteristics of the vehicles by using location and speed instead of their identities. The authors allow each vehicle to send data of others masquerading as its own location data, significantly decreasing the traceability of the pseudonym-base LPPM without breaching the data utility. Farouk et al. [2] outsource the location data to a cloud server that prevents the vehicles from sharing their location data with multiple entities. By only sharing data with the cloud server, the vehicle can obtain LBSs with a low delay. Tan et al. [165] use the prior knowledge of road networks to decrease the consumption of computational private information retrieval in preprocessing and communication.

*6) Virtual Node:* LPPMs play a crucial role in balancing location privacy and data utility, particularly for methods like *k*-anonymity that require a specific number of neighboring vehicles to provide strong privacy guarantees. In scenarios where physical vehicles are sparse, LPPMs can generate virtual nodes that act as placeholders for real vehicles, allowing the system to meet the required anonymity threshold without compromising the accuracy or utility of the data [125]. These lightweight virtual nodes occupy limited storage, making them an efficient solution for enhancing privacy while maintaining system performance. For instance, Wang et al. [138] employ virtual devices, such as triggers, to assist with pseudonym changes, thereby improving the privacy protection in pseudonym-based LPPMs. Similarly, Zhu et al. [155] introduce fixed mixing zones in road networks, where virtual nodes help prevent adversaries from linking pseudonyms, ensuring consistent location privacy even in low-traffic areas. By creating virtual entities, these schemes reduce the reliance on actual vehicle density, allowing the LPPMs to maintain high privacy-preserving capabilities without degrading the data utility needed for real-time services. However, the effectiveness of virtual nodes hinges on ensuring they do not distort the underlying data too much, as overuse can lead to inaccurate location-based services.

In summary, existing methods that attempt to balance location privacy and data utility are insufficient for the demands of future 5G/6G-enabled vehicular networks. For instance, Geo-I-based LPPMs inject controlled noise into location data and introduce inaccuracies that require additional data optimization. This degrades big data processing in 5G/6G networks, and even with fine-tuning, the data utility remains significantly reduced, making these methods unsuitable for 5G/6G service providers [216]. Similarly, virtual nodes-based methods struggle to balance privacy and utility. As future 5G/6G networks will be equipped with numerous sensors for real-time environmental monitoring, virtual nodes will no longer be necessary, further highlighting the need for more advanced privacy-preserving solutions that ensure both privacy protection and data utility in high-density, sensor-rich environments.

*7) Example of Localization Application:* COVID-19 tracking applications that utilize vehicular networks and Bluetooth-based proximity detection illustrate the complex challenge of balancing location privacy and data utility. During the pandemic, these systems were instrumental in monitoring the spread of the virus and ensuring compliance with health guidelines by collecting vast amounts of location data from vehicles and smartphones [217], [218]. While this data proved essential for public health efforts, it also raised significant privacy concerns. By integrating GPS data, driving logs, and Bluetooth-based encounters, such applications could reveal sensitive location details of individual drivers and others nearby, increasing the risk of trajectory reconstruction and identity inference.

The broader context of vehicular networks further complicates privacy protection. The increasing availability of location data from various sources—such as connected vehicles, smart devices, and social networks—provides a rich environment for big data mining. This creates the potential for re-identification attacks where adversaries can reconstruct individuals' trajectories and infer personal details, thus compromising location privacy. Moreover, vehicular networks are prone to cybersecurity threats, making them vulnerable to malicious attacks that exploit this data, especially when linked with broader applications like public health monitoring during the pandemic.

Different from the theoretical balance between location privacy and data utility, the applications in practice are more interested in data utility. Location data collection and exchange in the fight against COVID-19 is an excellent example of in-vehicle tracking. Existing applications for epidemic prevention can be classified into centralized and decentralized management. In centralized management cases, the medical information of confirmed cases is monitored by governments or institutions. Governments or institutions will only disclose the trajectory of the patient when a new patient appears. In decentralized management cases, the location data is managed by the drivers. Only if the drivers are infected, their trajectories are exposed to authorities. Strict management of such information can save the world from the virus, but the location privacy of the patient is exposed. Table VIII shows the benefits and limitations of related applications. In Table VIII, the existing COVID-19 applications are classified into centralized and decentralized applications.

TABLE VIII
BALANCE BETWEEN LOCATION PRIVACY AND DATA UTILITY OF IN-VEHICLE TRACKING IN PRACTICE: COVID-19 APPLICATIONS.

| Architecture | Applications | Required information | Limitations |
|---|---|---|---|
| Centralized | COVIDSafe (Australia) [188] | Encounter information | The personal information of the driver, e.g., identity, phone number, and email. Encounter information that is frequently recorded exposes the driver's trajectories. Private information can be inferred from the reported geographical information. The geographical information is disclosed to the public for epidemic prevention |
| | E-Tabib (Azerbaijan) [189] | | |
| | BeAware Bahrain (Bahrain) [190] | | |
| | Corona Tracer BD (Bangladesh) [191] | | |
| | Taiwan Social Distancing (China) [192] | | |
| | TousAntiCovid (France) [193] | | |
| | VirusRadar (Hungary) [194] | | |
| | Rakning C-19 (Iceland) [195] | | |
| | Smittestopp (Norway) [196] | | |
| | BlueTrace (Singapore) [197] | | |
| | Alipay (China) [198] | Geographical information | |
| | WeChat (China) [199] | | |
| | LeaveHomeSafe (China) [200] | | |
| Decentralized | Stopp Corona (Austria) [201] | Encounter information | The malicious drivers can infer the trajectories of patients and match the trajectories with identifiers by creating multiple accounts and recording multiple routes. The geographical information is disclosed to the public for epidemic prevention |
| | Stop COVID-19 (Croatia) [202] | | |
| | eRouška (Czechia) [203] | | |
| | Smittestop (Denmark) [204] | | |
| | Koronavilkku (Finland) [205] | | |
| | Corona-Warn-App (Germany) [206] | | |
| | Apturi Covid (Latvia) [207] | | |
| | Radar COVID (Spain) [208] | | |
| | SwissCovid (Switzerland) [209] | | |
| | NHS COVID-19 (United Kingdom) [210] | | |
| | DP-3T (European) [211] | | |
| | PACT (USA) [212] | | |
| | Private Kit: Safe Path (USA) [213] | Geographical information | |
| | South Korea system [214] | | |
| | HaMagen (Israel) [215] | | |

The centralized applications (e.g., TraceTogether (Singapore) [219], COVIDSafe (Australia) [188], and BlueTrace (Singapore) [197]) record the encrypted encounter history rather than the location information. Only the authority can decrypt the encounter history. The authority can obtain the driver's identification if the driver is infected. Other drivers can check their risk of infection by the encounter history. However, the patients' identities and trajectories are exposed to the authorities, which increases the risk of location privacy disclosure. The privacy-preserving capability of the applications can be decreased if other drivers collude to infer targets.

DP-3T (European) [211] and PACT (USA) [212] are two examples of decentralized applications. The two applications create a periodical key to generate several ephemeral identifiers. The ephemeral identifiers are broadcasted as a beacon message within a region. The two applications store the received beacon messages with extra information. The applications match the ephemeral identifiers based on the information from the authority. Nevertheless, malicious drivers can infer the trajectories of patients and match the trajectories with identifiers by creating multiple accounts and recording multiple routes. Private Kit: Safe Path (USA) [213] and South Korea system [214] are the other types of COVID-19 applications. They periodically record the drivers' trajectories with time-stamped logs and pseudonyms based on the GPS (the South

Korean system records detailed personal information). The drivers can report their trajectories to the authorities if they are infected. The authority can select information to be exposed to the public. The trajectories are disclosed to the public for epidemic prevention. Nevertheless, the applications cannot offer the expected location privacy-preserving capability, as everyone can view the trajectories.

The existing COVID-19 contact tracing applications balance location privacy protection capability and data utility by using anonymity [220]. The applications, e.g., TraceTogether (Singapore) [219] and COVIDSafe (Australia) [188], periodically broadcast random time-varying tokens as the driver's temporary IDs [221]. The applications record the encounter information of the drivers, which are employed to report infection risk. However, the authorized entities can still link the temporary IDs and the driver's personal information (e.g., trajectory, phone number), even if the applications do not collect personal information on purpose [221].

The correlation between vehicle location privacy and COVID-19 tracking offers crucial insights into the broader challenges of managing location privacy in vehicular networks. Both contexts involve the collection and analysis of vast amounts of real-time location data for public benefit—whether it's tracking vehicle movements for traffic management or monitoring disease spread during a pandemic. However, they also highlight the inherent tension between ensuring data utility for services like contact tracing and protecting individual privacy. In COVID-19 tracking, vehicle driving logs, GPS data, and proximity encounters were used to map infection risk, yet these same mechanisms posed significant risks of trajectory reconstruction and identity inference. Similarly, in vehicular networks, the aggregation of sensor data and social interactions can lead to unintended privacy breaches if not properly safeguarded.

### E. Key Takeaway

In this section, through a comprehensive analysis of user-side, server-side, and user-server-interface LPPMs, several fundamental insights emerge, contributing to the broader understanding of the challenges and opportunities in this field.

This section analyzes the intricate balance between location privacy and data utility. High privacy often necessitates advanced cryptographic techniques, data perturbation, or obfuscation strategies, which can significantly reduce the accuracy and timeliness of LBSs. Conversely, prioritizing data utility can lead to privacy vulnerabilities, particularly in dynamic vehicular networks where real-time information is crucial. For instance, blockchain-based LPPMs offer robust privacy protection without relying on centralized authorities, but their computational overhead can hinder real-time applications. Similarly, differential privacy and its variants achieve strong privacy guarantees but often struggle with maintaining high data utility in practical scenarios, such as edge computing or high-density environments []. This finding emphasizes the need for adaptive mechanisms that optimize privacy and utility based on the context and user requirements.

The analysis in this section highlights the limitations of existing LPPMs in addressing the evolving threats in 5G/6G

vehicular networks. Key challenges include the scalability of cryptographic approaches, the dependency on trusted third parties, and the computational demands of real-time privacy-preserving mechanisms. These constraints are exacerbated by the increasing complexity of vehicular networks, where high mobility, dense communication environments, and advanced adversarial capabilities demand robust, efficient, and scalable solutions. The review demonstrates that while techniques such as group signatures and homomorphic encryption offer promising pathways for secure data exchange, they require significant optimization to align with the low-latency and high-throughput requirements of 6G networks.

The comprehensive analysis of LPPMs in vehicular networks reveals several critical themes that define the current state and future potential of location privacy protection. By examining user-side, server-side, and user-server-interface mechanisms, this section identifies recurring strengths and limitations across existing approaches. The key observations of this section are as follows:

- **Trade-off Dynamics:** Effective LPPMs must balance privacy and utility, avoiding approaches that excessively compromise service quality or expose users to privacy risks. Techniques that dynamically adjust privacy parameters based on contextual factors offer a promising direction.
- **Scalability and Efficiency:** Existing LPPMs face scalability challenges, particularly in large-scale, high-density vehicular networks. Future solutions should prioritize lightweight cryptographic methods and decentralized frameworks to reduce dependency on resource-intensive operations.
- **Integration of Emerging Technologies:** The integration of technologies such as blockchain, edge computing, and artificial intelligence can enhance the robustness of LPPMs. However, their adoption requires careful consideration of their resource requirements and potential vulnerabilities.

## V. LESSONS LEARNED AND OPEN CHALLENGES FOR LPPM IN 5G/6G VEHICULAR NETWORKS

This section starts with illustrating the localization techniques in future vehicular networks, compared with their counterparts in current networks. By discussing the new requirements of location privacy and data utility introduced by the novel localization, we present the limitations of the existing LPPMs in 5G/6G vehicular networks. The potential location privacy issues under the time cross-layer tracking techniques introduced by new communication technologies are illustrated, and we offer solutions to these issues by improving the existing LPPMs.

### A. Advancement of Localization vs. Location Privacy

In 5G/6G-enabled vehicular networks, advancements in localization technologies bring both enhanced precision and increased privacy risks. Emerging techniques such as cooperative localization, machine learning-based tracking, and multipath exploiting localization offer superior accuracy by leveraging dense BS deployment, sensor fusion, and machine learning algorithms to pinpoint vehicle locations with unprecedented precision [222]. However, this heightened localization capability introduces new location privacy challenges. As localization, sensing, and communication systems converge, sharing time-frequency-spatial resources, integrating cross-layer tracking techniques enabled by 5G/6G opens up vulnerabilities to multi-source data mining and trajectory reconstruction. These risks stem from the extensive collection of high-resolution location data, which, if compromised, can reveal sensitive personal information about drivers. The existing LPPMs, designed for current networks, are inadequate in addressing the cross-layer tracking introduced by these novel localization techniques.

TABLE IX
DIFFERENT LOCALIZATION TECHNIQUES IN CURRENT AND 5G/6G VEHICULAR NETWORKS.

| Localization techniques | Features of 5G/6G vehicular networks | Corresponding techniques in current vehicular networks | Difference |
|---|---|---|---|
| Traditional localization | High data rates<br>High scalability | Cellular radio-based tracking | More BS and UE<br>More frequent signal transmission<br>More accurate localization |
| Cooperative localization | Ultra-low latency<br>High scalability | Sensing infrastructure-based tracking<br>Cellular radio-based tracking<br>Upper-layer message-based tracking | More sensors<br>More frequent signal transmission<br>Closer BS and UE |
| AI-based localization | High data rates<br>High scalability | Sensing infrastructure-based tracking<br>Optical vision-based tracking<br>Upper-layer message-based tracking | More sensors<br>More detailed dataset<br>Massive devices |
| Channel charting | High scalability | Cellular radio-based tracking<br>Sensing infrastructure-based tracking<br>Upper-layer message-based tracking | Considering channel state information |
| Driver tracking localization | High data rates<br>Ultra-low latency | Sensing infrastructure-based tracking<br>Upper-layer message-based tracking | More sensors<br>3D vehicular network |
| Multipath exploiting localization | Ultra-low latency<br>High scalability | Cellular radio-based tracking | More sensors |
| SLAM | High data rates<br>Ultra-low latency<br>High scalability | Vehicle driving log-based tracking<br>Upper layer message-based tracking<br>Cellular radio-based tracking | More sensors<br>Considering time-varying states<br>More fixed landmarks |
| Context-aware localization | High data rates<br>Ultra-low latency | Vehicle driving log-based tracking<br>Upper-layer message-based tracking | Intelligent prediction<br>Massive personal information |

*1) Traditional Localization:* While traditional localization techniques provide reliable positioning through geometric calculations, they present significant location privacy limitations in the context of 5G/6G vehicular networks. The assumption that the BS locations and UEs constraints are known inherently exposes the system to privacy risks. For instance, frequent communication between UEs and BSs, coupled with the increasing BS density in 5G/6G networks, allows for generating highly granular trajectory data, making users more susceptible to tracking and trajectory reconstruction attacks. Additionally, distance and angle measurement techniques require continuous signal exchange, which adversaries can intercept or analyze to infer sensitive movement patterns. Area measurement and hop-count methods also aggregate data over time, making it easier for malicious entities to map detailed movement paths across large areas. As these methods inherently prioritize localization precision, they often fail to incorporate robust privacy-preserving mechanisms, leaving gaps that could be exploited for unauthorized location tracking and compromising users' location privacy.

The traditional localization techniques can be classified into distance, angle, area, and hop-count measurement [223], as follows.

• **Distance Measurement:** Distance measurement techniques in localization rely on distance-related parameters such as TDoA, RSSI, and connectivity conditions to calculate the position of UE [224]. While these methods offer significant precision, they pose substantial location privacy risks due to their reliance on continuous signal exchanges between UEs and BSs. The transmission time and signal strength data, essential for calculating distances, can be easily intercepted by adversaries, allowing them to triangulate a vehicle's or user's exact position. This makes distance measurement techniques particularly vulnerable to eavesdropping and man-in-the-middle attacks, where malicious actors can exploit the time and strength of signals to reconstruct user trajectories. Additionally, as 5G/6G networks introduce higher BS densities and more frequent communication, the granularity of location data increases, further exacerbating the risk of tracking and profiling. Existing LPPMs may struggle to handle the fine-grained data generated by these measurements, as the continuous data exchange increases the chances of location inference attacks, making it crucial to develop advanced obfuscation and encryption strategies to protect user privacy in such high-precision localization environments.

- **Angle Measurement:** Angle measurement techniques rely on trigonometry and geometric calculations to estimate the position of a target, such as a vehicle, by using the angular information between the BSs and the target device [225]. While this method can achieve high accuracy, especially in dense vehicular networks, it presents several location privacy challenges. The precision of angle-based localization means that even minor variations in angle data can be used to track subtle movements of a vehicle, making it easier for adversaries to reconstruct continuous trajectories or predict future locations. Furthermore, the requirement for continuous communication with multiple BSs for accurate angle and distance data collection increases the frequency of data exchange, thereby amplifying the exposure of location data to potential eavesdropping. These systems are particularly vulnerable to cross-referencing attacks, where adversaries combine angular and distance information from multiple sources to pinpoint exact locations, heightening the risk of tracking and identity inference. As future networks become more interconnected and data-driven, the fine granularity of angle measurements poses a significant threat to location privacy, requiring advanced obfuscation and encryption methods to mitigate the risk of unauthorized access and exploitation of sensitive location data.

- **Area Measurement:** Area measurement localization relies on estimating a position by calculating the center intersection of overlapping coverage regions from multiple base stations. While this method offers improved precision by narrowing the target area, it presents significant location privacy challenges. As the precision increases with more restricted zones, more detailed location data is generated, inadvertently making the user's movements more traceable. The higher precision resulting from smaller, intersecting regions leads to the continuous generation of fine-grained trajectory data, which can be vulnerable to eavesdropping and tracking attacks if not adequately protected. Adversaries could exploit the precision of area measurements to reconstruct users' paths, potentially revealing sensitive locations such as home or work addresses. Moreover, in high-density network environments, where base stations are more abundant, the area measurement method provides even more detailed tracking opportunities, amplifying the risk of location inference and identity linkage.

- **Hop-count Measurement:** Hop-count measurement localization estimates the position of a target by analyzing the number of communication hops between UEs and BSs or other network nodes. While this method offers a simple and efficient way to determine location, especially in multi-hop wireless networks, it has significant location privacy limitations. The primary issue is that hop-count measurements require the continuous transmission of data across multiple nodes, each of which potentially logs and forwards location information. As the number of hops increases, so does the exposure of sensitive data, providing adversaries with more opportunities to intercept and analyze communications. Furthermore, hop-count measurement is susceptible to side-channel attacks, where attackers exploit temporal or traffic flow patterns to deduce users' trajectories over extended periods. In networks with high node density, the precision of hop-count localization improves, but this also increases the risk of granular tracking, as adversaries can reconstruct detailed movement paths by cross-referencing hop-count data across different network segments. Additionally, since hop-count localization typically lacks encryption and obfuscation, it is particularly vulnerable to eavesdropping and trajectory inference attacks, compromising users' location privacy in dynamic vehicular networks.

*2) Cooperative Localization:* While cooperative localization significantly improves positioning accuracy by leveraging Device-to-Device (D2D) communication, where UEs can measure distance and angular information directly, it introduces critical location privacy concerns in vehicular networks. The high precision achievable through closer proximity between UEs and the use of ultra-low latency D2D links increases the risk of privacy breaches. As more nodes are connected in 5G/6G-enabled networks, the vast amount of location data exchanged between UEs can be exploited by adversaries to reconstruct highly detailed movement trajectories [1]. The frequent and real-time communication between numerous devices makes it easier for unauthorized entities to intercept, monitor, or eavesdrop on D2D links, leading to significant vulnerabilities in location tracking. Moreover, while the high SNR improves position accuracy, it also increases the exposure of precise spatio-temporal data, which can be used to infer user behavior, monitor driving patterns, or even target specific individuals. Without robust privacy-preserving mechanisms in place, cooperative localization poses significant challenges in safeguarding location privacy due to the sheer volume of data generated and the fine granularity of location information shared among connected vehicles.

*3) AI-based Localization:* AI-based localization leverages machine learning to create highly detailed fingerprinting databases that store environmental channel parameters, such as CSI, to estimate vehicle positions based on reference points [222]. While this data-centric approach enhances localization precision in 5G/6G vehicular networks, it also introduces significant location privacy risks. The richness and granularity of the datasets required for AI-based localization, including real-time updates and integrating multi-layered environmental data, increase the potential for data leakage. Since machine learning models depend on extensive data collection and environmental profiling, adversaries could exploit this wealth of information to infer sensitive patterns, track vehicle movements, and even predict future trajectories. Additionally, the sheer volume of data generated by hierarchical coexistence, flexible storage, and flexible processing increases the risk of data breaches and unauthorized access. AI-based models often rely on centralized servers or cloud systems for training and processing, which can create vulnerabilities if data encryption and privacy-preserving measures are not robustly implemented. Furthermore, model inversion attacks—where adversaries attempt to reverse-engineer a machine learning model to extract sensitive information—pose a unique privacy challenge in AI-based localization, making it crucial to integrate advanced encryption and privacy-aware machine learning techniques to mitigate these risks.

| Localization | BS Density | Cost | Precision |
|---|---|---|---|
| Traditional | High | Low [226] | Low [227] |
| Cooperative | Low | Low [228] | Medium [229] |
| AI-based | Medium | High [230] | High [230] |
| Channel Charting | Depends | Depends | Depends |
| Driver Tracking | Medium | Low [231] | Medium [231] |
| Multipath Exploiting | Low | Medium [232] | High [233] |
| SLAM | Low | Depends | High [234] |
| Context-aware | High | Medium [235] | High [236] |

*4) Channel Charting:* Channel charting is an AI-driven localization technique that creates a virtual map based on CSI, allowing for the tracking of vehicles in vehicular net- works without relying on precise geographical coordinates. While this approach enhances data utility by offering real-time pseudo-locations as references, it poses significant location privacy challenges. Despite the fact that the generated virtual map does not reveal actual locations, the CSI-based tracking can still expose trajectory patterns, which adversaries can exploit for reconstructing user movements. The AI algorithms used in channel charting can also inadvertently learn and reveal correlations between the pseudo-locations and actual locations over time, especially as more data is gathered from repeated observations [237]. This increases the risk of location inference attacks, where attackers can link pseudo-locations with real-world identities, particularly in dense vehicular networks. Furthermore, the absence of robust encryption or obfuscation mechanisms within the channel charting process means that the privacy protections are weaker compared to traditional encryption-based methods, leaving the system vulnerable to data mining and pattern recognition techniques. Thus, while channel charting improves localization accuracy, its location privacy limitations must be addressed through enhanced LPPMs that prevent inference attacks and safeguard user anonymity.

*5) Driver tracking localization:* Driver tracking localization continuously infers the driver's position to smooth out the estimation errors [222]. This localization technique can predict the drivers' trajectories with the vehicle sensor data as follows.

. **Passive Sensing:** Passive sensing in 5G/6G vehicular net-works, also referred to as passive radar or passive coherent location, relies on detecting and processing reflected signals from targets, such as vehicles or pedestrians, to determine their locations [237]. While this method offers significant energy efficiency and non-intrusive tracking advantages, it poses substantial location privacy risks. The fundamental challenge with passive sensing is that it does not require active signal transmission from the target, making it dif-ficult for users to know when or how their location data is being collected. This raises concerns about involuntary surveillance, as the monitored targets often have no control over the captured data. In future sensor-rich environments, where dense sensing infrastructure will be prevalent, passive sensing will continuously gather location data, potentially without user consent, making it nearly impossible to prevent the leakage of sensitive information. Furthermore, aggre-gating high-resolution sensor data from multiple sources could enable adversaries to reconstruct detailed trajectories and identify individual movement patterns, compromising location privacy on a large scale.

. **Active Sensing:** Active sensing techniques, such as radar-based localization, are increasingly used in vehicular net-works for applications like adaptive cruise control and cross-traffic alerts, providing highly accurate 3D distance measurements. In 5G/6G-enabled vehicular networks, these sensors will share vast amounts of location-related data to support high-precision localization for real-time applica-tions. However, this data sharing poses significant location privacy challenges. The continuous exchange of precise sen-sor data across networks can lead to trajectory reconstruc-tion, where adversaries analyze the collected data to track vehicle movements in detail. Moreover, the real-time nature of active sensing increases data transmission frequency, thereby expanding the opportunities for eavesdropping and location inference attacks. The accuracy of 3D localization in such networks, while beneficial for safety and efficiency, paradoxically increases the risk of identity linkage and movement profiling, as even small discrepancies in vehicle positions can be exploited to identify individual users.

*6) Multipath-based Localization:* Multipath-based localiza-tion offers enhanced precision by leveraging multipath components that act as mirror reflections of a vehicle's surroundings, combining these with environmental geom-etry to disclose drivers' positions even in Non-Line-Of-Sight (NLOS) conditions [238]. While this method provides robust localization in complex environments like urban areas, it also introduces significant location privacy vul-nerabilities. Explaining reflected signals from various sur-faces enables multipath-based localization to capture more position-related data than traditional methods, leading to finer granularity in location tracking. In 5G/6G vehicular networks, where vehicles are equipped with multiple sensors emitting radar-like signals, the fusion of sensor data with multipath information further amplifies the risks of trajec-tory reconstruction and unauthorized tracking. The inherent limitation is that this technique continuously gathers detailed environmental data and radio reflections, making it easier for adversaries to collect and analyze side-channel information to infer exact vehicle positions. Moreover, the dense sensor networks in 5G/6G-enabled vehicular systems could make it increasingly difficult for traditional LPPMs to obfuscate such high-resolution data effectively, leaving users vulner-able to location inference attacks even in scenarios where direct line-of-sight tracking is not feasible.

*7) Simultaneous Localization And Mapping (SLAM):* Simultaneous Localization and Mapping (SLAM) is an advanced technique in 5G/6G vehicular networks, leveraging high data rates, ultra-low latency, and scalability to estimate vehicle positions and environmental landmarks in real time [239]. However, despite its benefits, SLAM presents significant location privacy challenges. In vision-based SLAM, which relies on cameras and sensors to capture the surroundings, the continuous collection and processing of detailed visual data can expose sensitive information about the environment and the vehicle's trajectory. This data can be intercepted or analyzed by adversaries to infer the precise location and movement patterns of vehicles, raising concerns about unauthorized surveillance and privacy breaches. Similarly, radar-based SLAM, which uses radio waves to detect objects and map environments, also generates precise location data that can be exploited for tracking and trajectory reconstruction. In both cases, the granular, real-time nature of SLAM data increases the risk of location inference attacks, especially in dense, highly connected vehicular networks.

In 5G/6G vehicular networks, the SLAM can be classified into vision-based SLAM and radar-based SLAM as follows.

- **Vision-based SLAM**: Vision-based SLAM relies on image sensors like cameras to detect landmarks and map the environment while tracking the vehicle's position. While this approach provides highly accurate localization in dynamic environments, it introduces substantial location privacy risks due to the nature of visual data collection. Cameras continuously capture images of the surroundings, including sensitive areas such as residential spaces or personal properties, which can inadvertently expose private information about users or bystanders. The data from these image sensors can be exploited by adversaries for location inference or even face and object recognition, leading to privacy breaches. Moreover, the storage and processing of large volumes of image data increase the risk of data leaks if robust encryption and anonymization mechanisms are not in place. The frequent need for real-time image processing also makes Vision-based SLAM vulnerable to cyberattacks, where adversaries could intercept and analyze visual data to reconstruct detailed trajectories and behavioral patterns.

- **Radar-based SLAM**: Radar-based SLAM, utilizing 3D LiDAR point clouds and laser sensors, offers higher localization accuracy than vision-based systems, making it a crucial component for future autonomous vehicular networks. However, the precision and richness of the 3D spatial data captured by these systems also introduce significant location privacy risks. LiDAR-generated point clouds provide detailed environmental mapping, which can easily be used to reconstruct a vehicle's exact movements and trajectories when combined with other vehicle and infrastructure data. This granularity in data exposes vehicles to continuous tracking, making it easier for adversaries to perform trajectory inference attacks and identify specific drivers or vehicles by analyzing movement patterns. Furthermore, the dense data generated by radar-based SLAM increases the likelihood of data breaches if proper encryption and privacy-preserving mechanisms are not implemented, as adversaries can exploit this detailed information for unauthorized tracking.

The entities in 5G/6G vehicular networks can collect location information with radar-like signals from other vehicles, fixed infrastructures, and sensors.

*8) Context-aware Localization:* Context-aware localization in 5G/6G-enabled vehicular networks leverages personal and public data to provide intelligent, multi-modal localization based on the driver's location and context [236]. While this approach enhances the accuracy and adaptability of localization by allowing drivers to dynamically switch communication channels and technologies depending on their surroundings, it introduces significant location privacy vulnerabilities. Integrating highly personal data (e.g., driving habits, frequently visited locations) with public data (e.g., traffic conditions, weather) creates a complex data environment where adversaries can exploit contextual information to infer sensitive details about the driver's behavior and movements. This is particularly problematic because context-aware localization involves continuously exchanging contextual signals through various channels, making it easier for attackers to conduct cross-layer tracking and data correlation attacks. Moreover, switching between communication technologies based on context can expose gaps in security protocols, leaving some channels more vulnerable to eavesdropping or message interception. The reliance on contextual data for localization also raises concerns about the long-term collection of personal information, which could be used to build comprehensive user profiles, further compromising location privacy.

### B. Limitations and Opportunities of LPPMs for Future Upper Layer Location Privacy Attacks

With the development of vehicular networks, there would be new challenges for the existing LPPMs. As discussed in Sections III and IV, the existing LPPMs mainly focus on upper-layer message-based tracking that the adversaries with other tracking techniques (i.e., sensing infrastructure-based, optical vision-based, vehicle driving log-based, and cellular radio-based) cannot be defended in the existing vehicular networks. Nevertheless, as shown in Table XI, the existing LPPMs could not provide acceptable location privacy protection under the upper-layer message-based attack scenario in 5G/6G vehicular networks.

*1) User-side LPPM:* Most of the existing user-side LPPMs cannot suit the low latency of the 5G/6G vehicular networks, i.e., pass-and-run, certificate, and secure computation. Although data perturbation can satisfy the latency requirement, the location data protected by data perturbation cannot provide high-precision LBSs for high-precision localization applications.

- **Pass-and-Run:** The pass-and-run communication model, which relies on vehicles LBSs through other nodes, introduces several limitations, particularly concerning location privacy. One key challenge is the high communication delay inherent in this method, as each vehicle forwards the data through intermediate nodes, increasing latency in delivering LBS requests. Additionally, identity verification within this multi-hop routing structure presents privacy risks. The need for legal cooperator authentication to ensure that data is routed securely can lead to increased communication delays, as each hop requires verification. This creates a scenario where frequent data exchanges between vehicles inadvertently improve the precision of signal-based localization techniques, such as basic localization and cellular radio-based localization. As vehicles communicate more frequently, adversaries may exploit the rich communication patterns to reconstruct granular trajectories and track user movements with higher accuracy, thus compromising the location privacy of drivers.

TABLE XI
CHALLENGE OF LOCATION PRIVACY IN 5G/6G VEHICULAR NETWORKS.

| Techniques | New challenges | LPPMs that are available | LPPMs that need to be improved |
|---|---|---|---|
| THz | Small coverage area<br>Spectrum penetration power<br>Centimeter-level precision via APs | Pass-and-run<br>Certificates for Privacy<br>Secure computation<br>Homomorphic encryption<br>Private information retrieval<br>Searchable encryption<br>Secure communication | Data perturbation<br>Statistical disclosure control<br>Trusted third party |
| VLC | Signal scatter<br>Observation signal<br>Effective transmission requirement | Certificates for Privacy<br>Secure communication<br>Trusted third party | Pass-and-run<br>Data perturbation<br>Secure computation<br>Statistical disclosure control<br>Private information retrieval<br>Homomorphic encryption<br>Searchable encryption |
| mmWave | Exchange CSI frequently<br>Low latency requirement<br>Eavesdrop channel easily | Certificates for Privacy<br>Secure computation<br>Data perturbation<br>Statistical disclosure control<br>Homomorphic encryption<br>Private information retrieval<br>Searchable encryption<br>Trusted third party<br>Secure communication | Pass-and-run |
| Sub-6 GHz | Flexible antenna design<br>Fake BS and malicious devices<br>In-vehicle tracking | Pass-and-run<br>Certificates for Privacy<br>Secure communication<br>Trusted third party | Secure computation<br>Data perturbation<br>Statistical disclosure control<br>Homomorphic encryption<br>Private information retrieval<br>Searchable encryption |
| Satellite Communication | Long propagation delay<br>Difficult to allocate MIMO<br>Communication resource allocation | Certificates for Privacy<br>Data perturbation<br>Statistical disclosure<br>Secure communication<br>Trusted third party | Pass-and-run<br>Secure computation<br>Homomorphic encryption<br>Private information retrieval<br>Searchable encryption |
| QC | Quantum channel<br>Quantum computer | Certificates for Privacy<br>Secure computation<br>Data perturbation<br>Statistical disclosure control<br>Secure communication<br>Trusted third party | Pass-and-run<br>Homomorphic encryption<br>Searchable encryption<br>Private information retrieval |

In 5G/6G-enabled vehicular networks, advanced sensor networks could mitigate some challenges by giving vehicles comprehensive knowledge of the road network and communication environment. With this information, drivers can select optimal channels and routing paths, potentially reducing transmission delays and enhancing system efficiency. However, the extensive use of sensors also raises concerns about data aggregation and privacy exposure, as multiple sensors across vehicles and infrastructure would continuously collect and transmit sensitive location data. Additionally, while blockchain technologies offer a promising solution for trust management, enabling decentralized authentication and secure certificate distribution, they come with high computational costs and potential scalability issues in managing vast networks of vehicles [240]. The transparency inherent in blockchain could also pose a privacy risk, as the immutability of the ledger may allow adversaries to trace vehicle interactions and infer user identities or trajectories over time. Thus, while blockchain can address trust issues, careful design is required to ensure that location privacy is not compromised by the verification and traceability mechanisms embedded in the system.

- **Certificates for Privacy:** Certificate-based LPPMs are widely used for authentication in vehicular networks by verifying certifications and signatures, ensuring secure communication between vehicles and network entities. However, these traditional certificate-based LPPMs face significant limitations in 5G/6G-enabled vehicular networks. One major limitation is the reliance on a trusted CA, which introduces centralization into a system that is otherwise highly decentralized and dynamic. In a large-scale vehicular network where thousands of vehicles and sensors communicate simultaneously, verifying certificates from a central authority can create bottlenecks, leading to unacceptable delays in real-time applications such as traffic management and accident prevention. Moreover, a single trusted authority presents a single point of failure, making the system vulnerable to attacks that can compromise the entire network's privacy.

  Another significant limitation of certificate-based LPPMs is the storage consumption associated with managing certificates across many nodes. In 5G/6G vehicular networks, the sheer number of vehicles, RSUs, and sensors generating and exchanging data will increase the complexity and storage requirements for certificate management. This could strain network resources and reduce the efficiency of communication. Additionally, as each certificate needs to be authenticated, the authentication delay will grow with the network size, negatively impacting the latency-sensitive services typical of vehicular networks. While blockchain technology offers potential solutions by providing immutable, decentralized, and secure consensus mechanisms for managing certificates, it introduces its challenges, such as increased computational costs and the energy consumption of maintaining the blockchain. Furthermore, collecting sensor data for authentication purposes raises privacy concerns, as adversaries could exploit this data for location tracking or behavioral analysis.

- **Secure Computation:** While secure computation techniques offer robust solutions for protecting location privacy in vehicular networks by allowing the processing of encrypted data without revealing the actual content, they suffer from significant computation delays. This delay is a major limitation, particularly in real-time vehicular environments where low-latency communication is critical for applications such as navigation, collision avoidance, and traffic management. The computational overhead associated with homomorphic encryption and other secure computation methods often results in increased processing time, which can lead to latency issues and affect the overall performance of the system [130]. In the context of 5G/6G-enabled vehicular networks, where data is generated and transmitted at extremely high rates, the current secure computation techniques may not scale efficiently, further exacerbating the time delays and limiting their practicality. Additionally, as vehicular networks evolve to handle more complex cross-layer tracking and big data analytics, the processing de-mands on secure computation methods are expected to grow, posing even greater challenges for maintaining a balance between location privacy and system efficiency. Therefore, while secure computation offers strong privacy guarantees, its limitations in computation delay and scalability necessitate further improvements to meet the real-time demands of modern vehicular networks.

- **Data Perturbation:** While data perturbation is a flexible and low-delay method for location privacy preservation, it faces significant limitations in the context of 5G/6G vehicular networks. Current data perturbation techniques typically provide meter-level precision, which may suffice for general applications but falls short of the centimeter-level or millimeter-level accuracy required in 5G/6G-enabled vehicular systems. As vehicular networks evolve, precise location data is crucial for critical services such as autonomous driving and collision avoidance, making any significant perturbation detrimental to data utility. Attempts to refine perturbation for higher precision risk significantly weaken the privacy-preserving capability of these mechanisms [15]. Specifically, perturbing data to such fine levels would allow adversaries to reverse-engineer the original positions with ease, thereby compromising location privacy. This tradeoff highlights the inadequacy of current data perturbation-based LPPMs for future networks. To address this, data perturbation must be combined with more robust privacy-preserving techniques, such as adaptive noise injection or multi-layered encryption, which can maintain high privacy protection while meeting the stringent accuracy demands of future vehicular applications.

*2) Server-Side LPPM:* The low latency requirement also challenges the existing server-side LPPMs, as the process on the server side could cause computation delays.

- **SDC:** SDC in vehicular networks relies on server-side LPPMs like data perturbation, secure computation, and anonymity techniques to mitigate the risk of sensitive location data exposure [30]. However, the same limitations faced by user-side LPPMs—such as computation cost, storage overhead, and data management complexity—also apply on the server side, and these challenges become magnified in the context of 5G/6G-enabled vehicular networks. As data transmission frequency and data volumes grow exponentially, driven by the ultra-dense connectivity of 5G/6G networks, the demand for processing and storing this massive influx of data significantly strains computational resources and storage infrastructures. This high-frequency data exchange, if not carefully controlled, can lead to vulnerabilities in location privacy, as adversaries may exploit frequent communications and large datasets to perform trajectory reconstruction and profile users over time.

  Although blockchain-based techniques offer the potential for optimizing data management and reducing storage consumption by decentralizing and securing data transactions, they do little to alleviate the computational overhead required to process location data for privacy protection. For instance, the computational cost of encryption, anonymization, and con-tinuous statistical disclosure control grows with the volume of data and the complexity of real-time localization services. Moreover, AI-based LPPMs hold promise in automating adaptive privacy protection and streamlining data analysis, but they introduce additional complexities in model training and resource allocation, especially when applied to real-time, high-volume data streams [241]. Furthermore, the introduction of trust management systems can help mitigate computational costs by bypassing unnecessary privacy protections for trusted entities, but these systems themselves require careful design to avoid trust exploitation or insider attacks.

- **Homomorphic Encryption:** HE is a widely used method for secure computations in preserving location privacy, allowing operations to be performed on encrypted data without revealing the plaintext. However, its use in server-side computations faces significant challenges due to its computational delay and high resource consumption. The complexity of HE operations, especially in FHE schemes, leads to heavy computational overhead during encryption, decryption, and processing, making it impractical for real-time applications in vehicular networks where low latency is critical. Moreover, the communication consumption in transferring large encrypted datasets can further exacerbate network congestion, especially in 5G/6G-enabled environments, where data transmission rates are high. These limitations create a balance between ensuring strong location privacy and maintaining system efficiency. To mitigate these issues, it is essential to develop simplified HE processes that reduce the computational burden while maintaining an adequate level of privacy protection. Techniques such as PHE or combining HE with edge computing could help alleviate the performance bottlenecks associated with fully encrypted computations, but they still fall short in addressing the scalability required for large-scale real-time vehicular networks.

- **Private Information Retrieval:** PIR is a promising technique for ensuring location privacy by allowing users to retrieve data from servers without revealing which data is being accessed. However, its practical implementation in vehicular networks faces significant challenges, particularly in terms of computational consumption and scalability. The high computational cost associated with processing PIR queries in the cloud makes it difficult to allocate the necessary resources efficiently in current networks. In 5G/6G-enabled vehicular networks, which have strict low-latency requirements for real-time applications such as autonomous driving and traffic management, the high overhead of PIR could result in unacceptable delays. Additionally, the increased data volume and frequency of queries in these networks exacerbate the problem, making the deployment of current PIR solutions impractical. While PIR provides a robust framework for enhancing location privacy, its computational inefficiency and resource demands pose significant limitations, particularly when it comes to scaling in high-density vehicular environments where fast data retrieval and real-time communication are critical.

- **Searchable Encryption:** SE provides a method for pro-

tecting location privacy by enabling drivers to search LBSs without revealing sensitive data, but it presents several limitations, particularly in 5G/6G vehicular networks. As these networks demand high-precision location data for real-time services such as navigation and traffic management, SE schemes often introduce accuracy errors in search results, which can degrade the data utility for users. Although combining SE with other LPPMs, such as certificate-based encryption or homomorphic encryption, can improve search accuracy and privacy, this integration significantly increases computational overhead and processing time, particularly in large-scale vehicular networks. The heightened computational burden may lead to delays in delivering real-time services, which is unacceptable for critical vehicular applications where low latency and high efficiency are essential. Furthermore, the complexity of managing cryptographic keys and ensuring secure query processing across distributed vehicular networks increases the risk of vulnerabilities, potentially exposing location data if any aspect of the system is compromised.

Information from sensors in 5G/6G vehicular networks could be referred to correct the result, as massive sensors would be allocated. Also, AI-based LPPMs can be introduced to optimize the result according to the driver's historical information, which would be fine-grained [11].

*3) User-Server-Interface LPPM:* The user-server-interface LPPMs would attract more researchers' attention in 5G/6G vehicular networks than in the current version because 5G/6G vehicular networks would ask the vehicles to share data through frequent V2X communication. Protecting location privacy in channels would be an approach to defend basic localization, channel charting localization, and driver tracking localization.

- **Secure Communication:** While secure communication through end-to-end encryption offers robust protection for location privacy in vehicular networks, it presents several limitations, particularly in the context of 5G/6G-enabled networks. The primary challenge lies in the computational overhead associated with encrypting and decrypting large volumes of real-time location data, which can significantly increase latency in communication. As 5G/6G vehicular networks are expected to handle high data rates and require low-latency communication for critical services like autonomous driving and real-time traffic management, the added delay from complex encryption protocols could degrade the data utility of LBSs. Additionally, the use of end-to-end encryption is currently rare in vehicular networks due to these performance constraints. As encryption techniques must secure sensitive location data at various points in transmission, including between vehicles, infrastructure, and cloud servers, they risk becoming a bottleneck, particularly in dense urban environments where data exchange is frequent and real-time response is essential. Moreover, encrypted communication, while protecting data confidentiality, can still be vulnerable to side-channel attacks where adversaries can infer user location through traffic analysis without needing to break the encryption directly.

- **Trusted Third Party (TTP):** Using a TTP in current vehicular networks is often considered ideal for managing location privacy, but it faces significant limitations, particularly in the context of 5G/6G vehicular networks. As vehicular networks expand with the integration of massive sensors and connected vehicles, the volume of data generated grows exponentially, increasing the risks of a TTP being compromised by adversaries. Even TTPs controlled by government entities are vulnerable to hijacking or data breaches, as evidenced by large-scale incidents involving the theft of sensitive data from state-controlled databases [3]. A compromised TTP in such a vast network could result in widespread privacy violations, exposing users' locations and movements to malicious actors. Furthermore, the sheer scale and complexity of future networks—characterized by the high number of sensors, vehicles, and nodes—challenges the efficiency and reliability of a centralized TTP model. The reliance on a single entity to manage trust across such a large network creates a single point of failure. To address these limitations, the TTP model must be enhanced by integrating additional privacy-preserving mechanisms such as encryption, blockchain, AI-based techniques, and data perturbation. These complementary approaches can decentralize trust management, reduce the dependency on a single entity, and improve the overall resilience of privacy protection strategies in 5G/6G-enabled vehicular networks.

## C. Location Privacy Challenges and Emerging Wireless Technologies

Different from the existing vehicular networks, the future version transmits data through new communication mediums. The new communication mediums improve the efficiency of vehicular networks but bring new cross-layer location privacy issues.

- **Sub-6 GHz:** The Sub-6 GHz spectrum, ranging from 0.45 GHz to 6 GHz, is widely adopted in vehicular networks due to its broad coverage, low cost, and ability to support V2X communications. However, the use of sub-6 GHz introduces several location privacy challenges. One key issue arises from the flexible antenna design, which enhances communication capabilities and makes it easier for adversaries to track vehicles. By leveraging malicious devices such as fake BSs and monitoring multiple sub-6 GHz links, attackers can gather distance and angular data to accurately calculate a vehicle's position [242]. This form of passive tracking significantly compromises the privacy of vehicles, as adversaries can exploit the open wireless nature of sub-6 GHz channels to obtain granular trajectory data. Additionally, malicious applications installed on drivers' mobile devices can amplify privacy concerns, as the location data from mobile phones often mirrors that of the vehicles, particularly when drivers carry their phones while driving [103].

These challenges necessitate stronger physical-layer LPPMs to mitigate the risks posed by the sub-6 GHz spectrum. Existing data perturbation and statistical disclosure control methods introduce new complexities in detecting illegal entities, making it harder to authenticate legitimate users and block eavesdropping attempts. While certificates and trusted third parties can be employed to identify and authenticate legal entities, these mechanisms may not fully address the stealthy nature of fake BSs and malicious apps that operate within the sub-6 GHz band. In particular, attackers can use signal triangulation techniques across multiple BSs to reconstruct precise vehicle locations, exposing vehicles to continuous tracking. The low latency and ubiquity of sub-6 GHz communications further exacerbate these privacy risks by allowing adversaries to intercept and analyze real-time vehicle data.

To safeguard vehicle location privacy within the sub-6 GHz spectrum, more advanced LPPMs that incorporate user-server interface and server-side privacy protections are necessary. These methods, in conjunction with AI and blockchain technologies, can provide enhanced privacy by ensuring that data is encrypted, obfuscated, or blocked from unauthorized entities before it is transmitted. For example, AI algorithms could detect anomalous behavior from malicious devices, and blockchain could provide a decentralized, tamper-proof system for verifying the authenticity of BSs and mobile devices. Additionally, drivers could hide semantic information in their location data or employ end-to-end encryption to prevent adversaries from reconstructing their movement patterns through sub-6 GHz channels.

- **mmWave:** In 5G/6G vehicular networks, mmWave technology, which operates in the 30 GHz to 100 GHz spectrum and offers high multi-gigabit transmission speeds, presents significant challenges for vehicle location privacy. The beam-based directional transmission of mmWave allows for precise communication, but this very precision increases the risk of location tracking. Due to the high frequency and narrow beam, the transmission patterns of mmWave signals can be easily mapped, allowing adversaries to detect the direction and origin of transmissions and, thus, infer the location of the vehicle [243]. In particular, in environments with dense BSs and frequent vehicle movement, the continuous exchange of CSI becomes necessary to maintain network connectivity, which inadvertently provides a wealth of data that could be exploited for location-based attacks. Moreover, the highly dynamic topology of 5G/6G vehicular networks, driven by frequent changes in vehicle positions, exacerbates privacy risks. The need for low-latency V2X communication forces vehicles, RSUs, and sensors to frequently exchange CSI, which increases the communication overhead and leaves more opportunities for adversaries to eavesdrop on the communication and extract location information [244]. While data perturbation techniques could be employed to anonymize location data, they are often insufficient when the communication channels themselves are vulnerable. Eavesdropping on mmWave signals, particularly in high-mobility scenarios like highways or urban areas, allows adversaries to triangulate vehicle positions through repeated signal observation, even if the actual content of the message is encrypted [245]. This increases the granularity of tracking in vehicular networks, heightening the need for robust LPPMs.

One promising solution to mitigate these risks lies in leveraging the Doppler shift inherent in mmWave transmissions caused by the high relative speeds of vehicles. By combining the Doppler effect with data perturbation and anonymity techniques, the transmitted signal characteristics can be altered in ways that obfuscate location data, making it more difficult for adversaries to infer real-time positions from intercepted communications [246]. Additionally, advanced secure computation techniques such as homomorphic encryption, PIR, and searchable encryption can be deployed to prevent adversaries from accessing useful data even if they manage to intercept the signal. These cryptographic methods allow vehicles to perform computations on encrypted data or retrieve information from the network without revealing their exact location or identity. However, such techniques introduce their own challenges, as they often increase computational and communication delays, which may not be compatible with the low-latency demands of mmWave-based vehicular networks.

. **Terahertz (THz) communication:** THz communication introduces both opportunities and challenges for vehicular networks, particularly concerning vehicle location privacy. With a spectrum range from 100 GHz to 10 THz and extremely short wavelengths (30–3000 $\mu$m), THz communication offers the potential for high-speed data transmission and centimeter-level localization precision [247]. However, the small coverage area of less than 50 meters and limited penetration power of THz waves create significant privacy concerns. The precise localization capabilities in THz-based vehicular networks allow adversaries to track vehicle movements with unprecedented accuracy, posing severe risks to location privacy. Even though the limited range of THz signals may appear to enhance privacy by restricting communication to shorter distances, the high data resolution and frequent handovers between base stations or APs could still expose fine-grained location data, enabling adversaries to reconstruct continuous trajectories of vehicles. The small THz coverage area increases the likelihood of relying on multiple APs for seamless communication. These APs, functioning as trusted third parties, are required to process large volumes of vehicle data, including whereabouts, driving patterns, and road conditions, which significantly heightens the risk of location privacy breaches. The centralized nature of these APs, while essential for maintaining network connectivity, could become a target for cyberattacks or data interception, as adversaries may exploit vulnerabilities in APs to gain access to sensitive location information. Certificate-based LPPMs, which rely on cryptographic credentials, will likely become more prevalent in THz vehicular networks to address trust issues, but these solutions introduce management overhead and computational complexity, particularly in high-density urban environments with frequent handovers [248].

In addition to the trust management and cryptographic challenges, existing LPPMs, such as DP and data perturbation techniques, may not be effective in THz-based communication. DP-based LPPMs on the user side, which rely on adding controlled noise to location data, could struggle to maintain privacy in high-precision THz networks due to the limited number of perturbation candidates. As THz networks provide centimeter-level localization, even small perturbations may not sufficiently obscure a vehicle's true location, allowing adversaries to infer movement patterns with high accuracy [16]. This limitation suggests a shift towards server-side DP-based LPPMs, where more advanced privacy techniques, such as context-aware obfuscation and blockchain-enhanced trust management, could better protect location privacy without sacrificing data utility.

. **Visible Light Communication (VLC):** VLC in vehicular networks presents both promising opportunities and significant challenges to location privacy. VLC employs Light-Emitting Diode (LED) installations, such as traffic lights, vehicle headlights, and roadside lights, to transmit data, offering low energy consumption, high efficiency, and access to a large spectrum range from 400 THz to 800 THz [249]. In vehicular environments, VLC can facilitate V2X communication, improving data transfer between vehicles and infrastructure. However, despite its advantages, VLC is inherently vulnerable to location privacy breaches due to the physical characteristics of light transmission. Unlike radio waves, light signals are prone to scattering, especially in outdoor environments, which makes it possible for adversaries to eavesdrop on the transmitted signals. Even when advanced techniques such as Non-Orthogonal Multiple Access (NOMA) and pseudo surface waves are applied to mitigate signal attenuation, they cannot entirely eliminate the risk of signal interception [250].

The outdoor deployment of VLC exacerbates the location privacy challenges because eavesdroppers can easily intercept the scattered light signals. Since VLC relies on line-of-sight communication, the physical visibility of the transmitting source (such as vehicle lights) allows adversaries to infer location data by simply observing or intercepting the scattered signals. This presents a unique threat to vehicular location privacy, as eavesdroppers can potentially track vehicle movements without sophisticated hacking methods. Traditional LPPMs, such as data perturbation or statistical disclosure control, become ineffective in these scenarios because the adversary can directly observe the light source, thereby bypassing server-side protections [251]. The inherent physical visibility of VLC signals limits the scope of digital obfuscation techniques, exposing vehicle locations to real-world tracking threats.

To address these challenges, certification-based authentication methods and user-server-interface LPPMs offer better location privacy protection in VLC systems. In short-range VLC, certification-based methods can effectively block unauthorized receivers from accessing the communication channel by ensuring that only certified devices participate in the data exchange. This can help mitigate the risk of eavesdropping by restricting the number of nodes capable of receiving VLC signals [251]. However, in long-range or remote communication scenarios, additional measures are necessary. Techniques like cooperative jamming and CSI estimation are viable physical-layer LPPMs that can prevent eavesdroppers from obtaining sensitive location data [252]. Cooperative jamming introduces interference signals from multiple legitimate nodes, thereby blocking the eavesdropper's ability to access the communication channel. Meanwhile, CSI estimation allows legal nodes to monitor the state of both legitimate and eavesdropping channels, dynamically adjusting the transmission to protect location information.

- **Quantum Communication (QC):** The QC utilizes the quantum states of lights, which achieves secure communication by using microscopic particles to carry quantum information.

   While Quantum Key Distribution (QKD) provides unparalleled security in QC by leveraging quantum mechanics principles such as Heisenberg's uncertainty principle and the no-cloning theorem, it presents significant challenges in protecting vehicle location privacy within vehicular networks [253], [254]. Although QKD ensures eavesdropping detection and protects the secrecy of the cryptographic keys used in communication, it does not inherently safeguard location privacy. Specifically, the laser signals used in QKD-based communication protocols can be exploited to infer the direction and relative positions of both the source and destination vehicles. In vehicular networks, where continuous real-time communication is essential for services like navigation and collision avoidance, the trajectory and location of vehicles may be exposed through analysis of the communication patterns and the physical properties of the quantum signals being transmitted.

   This limitation creates a significant privacy risk, as adversaries could use side-channel attacks to trace vehicles' locations based on the timing, direction, and intensity of the laser signals used in QKD. While QKD is highly effective for protecting data integrity and preventing eavesdropping, it lacks the location anonymization mechanisms necessary to prevent vehicle tracking in high-precision environments. To mitigate these risks, researchers have proposed advanced quantum cryptographic methods like quantum homomorphic encryption, quantum searchable encryption, and quantum private information retrieval. These technologies allow encrypted data to be processed, searched, and retrieved without revealing the actual contents or query information, addressing the data privacy concerns of QC. However, these methods do not fully address the specific location privacy challenges QKD poses in vehicular networks, where physical signal leakage may still compromise location confidentiality.

   Introducing Quantum Teleportation (QT) into vehicular networks presents significant challenges to vehicle location privacy, especially when long-range communication is required. QT facilitates the transmission of quantum information by utilizing both classical and quantum channels, ensuring the destruction of the sender's message during transmission. While this method holds promise for secure communication in high-density vehicular networks, it simultaneously disrupts traditional LPPMs. For instance, routing-based techniques like pass-and-run, which rely on intermediate nodes to transmit location information securely, are rendered ineffective in quantum communication settings, as QT bypasses the need for such routing mechanisms by directly transferring quantum data through entangled quantum states [255]. This creates a new paradigm where traditional hop-based encryption or obfuscation techniques cannot be applied, leaving vehicle location data vulnerable. The core privacy challenge with QT in vehicular networks lies in the quantum measurement process. Once quantum information is teleported, the original message is destroyed, and any interference with the transmission—such as eavesdropping or data interception—would result in quantum decoherence, immediately alerting both parties. However, the reliance on quantum entanglement for communication across vast distances could expose location data through quantum channel vulnerabilities or side-channel attacks if adversaries gain access to the classical channel accompanying the quantum data transfer. Furthermore, the potential use of quantum repeaters to extend the communication range poses risks of data leakage, as adversaries might exploit the classical channel to infer real-time location information by monitoring the flow of traditional data associated with the teleportation process. These issues suggest that existing LPPMs, tailored to classical communication frameworks, will need substantial adaptations or replacements to ensure location privacy in a quantum communication landscape.

The introduction of Quantum Identity Authentication (QIA) and Quantum Signature (QS) into vehicular networks presents new opportunities for secure communication but also introduces significant challenges to vehicle location privacy. QIA leverages the principles of quantum cryptography to provide certification-based objective trust, which enhances security by making it virtually impossible for adversaries to forge identities or eavesdrop on quantum channels. However, the increased precision and granularity of quantum communications, combined with certifiable evidence required for objective trust, pose potential risks to location privacy. Each authentication instance generates detailed data about the vehicle's identity and communication endpoints, which, if not properly managed, could be exploited for location tracking and trajectory inference. Adversaries could leverage quantum-secure channels to collect certifiable location data, linking it to real-time communications and compromising the anonymity of vehicles and drivers.

QS, designed to ensure the integrity and authenticity of messages, also introduces potential vulnerabilities. While QS offers subjective trust through TTP-based groups, where group-specific behaviors and characteristics are leveraged, the reliance on group-based trust models could inadvertently expose location data. For instance, TTPs managing subjective trust might aggregate and process significant amounts of location-related metadata to ensure authenticity, which could lead to statistical disclosure risks if patterns in vehicular movements are detected. The combination of quantum signatures and statistical disclosure control could allow adversaries to analyze the frequency and context of communication patterns, identifying potential weak points in privacy protections. Therefore, while QIA and QS offer enhanced trust management and security, they also necessitate the development of more advanced LPPMs to counteract the growing threats posed by quantum-level data collection and ensure the anonymity and location privacy of vehicles in these increasingly secure yet transparent communication environments.

. **Reconfigurable Intelligent Surface (RIS):** The deployment of Reconfigurable Intelligent Surface (RIS) in vehicular networks introduces novel challenges to vehicle location privacy. While RIS, with its metamaterials and passive reflect arrays, can enhance wireless communication by manipulating electromagnetic waves and offering full-duplex and full-band communication, its widespread use in vehicle-road-human integrated networks raises significant privacy concerns [256]. The dynamic and programmable nature of RIS allows for fine-grained control over signal reflections, potentially enabling precise tracking of vehicles by leveraging the reflections and interactions between vehicles and the environment. As RIS can be easily installed on building facades and billboards near roadways, this pervasive infrastructure could be exploited by adversaries to infer vehicle trajectories through continuous monitoring and signal triangulation. The low computational and energy consumption of RIS facilitates large-scale deployment, which further increases the surface area for tracking attacks. Additionally, the mobility of vehicles and the frequent disconnections in vehicular networks complicate the ability to secure location data, as RIS cannot easily adapt to the dynamic topology or gather reliable feedback from vehicles to implement effective privacy-preserving measures [257].

*D. Challenges Arising from Networks Convergence*

In the integrated vehicular networks, the vehicles can exchange information and communicate with other vehicles, roadside infrastructure, and pedestrians automatically through real-time V2X communications. With the support of V2X technology, traffic information (e.g., vehicle status, live road conditions, and pedestrian information) enable the formation of the integrated vehicle-road-human network [3]. The components with 5G/6G characteristics in the integrated vehicle-road-human network are illustrated as follows.

. **Integrated Satellite and Terrestrial Network (ISTN):** The ISTN in 5G/6G vehicular networks introduces several challenges to vehicle location privacy, particularly due to the unique characteristics of satellite communication. While satellite communication enables 3D vehicular services by extending coverage to remote and hard-to-reach areas, it faces significant limitations such as long propagation delays, high transmission latency, and challenges in efficiently managing Multiple-Input Multiple-Output (MIMO) networks and communication resource allocation [3]. These limitations not only affect the quality of service but also increase the vulnerability of location privacy by exposing more detailed trajectory data to adversaries over extended periods. The increased delay in transmission, combined with the distributed nature of satellite networks, makes real-time privacy protection more difficult, as traditional LPPMs may not be optimized for such environments.

Moreover, the pass-and-run method, which relies on quick transmission through neighboring vehicles, is less effective in satellite networks due to the inherent transmission delay and communication lag in satellite systems. Similarly, advanced LPPMs like homomorphic encryption and secure communication protocols, which are already computationally intensive, become even more challenging in ISTN due to the increased latency and resource constraints [258]. These factors significantly limit the scalability and efficiency of current privacy mechanisms. To mitigate these challenges, the recourse consumption of these methods could be offloaded to trusted third parties, where data perturbation techniques, such as adding noise to location data, and statistical disclosure control mechanisms could be used to better protect location privacy. By reducing the reliance on real-time computation and shifting towards more batch-processing privacy strategies, it may be possible to effectively balance the need for privacy with the high latency of satellite-based communications, providing a more robust defense against location privacy breaches in ISTN-enabled vehicular networks.

. **Human Interaction:** Human interaction in vehicular networks, particularly through mobile devices carried by pedestrians, drivers, and passengers, introduces significant location privacy challenges. As mobile devices bring enhanced communication and data processing capabilities, frequent vehicle-to-human interactions are becoming a core part of 5G/6G vehicular networks, allowing vehicles to gather real-time data and overcome sensor limitations like blind spots [259]. While this data-sharing enables critical services such as situation reporting and early warnings to reduce accidents [260], it also amplifies the risk of cross-layer privacy attacks. The frequent communication between vehicles and pedestrians can lead to trajectory inference, where adversaries combine encounter information from different layers (e.g., mobile data and vehicle logs) to track vehicles with greater precision. Additionally, in-vehicle sensor data, including driving logs and passenger information, can be leveraged to infer the vehicle's location or movements. This makes trust management-based LPPMs essential to protect against adversaries who exploit these multi-source data interactions to localize drivers and compromise their privacy. As human interaction grows in vehicular ecosystems, these risks necessitate more robust cross-layer privacy protections to safeguard against increasingly sophisticated tracking methods.

*E. Key Takeaway*

The exploration of location privacy challenges and protective mechanisms in 5G/6G vehicular networks has revealed critical gaps in existing LPPMs and underscored the complexities introduced by advanced communication technologies and integrated systems. The convergence of multi-modal communication platforms such as satellite-terrestrial networks, V2X systems, and RIS amplifies the precision and scale of localization capabilities, consequently exposing vehicles to novel privacy threats. These threats are compounded by the ultra-low latency, high data rates, and extensive device interconnectivity characteristic of 5G/6G-enabled vehicular environments.

One of the primary challenges lies in addressing the cross-layer privacy vulnerabilities introduced by integrating sensing infrastructure, optical vision, and channel-based tracking techniques. Existing LPPMs often focus on a single-layer threat model, rendering them insufficient in defending against multi-source data correlation and trajectory reconstruction attacks. Additionally, while technologies such as homomorphic encryption and quantum communication offer promising solutions, their computational demands and latency constraints often conflict with the stringent real-time requirements of vehicular applications.

Key observations from this section include:

- **Scalability and Adaptability:** Current LPPMs lack the scalability required to handle the exponential growth in vehicular communication traffic and connected devices. Future solutions must incorporate decentralized frameworks, such as blockchain, and adaptive mechanisms that dynamically adjust privacy protections based on network conditions and application requirements.
- **Integration of Advanced Technologies:** Emerging technologies, including quantum communication, AI-based localization, and RIS, introduce new dimensions to location privacy. However, the integration must be carefully managed to prevent the creation of unintended vulnerabilities, such as adversarial attacks on machine learning models or quantum-side channel exploits.
- **Cross-Layer Threat Mitigation:** Future 5G/6G vehicular networks demand holistic privacy strategies that address cross-layer interactions. Multi-layered LPPMs capable of simultaneously obfuscating user-side data, securing server-side processes, and protecting communication interfaces are essential to mitigate privacy risks comprehensively.

## VI. Conclusion

In this paper, we have comprehensively investigated the development of LPPMs and the threats to location privacy in vehicular networks. By reviewing the localization methods in vehicular networks, we have illustrated the threats to location privacy introduced by localization and classified LPPMs into user-side, server-side, and user-server-interface LPPMs. Our analysis has evaluated the performance of existing LPPMs under various localization and communication scenarios and provided methods for balancing location privacy preservation and data utility to improve their effectiveness.

In addition, we have identified potential threats introduced by new communication technologies in future vehicular networks that have been overlooked by existing studies. Our findings provide potential directions for the existing LPPMs to assist future research, which is important for LPPMs design. By considering the challenges and opportunities related to location privacy in vehicular networks, we can develop effective LPPMs that balance data utility and privacy protection to enhance the security and safety of drivers and passengers.